\documentclass[12pt,leqno]{article}
\title{Anomalous diffusion and collapse of self-gravitating \\
 Langevin particles in $D$ dimensions}

\usepackage{amsthm,amsmath,amssymb,a4wide,psfig,epsf}
cm \hsize=18 true cm \topmargin=-2cm \oddsidemargin=-0.5cm
\def\mb#1{\setbox0=\hbox{$#1$}\kern-.025em\copy0\kern-\wd0
\kern-0.05em\copy0\kern-\wd0\kern-.025em\raise.0233em\box0}

\begin{document}

\author{Pierre-Henri Chavanis and Cl\'ement Sire}
\maketitle
\begin{center}
Laboratoire de Physique Th\'eorique (FRE 2603 du CNRS), Universit\'e
Paul Sabatier,\\ 118, route de Narbonne, 31062 Toulouse cedex 4, France\\
E-mail: {\it chavanis{@}irsamc.ups-tlse.fr  ~\&~
clement.sire{@}irsamc.ups-tlse.fr}

%\date{}
\vspace{0.5cm}
\end{center}

\begin{abstract}

We address the generalized thermodynamics and the collapse of a system
of self-gravitating Langevin particles exhibiting anomalous diffusion
in a space of dimension $D$. This is a basic model of stochastic
particles in interaction. The equilibrium states correspond to
polytropic configurations similar to stellar polytropes and polytropic
stars. The index $n$ of the polytrope is related to the exponent of
anomalous diffusion. We consider a high-friction limit and reduce the
problem to the study of the nonlinear Smoluchowski-Poisson system.  We
show that the associated Lyapunov functional is the Tsallis free
energy. We discuss in detail the equilibrium phase diagram of
self-gravitating polytropes as a function of $D$ and $n$ and determine
their stability by using turning points arguments and analytical
methods. When no equilibrium state exists, we investigate self-similar
solutions of the nonlinear Smoluchowski-Poisson system describing the
collapse. Our stability analysis of polytropic spheres can be used to
settle the generalized thermodynamical stability of self-gravitating
Langevin particles as well as the nonlinear dynamical stability of
stellar polytropes, polytropic stars and polytropic vortices.

\end{abstract}

\section{Introduction}
\label{sec_introduction}

In preceding papers of this series \cite{csr,crs,sc,sc2}, we have
studied a model of self-gravitating Brownian particles enclosed
within a spherical box of radius $R$ in a space of dimension $D$.  This
model can be considered as a prototypical dynamical model of
systems with long-range interactions possessing a rich
thermodynamical structure. For simplicity, we considered the
Smoluchowski-Poisson (SP) system which is deduced from the
Kramers-Poisson (KP) system in a high friction limit (or for large
times). These Fokker-Planck-Poisson equations were first proposed
in \cite{csr} as a simplified dynamical model of self-gravitating
systems. Their relation with thermodynamics (first and second
principles) was clearly established in terms of a maximum entropy
production principle (MEPP) and their rich properties
(self-organized states or collapse) were described qualitatively
in this preliminary work. A thorough study of this system of
equations was undertaken more recently in \cite{crs,sc,sc2},
complemented by rigorous mathematical results (see
\cite{herrero,dolbeault,rosier,biler,guerra} and references
therein).

The Smoluchowski equation is a particular Fokker-Planck equation
involving a diffusion and a drift \cite{risken}. In our model, the
drift is directed toward the region of high densities due to the
gravitational force which is generated by the particles
themselves. This retroaction leads to a situation of collapse when
attraction prevails over diffusion \cite{crs,sc,sc2}. The SP system also
provides a simplified model for the chemotactic aggregation of
bacterial populations \cite{murray} and for the formation of
large-scale vortices in two-dimensional hydrodynamics
\cite{rsmepp,drift,kineuler,kin,houches,ch}. It can be shown
that the SP system decreases continuously a free energy constructed
with the Boltzmann entropy
\cite{csr,crs}. Accordingly, the stationary solutions of the SP
system are given by the Boltzmann distribution which minimizes
Boltzmann's free energy at fixed mass. The equilibrium state is then
determined by solving the Boltzmann-Poisson (or Emden) equation, as in
the case of isothermal gaseous stars and isothermal stellar systems
\cite{chandra,bt}. However, depending on the value of the temperature $T$, 
an equilibrium solution does not always exist and the system can
undergo a catastrophic collapse.  We determined analytically and
numerically self-similar solutions leading to a finite time
singularity \cite{crs,sc}. In \cite{sc,sc2}, we showed that the
evolution continues after the collapse until a Dirac peak is formed.

In this paper, we propose to extend our study to a generalized class
of Smoluchowski equations \cite{Ctsallis}.  They can be obtained from
the familiar Smoluchowski equation by assuming that the diffusion
coefficient depends on the density while the drift coefficient is
constant (or the opposite). They can also be obtained from standard
stochastic processes by considering a special form of multiplicative
noise \cite{borland,Ctsallis} or an extended class of transition
probabilities \cite{kaniadakis,Ctsalliskin}. These equations are
consistent with a generalized maximum entropy production principle
\cite{Ctsallis}.  For simplicity, we shall assume that the diffusion
coefficient is a {power law} of the density. In the absence of drift,
this would lead to anomalous diffusion. If we take into account a
drift term and a self-attraction, we have to solve the nonlinear
Smoluchowski-Poisson (NSP) system. It can be shown
\cite{Ctsallis,shiino} that the NSP system decreases continuously
a free energy associated with Tsallis entropy \cite{tsallis}.
Accordingly, the stationary solutions of the NSP system are given
by a polytropic distribution which minimizes Tsallis' free energy
at fixed mass. The equilibrium state is then determined by solving
the Lane-Emden equation, as in the case of polytropic stars and stellar polytropes \cite{chandra,bt}. Depending on the value of the
control parameter and on the index $n$ of the polytrope, three
situations can occur: (i) the NSP system can relax toward an
incomplete polytrope maintained by the walls of the confining box
(ii) the NSP system can relax toward a stable complete polytrope of
radius $R_*<R$, unaffected by the box (iii) the NSP system can
undergo a catastrophic collapse leading to a finite time
singularity.

The paper has two parts that are relatively independent. In the first
part of the paper (Secs. \ref{sec_dyn} and \ref{sec_statics}), we
study the {dynamical} stability of stellar systems, gaseous stars
and two-dimensional (2D) vortices by using a {\it thermodynamical
analogy} \cite{grand,Ctsallis,Ctsalliskin}. Stellar polytropes
maximize Tsallis entropy (considered as a $H$-function) at fixed mass
and energy. This is a condition of nonlinear dynamical stability via
the Vlasov equation. Polytropic stars minimize Tsallis free energy
(related to the star energy functional) at fixed mass. This is a
condition of nonlinear dynamical stability via the Euler-Jeans
equations. Polytropic vortices maximize Tsallis entropy (considered as
a $H$-function) at fixed circulation and energy. This is a condition
of nonlinear dynamical stability via the 2D Euler equation. We perform
an exhaustive study of the structure and stability of polytropic
spheres by determining whether they are maxima or minima (or saddle
points) of Tsallis functional. For sake of generality, we perform our
study in a space of dimension $D$. We shall exhibit particular
dimensions $D=2$, $D=4$, $D=2(1+\sqrt{2})$ and $D=10$ which play a
special role in our problem. The dimension $D=2$ is {\it critical}
because the results established for $D>2$ cannot be directly extended
to $D=2$ \cite{sc}. On the other hand, the nature of the caloric curve
changes for $D=4$ and $D=10$.  This extends the study performed by
Taruya \& Sakagami \cite{taruya1,taruya2} and Chavanis
\cite{poly,grand} for $D=3$.

In the second part of the paper (Sec. \ref{sec_dynamics}), we study
the dynamics and thermodynamics of self-gravitating Langevin particles
experiencing anomalous diffusion.  Their equilibrium distribution
minimizes Tsallis free energy at fixed mass. This is a condition of
{thermodynamical} stability in a generalized sense. This is also a
condition of linear dynamical stability via generalized Fokker-Planck
equations (in the present context the NSP
system)\cite{crs,shiino,Ctsallis}. Thus, the stability analysis of
Sec. \ref{sec_statics} can also be used in that context. When the
static solution does not exist or is unstable, the system undergoes a
catastrophic collapse. In Sec. \ref{sec_dynamics}, we show that the
NSP system admits self-similar solutions describing the collapse and
leading to a finite time singularity. The density decreases at large
distances as $\rho\sim r^{-\alpha}$. In the canonical situation (fixed
$T$), the scaling exponent is $\alpha_{n}=2n/(n-1)$ where $n$ is the
polytropic index. We also consider a microcanonical situation in which
the generalized temperature $T(t)$ varies in time so as to rigorously
conserve energy. In that case, the scaling equation has solutions for
$\alpha_{n}\le \alpha\le \alpha_{\rm max}(n,D)$ where $\alpha_{\rm
max}(n,D)$ is a non-trivial exponent. The value of $\alpha$
effectively selected by the system is determined by the dynamics.  In
Sec. \ref{sec_num}, we perform direct numerical simulations of the SP
and NSP systems with higher accuracy than in Ref. \cite{crs}. We
confirm the scaling regime and discuss the value of the scaling
exponent. In the microcanonical situation, we give numerical evidence
that $\alpha=\alpha_{n}$ so that the temperature $T(t)$ is {\it
finite} at the collapse time (this observation has been made
independently by \cite{guerra} for the SP system). However, the
convergence to this value is so slow that the evolution displays a
pseudo-scaling regime with $\alpha_{n}\le \alpha\le \alpha_{\rm max}$
for the densities achieved. We explain the physical reason for this
behavior and we conjecture that $\alpha\simeq \alpha_{\rm max}$ will
be reached in more realistic models with non-uniform temperature
\cite{sc3}.  This paper closely follows the style and presentation of
our companion paper for isothermal spheres
\cite{sc}. These two papers complete the classical monograph of
Chandrasekhar on self-gravitating isothermal and polytropic
spheres in $D=3$ \cite{chandra}.

\section{Dynamical stability of systems with long-range interactions}
\label{sec_dyn}

\subsection{Stellar systems}
\label{sec_ss}

Let us consider a collection of $N$ stars with mass $m$ in
gravitational interaction. They form a Hamiltonian $N$-body system
with long-range (Newtonian) interactions. We work in a space of
dimension $D$ and enclose the system within a spherical box of radius
$R$. Let $f({\bf r},{\bf v},t)$ denote the distribution function of
the system, i.e. $f({\bf r},{\bf v},t)d^{D}{\bf r}d^{D}{\bf v}$ gives
the mass of stars whose position and velocity are in the cell
$({\bf r},{\bf v};{\bf r}+d^{D}{\bf r},{\bf v}+d^{D}{\bf v})$ at time
$t$. The integral of $f$ over the velocity determines the spatial
density
\begin{equation}
\rho=\int f \,d^{D}{\bf v}. \label{maxent1}
\end{equation}
The total mass of the configuration is
\begin{equation}
M=\int \rho \,d^{D}{\bf r}. \label{maxent2}
\end{equation}
In the mean-field approximation, the total energy of the system
can be expressed as
\begin{equation}
E={1\over 2}\int f v^{2}\,d^{D}{\bf r}d^{D}{\bf v}+{1\over
2}\int\rho\Phi \,d^{D}{\bf r}=K+W, \label{maxent3}
\end{equation}
where $K$ is the kinetic energy and $W$ the potential energy. The
gravitational potential $\Phi$ is related to the density by the
Newton-Poisson equation
\begin{equation}
\Delta\Phi=S_{D}G\rho, \label{maxent4}
\end{equation}
where $S_{D}$ is the surface of a unit sphere in a $D$-dimensional
space and $G$ is the constant of gravity.

For fixed $N\gg 1$ and $t\rightarrow +\infty$, the system is expected
to reach a statistical equilibrium state described by the classical
Boltzmann entropy $S_{B}[f]=-\int f\ln f d^{D}{\bf r}d^{D}{\bf
v}$. However, the relaxation time $t_{relax}$ due to ``collisions''
(more properly close encounters) is in general considerably larger
than the dynamical time $t_{D}$ so that this statistical equilibrium
state is often not physically relevant \cite{ny,dubrovnik}. This is
the case in particular for elliptical galaxies where $t_{relax}\sim
{N\over\ln N}t_{D}$ with $N\sim 10^{12}$, while their age is $\sim
100t_{D}$ \cite{bt}. For $t\ll t_{relax}$ and $N\rightarrow +\infty$,
the dynamics of stars is described by the Vlasov equation
\begin{equation}
{\partial f\over\partial t}+{\bf v}{\partial f\over\partial {\bf r}}+{\bf F}{\partial f\over\partial {\bf v}}=0,\label{ss1}
\end{equation}
where ${\bf F}=-\nabla\Phi$ is the gravitational force determined  by the Poisson equation (\ref{maxent4}). For any $H$-function
\begin{equation}
S[f]=-\int C(f)d^{D}{\bf r}d^{D}{\bf v},\label{ss2}
\end{equation}
where $C$ is convex, i.e. $C''>0$, it can be shown that the variational problem
\begin{equation}
{\rm Max} \quad S[f] \quad {\rm at\ fixed} \quad E[f], M[f],
\label{ss3}
\end{equation}
determines a stationary solution of the Vlasov equation with strong
(nonlinear) dynamical stability properties
\cite{ipser,tremaine}. Such solutions can result from a process
of (possibly incomplete) violent relaxation \cite{grand}. Introducing
Lagrange multipliers, the first order variations $\delta S-\beta\delta
E-\alpha\delta M=0$ lead to
\begin{equation}
C'(f)=-\beta\biggl ({v^{2}\over 2}+\Phi\biggr )-\alpha.
\label{ss4}
\end{equation}
Therefore, $f=f(\epsilon)$ where $\epsilon={v^{2}\over 2}+\Phi$ is the
energy of a star by unit of mass. These distribution functions,
depending only on the energy, form a particular class of stationary
solutions of the Vlasov equation. Other solutions can be constructed
with the Jeans theorem \cite{bt} but their stability is more difficult
to investigate. The conservation of angular momentum can be easily
included in the foregoing discussion \cite{rieutord}.

\subsection{Barotropic stars}
\label{sec_bs}

Let us now consider a self-gravitating gaseous system described by the
Euler-Jeans equations
\begin{equation}
{\partial\rho\over\partial t}+\nabla (\rho {\bf u})=0,
\label{bs1}
\end{equation}
\begin{equation}
{\partial {\bf u}\over\partial t}+({\bf u}\cdot \nabla) {\bf u}=-{1\over\rho}\nabla p-\nabla\Phi.
\label{bs2}
\end{equation}
We assume that the gas is barotropic with an equation of state
$p=p(\rho)$. The most important examples of barotropic fluids are
those that are isentropic or adiabatic, that is those whose specific
entropy is constant. If $s={\rm Cst.}$, the first principle of
thermodynamics $du=-pdv+Tds$ (where $v=1/\rho$) reduces to
$du={p\over\rho^{2}}d\rho$, where $u$ is the internal energy by unit
of mass. The total energy of the fluid is therefore
\begin{equation}
{\cal
W}[\rho]=\int\rho\int_{0}^{\rho}{p(\rho')\over\rho^{'2}}d\rho'd^{D}{\bf
r}+{1\over 2}\int \rho\Phi d^{D}{\bf r}+\int \rho {{\bf u}^{2}\over
2}d^{D}{\bf r}.
\label{bs3}
\end{equation}
The first term is the internal energy, the second the gravitational
energy and the third the kinetic energy associated with the mean
motion. As argued in \cite{grand}, the variational problem
\begin{equation}
{\rm Min} \quad {\cal W}[\rho] \quad {\rm at\ fixed} \quad  M[\rho],
\label{bs4}
\end{equation}
determines a stationary solution of the Euler-Jeans equations with
strong (nonlinear) dynamical stability properties.  We shall not prove
this result here but it is expected to follow from relatively standard
methods of stability theory.  The solutions of this variational
problem satisfy the condition of hydrostatic balance
\begin{equation}
\nabla p=-\rho\nabla\Phi,
\label{eqhydro}
\end{equation}
between pressure and gravity. The foregoing results can be extended to
barotropic stars rotating rigidly with angular velocity ${\mb\Omega}$
\cite{rieutord}.

\subsection{Two-dimensional vortices}
\label{sec_vortex}

Let us finally consider a collection of $N$ point vortices with
circulation $\gamma$ in 2D hydrodynamics. They form a Hamiltonian
system with long-range (logarithmic) interactions. We call $\omega$
the vorticity, $\psi$ the stream function and ${\bf u}=-{\bf
z}{\times} \nabla\psi$ the velocity field.  We also note
$\Gamma=\int\omega d^{2}{\bf r}$ the circulation and $E={1\over
2}\int\omega\psi d^{2}{\bf r}$ the energy. For fixed $N\gg 1$ and
$t\rightarrow +\infty$, the system is expected to reach a statistical
equilibrium state described by the classical Boltzmann entropy
$S_{B}[\omega]=-\int \omega\ln
\omega d^{2}{\bf r}$ \cite{jm}. However, the relaxation
time $t_{relax}$ due to ``collisions'' is in general considerably
larger than the dynamical time $t_{D}$ so that this statistical
equilibrium state is often not physically relevant \cite{kin,houches}. For
$t\ll t_{relax}$ and $N\rightarrow +\infty$, the dynamics of point vortices
is described by the 2D Euler-Poisson system
\begin{equation}
{\partial \omega\over\partial t}+{\bf u}\cdot \nabla\omega=0,\label{vortex1}
\end{equation}
\begin{equation}
\omega=-\Delta\psi.\label{vortex2}
\end{equation}
The Euler equation also governs the dynamics of 2D incompressible and
inviscid continuous vorticity fields. For any $H$-function
\begin{equation}
S[\omega]=-\int C(\omega)d^{2}{\bf r},\label{vortex3}
\end{equation}
where $C$ is convex, i.e. $C''>0$, it can be shown that the variational problem
\begin{equation}
{\rm Max} \quad S[\omega] \quad {\rm at\ fixed} \quad E[\omega], \Gamma[\omega],
\label{vortex4}
\end{equation}
determines a stationary solution of the 2D Euler equation with strong (nonlinear) dynamical stability
properties \cite{ellis}. Such solutions can result from a process of
(possibly incomplete) violent relaxation \cite{ch}. Introducing Lagrange
multipliers, the condition $\delta S-\beta\delta E-\alpha\delta
\Gamma=0$ leads to
\begin{equation}
C'(\omega)=-\beta\psi-\alpha.
\label{vortex5}
\end{equation}
Therefore, $\omega=\omega(\psi)$, which is the general form of stationary
solutions of the 2D Euler equation for domains with no specific
symmetries. The conservation of angular momentum and impulse can be
easily included in the foregoing discussion \cite{Ctsallis}.

\subsection{Thermodynamical analogy}
\label{sec_analogy}

Since the variational problems (\ref{ss3}), (\ref{bs4}) and
(\ref{vortex4}) are {similar} to the usual variational problems that
arise in thermodynamics (with the Boltzmann entropy), we can develop a
{\it thermodynamical analogy} to analyze the dynamical stability of
stellar systems, gaseous stars and 2D vortices
\cite{grand,Ctsallis,Ctsalliskin,ch}. In this analogy, the functional $S$
plays the role of a generalized entropy, $\beta$ is a generalized
inverse temperature, $\beta(E)$ a generalized caloric curve etc... The
variational problem (\ref{ss3}) is similar to a condition of
microcanonical stability.  We can also introduce a generalized free
energy $F[f]=E[f]-T S[f]$ which is the Legendre transform of
$S[f]$. The minimization of $F[f]$ at fixed $T$ and $M[f]$ is similar
to a condition of canonical stability. This is equivalent to first
minimize $F[f]$ at fixed $\rho({\bf r})$ to get $f_{*}({\bf r},{\bf
v})$ and then to minimize $F[\rho]=F[f_{*}]$, calculated with $f_{*}$,
at fixed $M[\rho]$. Now, it can be shown
\cite{grand} that $F[\rho]$ is precisely the functional (\ref{bs3})
with ${\bf u}={\bf 0}$. Therefore, the variational problem (\ref{bs4})
is similar to a condition of canonical stability. Since canonical
stability implies microcanonical stability (but not the converse)
\cite{Ctsallis}, we conclude that ``stellar systems are stable
whenever corresponding barotropic stars are stable'' which provides a
new interpretation of Antonov's first law \cite{grand}.

In 2D hydrodynamics, the variational problem (\ref{vortex4}) is
similar to a condition of microcanonical stability. It is stronger
than the maximization of $J[\omega]=S[\omega]-\beta E[\omega]$ at
fixed $\beta$ and $\Gamma[\omega]$ (canonical stability), which is
just a sufficient condition of nonlinear dynamical stability. It is
not a necessary condition of stability if the ensembles are
inequivalent (i.e. if the ``caloric curve'' presents bifurcations or
turning points). The Arnold's theorems just provide sufficient
conditions of canonical stability (see, e.g.,
\cite{Ctsallis}). Therefore, in the domain of inequivalence
(corresponding to a region of ``negative specific heats'') a flow can
be nonlinearly dynamically stable while it violates Arnold's
theorems. This has important implications in jovian fluid dynamics
\cite{ellis,bouchet}.

The preceding arguments also apply to other systems with long-range
interactions such as the HMF model for example \cite{yamaguchi}.  This
opens the route to many generalizations by changing the potential of
interaction and the ``generalized entropy'' (H-function).  In this
paper, we shall specialize in the case of particles interacting via a
Newtonian potential (e.g., self-gravitating systems, 2D
vortices,...). We shall also consider a special form of $H$-function,
known as Tsallis entropy, leading to power-laws distributions
(polytropes).

\section{Equilibrium structure of polytropic spheres in dimension $D$}
\label{sec_statics}

\subsection{Stellar polytropes}
\label{sec_sp}

Let us consider a particular class of stationary solutions of the
Vlasov equation called {\it stellar polytropes}
\cite{bt}. Nonlinearly dynamically stable solutions  maximize the Tsallis
entropy
\begin{equation}
S_{q}=-{1\over q-1}\int (f^{q}-f) \,d^{D}{\bf r}d^{D}{\bf v},
\label{maxent5}
\end{equation}
at fixed mass $M$ and energy $E$, where $q$ is a real number. For
$q\rightarrow 1$, Eq. (\ref{maxent5}) reduces to the ordinary
Boltzmann entropy describing isothermal stellar systems. We emphasize
that, in the present context, Tsallis and Boltzmann entropies are
particular $H$-functions (not true entropies) that are related to the
dynamics, not to the thermodynamics. Still, due to the thermodynamical
analogy discussed in Sec. \ref{sec_analogy}, we shall use a
thermodynamical langage to study the dynamical stability problem. In
this analogy, the dynamical stability criterion for stellar polytropes
corresponds to a {\it microcanonical} stability condition.

The critical points of entropy at fixed mass and energy satisfy
the condition
\begin{equation}
\delta S_{q}-\beta\delta E-\lambda\delta M=0, \label{maxent7}
\end{equation}
where $\beta=1/T$ and $\lambda$ are Lagrange multipliers ($T$ is the
temperature and $\lambda$ the chemical potential in the generalized
sense). The variational principle (\ref{maxent7}) leads to the
polytropic distribution function
\begin{equation}
f({\bf r},{\bf v})=\biggl\lbrace \mu-{(q-1)\beta\over q}\biggl \lbrack {v^{2}\over 2}+\Phi({\bf r})\biggr \rbrack \biggr \rbrace^{1\over
q-1}, \label{nDF}
\end{equation}
where $\mu=\lbrack 1-(q-1)\lambda\rbrack /q$. We define the polytropic index $n$ by the relation
\begin{equation}
n={D\over 2}+{1\over q-1}. \label{maxent10}
\end{equation}

We first consider the case $(q-1)\beta/q>0$ and allow $\beta$ to take negative values. Then, Eq. (\ref{nDF}) can be rewritten
\begin{equation}
f=A\biggl\lbrack \alpha-\Phi-{v^{2}\over 2}\biggr \rbrack^{n-{D\over 2}}, \label{maxent8}
\end{equation}
where
\begin{equation}
A=\biggl\lbrack {(q-1)\beta\over q}\biggr \rbrack^{1\over q-1},
\qquad \alpha={1-(q-1)\lambda\over (q-1)\beta}. \label{maxent9}
\end{equation}
If $n>D/2$ (i.e., $q>1$ and $\beta>0$), $f'(\epsilon)<0$ where
$\epsilon={v^{2}\over 2}+\Phi$ is the stellar energy. Therefore, high
energy particles are less probable than low energy particles, which
corresponds to the physical situation. Equation (\ref{maxent8}) is
valid for $v<v_{max}=\sqrt{2(\alpha-\Phi)}$.  If
$v>v_{max}=\sqrt{2(\alpha-\Phi)}$, we set $f=0$. This distribution
function describes stellar polytropes which were first introduced by
Plummer \cite{plummer}. If $n=D/2$ (i.e., $q\rightarrow \infty$), the
distribution $f(\epsilon)$ is a step function. This corresponds to the
self-gravitating Fermi gas at zero temperature (white dwarfs).  In
$D=3$, classical white dwarf stars are equivalent to polytropes with
index $n=3/2$
\cite{chandra}. In $D$-dimensions, classical ``white dwarf stars'' are equivalent to polytropes with index
\begin{equation}
n_{3/2}={D\over 2}.\label{new32}
\end{equation}
If $n\rightarrow +\infty$ (i.e. $q\rightarrow 1$), we recover
isothermal distibution functions.  If $n<D/2$ (i.e., $q<1$ and
$q\beta<0$), high energy particles are more probable than low energy
particles: $f'(\epsilon)>0$. This situation is unphysical but it can
be considered at a formal level. The distribution function diverges
like $(1-v/v_{max})^{n-D/2}$ as $v\rightarrow v_{max}$. The moments of
$f$ converge if and only if $n>D/2-1$. Therefore, if $D/2-1<n<D/2$
(i.e., $q<0$ and $\beta>0$), stellar polytropes exist mathematically
but they are not physical. If $n<D/2-1$ (i.e., $0<q<1$ and $\beta<0$),
stellar polytropes do not exist.

We now consider the case $(q-1)\beta/q<0$. Then, Eq. (\ref{nDF}) can be rewritten
\begin{equation}
f=A\biggl\lbrack \alpha+\Phi+{v^{2}\over 2}\biggr \rbrack^{n-{D\over 2}}, \label{newmaxent8}
\end{equation}
where
\begin{equation}
A=\biggl\lbrack {(1-q)\beta\over q}\biggr \rbrack^{1\over q-1},
\qquad \alpha={1-(q-1)\lambda\over (1-q)\beta}. \label{newmaxent9}
\end{equation}
If $n>D/2$ (i.e., $q>1$ and $\beta<0$), the model is ill-posed because  $f(v)$ diverges for $v\rightarrow +\infty$. If  $n<D/2$, the distribution function goes to zero like $v^{-(D-2n)}$ for  $v\rightarrow +\infty$. If $0<n<D/2$ (i.e., $q<1-2/D$), the density $\rho=\int f d^{D}{\bf v}$ does not exist. If $-1<n<0$ (i.e., $1-2/D<q<D/(D+2)$), the density  exists but not the pressure $p={1\over D}\int f v^{2}d^{D}{\bf v}$. If $n<-1$ (i.e., $D/(D+2)<q<1$ and $\beta>0$), the density and the pressure exist.

In conclusion, only positive temperatures states are physical. If
$n\ge D/2$ (i.e. $q>1$), the system is described by the distribution
function (\ref{maxent8}). If $n<-1$ (i.e., $D/(D+2)<q<1$), the system
is described by the distribution function (\ref{newmaxent8}). In this
paper, we shall only consider stellar polytropes with index $n\ge
D/2$. Using Eq. (\ref{maxent8}), the spatial density ${\rho}=\int f
d^{D}{\bf v}$ and the pressure $p={1\over D}\int fv^{2}d^{D}{\bf v}$
can be expressed as
\begin{equation}
{\rho}=2^{D/2-1} A S_{D} (\alpha-\Phi)^{n}B ({D/2},n+1-{D/2}),
\label{maxent11}
\end{equation}
\begin{equation}
{p}={1\over n+1} 2^{D/2-1} AS_{D} (\alpha-\Phi)^{n+1} B
({D/2},n+1-{D/2} ), \label{maxent11b}
\end{equation}
with $B(a,b)$ being the beta-function. In obtaining Eq.
(\ref{maxent11b}), we have used the identity $B(m+1,n)= m
B(m,n)/(m+n)$. Eliminating the gravitational
potential between these two relations, we recover the well-known
fact that stellar polytropes satisfy the equation of state
\begin{equation}
p=K\rho^{\gamma},\qquad \gamma=1+{1\over n}, \label{maxent12}
\end{equation}
like gaseous polytropes (see below). In the present context, the polytropic
constant is given by
\begin{equation}
K={1\over (n+1)}\biggl\lbrace 2^{D/2-1}S_{D} A B
({D/2},n+1-{D/2})\biggr\rbrace^{-1/n}. \label{maxent13}
\end{equation}
Using Eqs. (\ref{maxent9}) and (\ref{maxent11}), the distribution
function (\ref{maxent8}) can be written as a function of the
density as
\begin{eqnarray}
{f}={1\over Z}\biggl \lbrack \rho^{1/n}-{v^{2}/2\over
(n+1)K}\biggr \rbrack^{n-D/2}, \label{maxent14}
\end{eqnarray}
with
\begin{eqnarray}
Z={2^{D/2-1}S_{D}B(D/2,n+1-{D/2})}{\lbrack K(n+1)\rbrack^{D/2}}.
\label{maxent15}
\end{eqnarray}
The distribution function (\ref{maxent14}) can also be obtained by
maximizing $S_{q}[f]$ at fixed $M$, $E$ {\it and} $\rho({\bf r})$, or
equivalently at fixed $K={1\over 2}\int f v^{2}d^{D}{\bf r}d^{D}{\bf
v}$ and $\rho({\bf r})$. It is then possible to express the energy
(\ref{maxent3}) and the entropy (\ref{maxent5}) in terms of $\rho({\bf
r})$ and $T$. Using Eq. (\ref{maxent14}),
it is easy to show that
\begin{equation}
E={D\over 2}\int p d^{D}{\bf r}+{1\over 2}\int \rho\Phi d^{D}{\bf
r}, \label{maxent16}
\end{equation}
\begin{equation}
S_{q}[\rho]=-\biggl (n-{D\over 2}\biggr )\biggl (\beta \int p d^{D}{\bf
r}-M\biggr ). \label{maxent17}
\end{equation}
In arriving at Eq. (\ref{maxent17}), we have used the identity
$B(m,n+1)= n B(m,n)/(m+n)$. Proceeding carefully, we can check that
for $q\rightarrow 1$, Eq. (\ref{maxent17}) reduces to the Boltzmann
entropy expressed in terms of hydrodynamical variables (see Eq. (9) of
\cite{sc}). The problem of the stability of stellar polytropes  
now amounts to determining  {\it maxima} of $S_{q}[\rho]$ at 
fixed $E[\rho]$ and $M[\rho]$.

\subsection{Gaseous polytropes}
\label{sec_gp}

We shall consider a particular class of barotropic stars called
{\it gaseous polytropes}. They are characterized by an equation of
state of the form
\begin{equation}
p=K\rho^{\gamma}, \qquad \gamma=1+{1\over n},
\label{neweqstqte}
\end{equation}
where $K$ is a constant. We recall that gaseous polytropes are
described by a local thermodynamical equilibrium condition and {\it
not} by a distribution function of the form (\ref{nDF}), except in the
isothermal case $n\rightarrow +\infty$ (see \cite{poly}). Their energy
(\ref{bs3}) is
\begin{equation}
{\cal W}[\rho]=n\int p d^{D}{\bf r}+{1\over 2}\int \rho\Phi d^{D}{\bf r}+\int {{\bf u}^{2}\over 2}d^{D}{\bf r}.
\label{newnrj}
\end{equation}
Using Eqs. (\ref{maxent16}) and (\ref{maxent17}), we note that the
free energy of stellar polytropes $F_{q}=E-TS_{q}$ is
\begin{equation}
F_{q}[\rho]={1\over 2}\int \rho\Phi d^{D}{\bf r}+n\int p d^{D}{\bf r},
\label{newf}
\end{equation}
within an additional constant. Taking ${\bf u}={\bf 0}$, we check on
this specific example that $F_q[\rho]={\cal W}[\rho]$. In fact, this
relation is general as shown in \cite{grand}. Therefore, the dynamical
stability of gaseous polytropes can be settled by studying the
minimization of $F_q[\rho]$ at fixed mass. In the thermodynamical
analogy of Sec. \ref{sec_dyn}, this corresponds to a {\it canonical}
description. We note finally that the free energy (\ref{newf}) can be
written explicitly
\begin{equation}
F_{q}[\rho]={1\over 2}\int \rho\Phi d^{D}{\bf r}+{K\over\gamma-1}\int (\rho^{\gamma}-\rho) d^{D}{\bf r}.
\label{newftsallis}
\end{equation}
This can be viewed as a free energy $F_{\gamma}=E-K S_{\gamma}$ associated with a Tsallis entropy in position space $S_{\gamma}=-{1\over \gamma-1}\int (\rho^{\gamma}-\rho)d^{D}{\bf r}$, where $\gamma$ plays the role of the $q$ parameter and $K$ the role of a temperature. The parameters $q$ and $\gamma$ are related to each other by $\gamma=1+2(q-1)/\lbrack 2+D(q-1)\rbrack$. The equilibrium distribution can be written
\begin{equation}
\rho=\biggl\lbrack \lambda-{\gamma-1\over K\gamma}\Phi\biggr \rbrack^{1\over\gamma-1},
\label{tp}
\end{equation}
which is equivalent to Eq. (\ref{maxent11}). When considering gaseous
polytropes, we shall allow for arbitrary value of the index $n\ge 0$.

\subsection{The $D$-dimensional Lane-Emden equation}
\label{sec_emden}

The configuration of a stellar polytrope is obtained by substituting
Eq.  (\ref{maxent8}) in the Poisson equation (\ref{maxent4}) using Eq.
(\ref{maxent1}). This yields a self-consistent mean-field equation for
the gravitational potential $\Phi$. An equivalent equation can be
obtained by substituting the equation of state (\ref{maxent12}) in the
condition of hydrostatic equilibrium (\ref{eqhydro}), as for gaseous
polytropes (the equivalence between these two approaches is shown in
\cite{grand}). Using the Gauss theorem
\begin{equation}
{d\Phi\over dr}={GM(r)\over r^{D-1}}, \label{maxent19}
\end{equation}
where $M(r)\equiv \int_{0}^{r}\rho S_{D}r^{'D-1}dr'$ is the mass
within the sphere of radius $r$, we can rewrite Eq.
(\ref{eqhydro}) in the form
\begin{equation}
{1\over r^{D-1}}{d\over dr}\biggl ({r^{D-1}\over\rho}{dp\over
dr}\biggr )=-S_{D}G\rho, \label{maxent20}
\end{equation}
which is the fundamental equation of hydrostatic equilibrium in
$D$-dimensions. For the polytropic equation of state
(\ref{maxent12}), we have
\begin{equation}
K(n+1){1\over r^{D-1}}{d\over dr}\biggl
({r^{D-1}}{d\rho^{1/n}\over dr}\biggr )=-S_{D}G\rho.
\label{maxent21}
\end{equation}
The case of isothermal spheres with an equation of state $p=\rho
T$ is recovered in the limit $n\rightarrow +\infty$.
To determine the structure of polytropic spheres, we set
\begin{equation}
\rho=\rho_{0}\theta^{n}, \qquad \xi=\biggl\lbrack {S_{D}
G\rho_{0}^{1-1/n}\over K(n+1)}\biggr\rbrack^{1/2}r, \label{emden1}
\end{equation}
where $\rho_{0}$ is the central density. Then, Eq.
(\ref{maxent21}) can be put in the form
\begin{equation}
{1\over \xi^{D-1}}{d\over d\xi}\biggl (\xi^{D-1}{d\theta\over
d\xi}\biggr )=-\theta^{n}, \label{emden2}
\end{equation}
which is the $D$-dimensional generalization of the Lane-Emden
equation \cite{chandra}. For $D>2$ and
\begin{equation}
n>{D\over D-2}\equiv n_{3}, \label{emden3}
\end{equation}
Eq.~(\ref{emden2}) has a simple explicit solution, the singular
polytropic sphere
\begin{equation}
\theta_{s}=\biggl\lbrace {2\lbrack (D-2)n-D\rbrack\over
(n-1)^{2}}\biggr \rbrace^{1\over n-1}\xi^{-{2\over n-1}}.
\label{emden4}
\end{equation}
The regular solutions of Eq.~(\ref{emden2}) satisfying the
boundary conditions
\begin{equation}
\theta=1, \qquad \theta'=0 \qquad {\rm at}\qquad \xi=0,
\label{emden4b}
\end{equation}
must be computed numerically. For $\xi\rightarrow 0$, we can
expand the solutions in Taylor series and we find that
\begin{equation}
\theta=1-{1\over 2D}\xi^{2}+{n\over 8D(D+2)}\xi^{4}+...
\label{emden5}
\end{equation}
To obtain the asymptotic behavior of the solutions for
$\xi\rightarrow +\infty$, we note that the transformation
$t=\ln\xi$, $\theta=\xi^{-2/(n-1)}z$ changes Eq.~(\ref{emden2}) in
\begin{equation}
{d^{2}z\over dt^{2}}+{(D-2)n-(D+2)\over n-1}{dz\over
dt}=-z^{n}-{2\lbrack D+(2-D)n\rbrack\over (n-1)^{2}}z.
\label{emden6}
\end{equation}
For $D\le 2$ or for $D>2$ and
\begin{equation}
n<{D+2\over D-2}\equiv n_{5}, \label{emden7}
\end{equation}
the density falls off to zero at a finite radius $R_{*}$. This
defines a {\it complete polytrope } of radius $R_{*}$. If we
denote by $\xi_{1}$ the value of the normalized distance at which
$\theta=0$, then, for $\xi\rightarrow\xi_{1}$, we have
\begin{equation}
\theta=-\xi_{1}\theta'_{1}\biggl \lbrack
{\xi_{1}-\xi\over\xi_{1}}+{D-1\over 2}\biggl
({\xi_{1}-\xi\over\xi_{1}}\biggr )^{2}+{D(D-1)\over 6}\biggl
({\xi_{1}-\xi\over\xi_{1}}\biggr )^{3}+...\biggr\rbrack.
\label{emden8}
\end{equation}
On the other hand, for $D>2$ and $n>n_{5}$, Eq. (\ref{emden6})
corresponds to the damped motion of a fictitious particle in a
potential
\begin{equation}
V(z)={D+(2-D)n\over (n-1)^{2}}z^{2}+{1\over n+1}z^{n+1},
\label{emden9}
\end{equation}
where $z$ plays the role of position and $t$ the role of time. For
$t\rightarrow +\infty$, the particle will come at rest at the
bottom of the well at position $z_{0}=\lbrace {2\lbrack
(D-2)n-D\rbrack/ (n-1)^{2}}\rbrace^{1\over n-1}$. Returning to
original variables, we find that
\begin{equation}
\theta\rightarrow \biggl\lbrace {2\lbrack (D-2)n-D\rbrack\over
(n-1)^{2}}\biggr \rbrace^{1\over n-1}\xi^{-{2\over
n-1}}=\theta_{s}, \qquad {\rm for}\qquad \xi\rightarrow +\infty.
\label{emden10}
\end{equation}
Therefore, the regular solutions of the Lane-Emden equation
(\ref{emden2}) behave like the singular solution for
$\xi\rightarrow +\infty$. To determine the next order correction,
we set $z=z_{0}+z'$ with $z'\ll 1$. Keeping only terms that are
linear in $z'$, Eq.~(\ref{emden6}) becomes
\begin{equation}
{d^{2}z'\over dt^{2}}+{(D-2)n-(D+2)\over n-1}{dz'\over
dt}+{2\lbrack (D-2)n-D\rbrack\over n-1}z'=0. \label{emden11}
\end{equation}
The discriminant of the second order polynomial associated with this
equation is
\begin{equation}
\Delta(n)={-(D-2)(10-D)n^{2}-2(D^{2}-8D+4)n+(D-2)^{2}\over
(n-1)^{2}}. \label{emden12}
\end{equation}
For $n\rightarrow +\infty$, $\Delta(n)\sim -(D-2)(10-D)$.
Furthermore, $\Delta(n)=0$ for
\begin{equation}
n_{\pm}={-D^{2}+8D-4\pm 8\sqrt{D-1}\over (D-2)(10-D)}. \label{emden13}
\end{equation}
For $2<D<10$, it is straightforward to check that
$n_{-}<n_{+}<n_{5}$. Therefore, for $n>n_{5}$, $\Delta$ has the
sign of $-(D-2)(10-D)$ which is negative. Thus
\begin{equation}
\theta=\theta_{s}\biggl\lbrace 1+{C\over \xi^{b/2}}\cos\biggl
({\sqrt{-\Delta}\over 2}\ln\xi+\delta\biggr )\biggr\rbrace, \qquad
(\xi\rightarrow +\infty), \label{emden14}
\end{equation}
where $b=\lbrack (D-2)n-(D+2)]/(n-1)$.  The density profile
(\ref{emden14}) intersects the singular solution (\ref{emden4})
infinitely often at points that asymptotically increase
geometrically in the ratio $1:{\rm exp}\lbrace
2\pi/\sqrt{-\Delta}\rbrace$ (see, e.g., Fig.~1 of Ref. \cite{poly}
for $D=3$). For $D>10$, we have $n_{+}<n_{5}<n_{-}$. Therefore, if
$n_{5}<n<n_{-}$, $\Delta$ has the sign of $(D-2)(10-D)$ which is
negative and the asymptotic behavior of the solutions is still
given by Eq. (\ref{emden14}). However, for $n>n_{-}$, $\Delta(n)$
is positive and therefore
\begin{equation}
\theta=\theta_{s}\biggl\lbrace 1+{1\over \xi^{b/2}}\biggl
(A\xi^{\sqrt{\Delta}\over 2}+B\xi^{-\sqrt{\Delta}\over 2}\biggr )
\biggr\rbrace, \qquad (\xi\rightarrow +\infty). \label{emden15}
\end{equation}
Finally, for $n=n_{-}$, $\Delta=0$ and we have
\begin{equation}
\theta=\theta_{s}\biggl\lbrace 1+{1\over \xi^{b/2}}(A\ln\xi+B)
\biggr\rbrace, \qquad (\xi\rightarrow +\infty). \label{emden16}
\end{equation}
For $D=10$, $n_{-}\rightarrow +\infty$ so that the asymptotic
behavior of $\theta$ is given by Eq. (\ref{emden14}) for
$n<+\infty$. Since $\theta_{s}^n\sim \xi^{-2n/(n-1)}$ at large
distances, the configurations described previously have an
``infinite mass'' which is clearly unphysical. In the following,
we shall confine these configurations to a ``box" of radius $R$ as
for isothermal spheres \cite{sc}. Such configurations will be
called {\it incomplete polytropes}. We note that the
self-gravitating Fermi gas at zero temperature forms a complete
polytrope only if $n_{3/2}<n_{5}$, i.e. $D<2(1+\sqrt{2})\simeq
4.83$.

The Lane-Emden equation can be solved analytically for some
particular values of the polytropic index. For $n=0$, which
corresponds to a body with constant density, we have
\begin{equation}
\theta=1-{1\over 2D}\xi^{2}, \qquad \xi_{1}=\sqrt{2D}.
\label{emden17}
\end{equation}
For $n=1$, Eq. (\ref{emden2}) reduces to the $D$-dimensional
Helmholtz equation. For $D=1$,
\begin{equation}
\theta=\cos\xi, \qquad \xi_{1}={\pi\over 2}. \label{emden18}
\end{equation}
For $D=2$,
\begin{equation}
\theta=J_{0}(\xi), \qquad \xi_{1}=2.40482... \label{emden19}
\end{equation}
Finally, for $D=3$, performing the change of variables
$\theta=\chi/\xi$, we get
\begin{equation}
\theta={\sin\xi\over\xi}, \qquad \xi_{1}=\pi. \label{emden20}
\end{equation}
The Lane-Emden equation can also be
solved analytically in any dimension of space $D>2$ for the
particular index value $n_{5}$. The solution is
\begin{equation}
\theta_{5}={1\over \bigl (1+{\xi^{2}\over D(D-2)}\bigr )^{D-2\over
2}}, \label{emden21}
\end{equation}
as can be checked by a direct substitution in Eq. (\ref{emden2}).
For $D=3$, we recover the Schuster solution \cite{chandra}. We
note that $\theta_{5}\sim \xi^{2-D}$ for $\xi\rightarrow +\infty$
implying a finite mass. This contrasts with the asymptotic
behavior (\ref{emden10}) of the solutions of the Lane-Emden
equation with index $n>n_{5}$. For $D=2$, $n_{5}\rightarrow
+\infty$ and we are lead back to the isothermal case where an
analytical solution is also known \cite{sc}.

Finally, for $D=1$, the Lane-Emden equation reduces to the form
\begin{equation}
{d^{2}\theta\over d\xi^{2}}=-\theta^{n}. \label{emden22}
\end{equation}
This equation corresponds to the motion of a fictitious particle
in a potential $V(\theta)=\theta^{n+1}/(n+1)$, where $\theta$
plays the role of position and $\xi$ the role of time. The first
integral of motion is
\begin{equation}
E={1\over 2}\biggl ({d\theta\over d\xi}\biggr
)^2+{\theta^{n+1}\over n+1}. \label{first}
\end{equation}
The ``energy'' $E$ is determined by the boundary condition (\ref{emden4b})
yielding $E=1/(n+1)$.  Thus, the solution $\theta(\xi)$ can be
written in integral form as
\begin{equation}
\int_{\theta}^{1}{dx\over (1-x^{n+1})^{1/2}}=\biggl ({2\over
n+1}\biggr )^{1/2}\xi. \label{emden23}
\end{equation}
Except for $n=0$ and $n=1$, it seems not possible to obtain
$\theta(\xi)$ in a closed form. However, the zero $\xi_{1}$ of
$\theta$ is explicitly given by
\begin{equation}
\xi_{1}=\biggl ({n+1\over 2}\biggr )^{1/2}\sqrt{\pi}{\Gamma
({2+n\over 1+n})\over \Gamma ({3+n\over 2(1+n)})} \label{emden24}
\end{equation}
Therefore, $\xi_1\sim \sqrt{n/2}$ for $n\rightarrow +\infty$.
Furthermore, according to Eq. (\ref{first}), we have
$\theta'(\xi_1)=-(2/(n+1))^{1/2}$.

\subsection{The Milne variables}
\label{sec_milne}

\begin{figure}
\centerline{ \psfig{figure=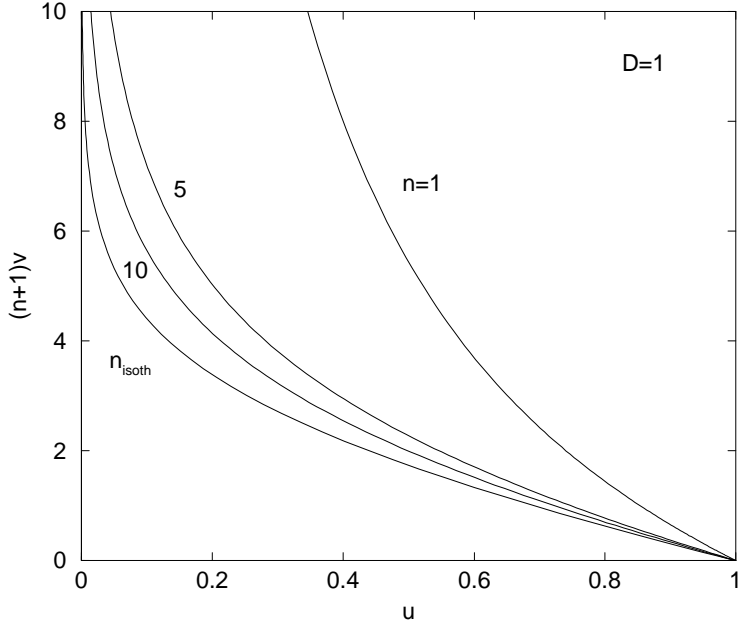,angle=0,height=8.5cm}}
\caption{Phase portrait of the Lane-Emden equation in the $(u,v)$
plane for $D<2$ (specifically $D=1$). The value of the polytropic
index is indicated on each curve. For $n\rightarrow +\infty$, denoted
$n_{\rm isoth}$, we recover the phase portrait of isothermal spheres.}
\label{uvD1}
\end{figure}

\begin{figure}
\centerline{ \psfig{figure=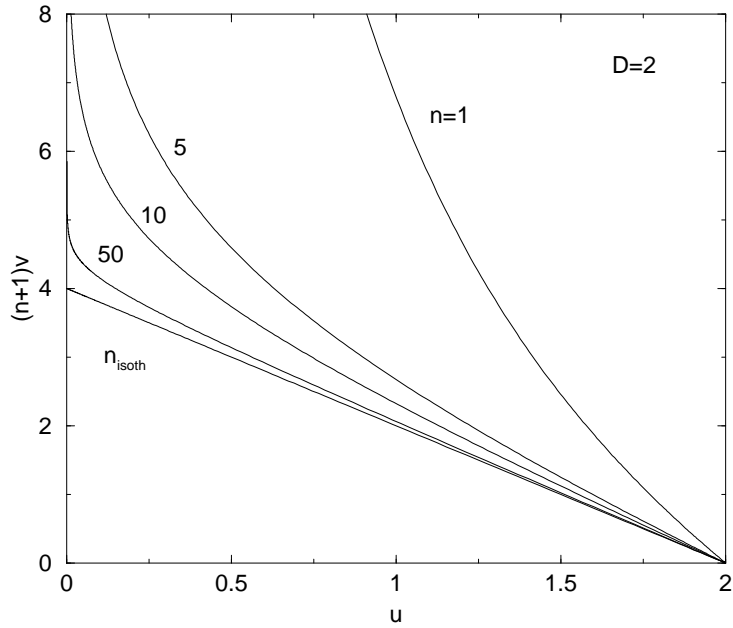,angle=0,height=8.5cm}}
\caption{Phase portrait of the Lane-Emden equation in the $(u,v)$
plane for $D=2$. At this dimension $n_{5}\rightarrow +\infty$, so
that the $2D$-Schuster solution (\ref{emden21}) becomes equivalent
to the $2D$-isothermal solution \cite{sc}. In the $(u,v)$ plane
this corresponds to a straight line.} \label{uvD2}
\end{figure}

\begin{figure}
\centerline{ \psfig{figure=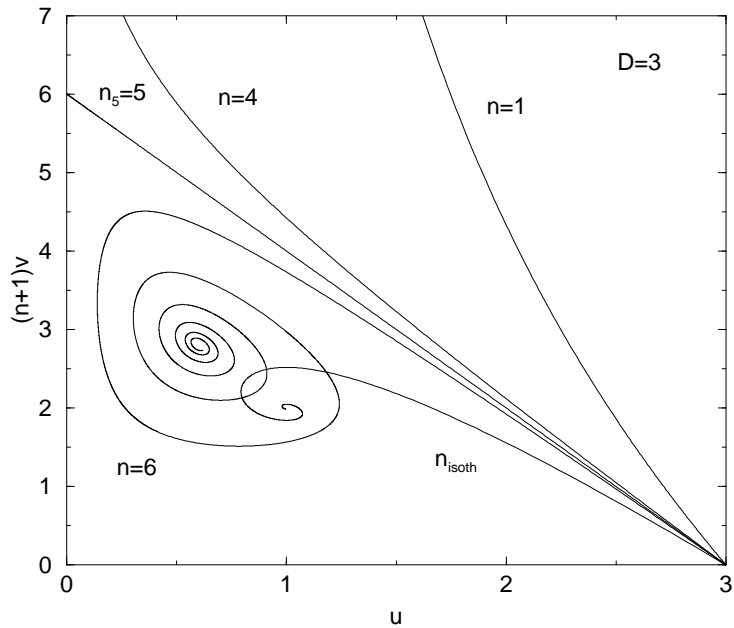,angle=0,height=8.5cm}}
\caption{Phase portrait of the Lane-Emden equation in the $(u,v)$
plane for $2<D<10$ (specifically $D=3$).  } \label{uvD3}
\end{figure}

\begin{figure}
\centerline{ \psfig{figure=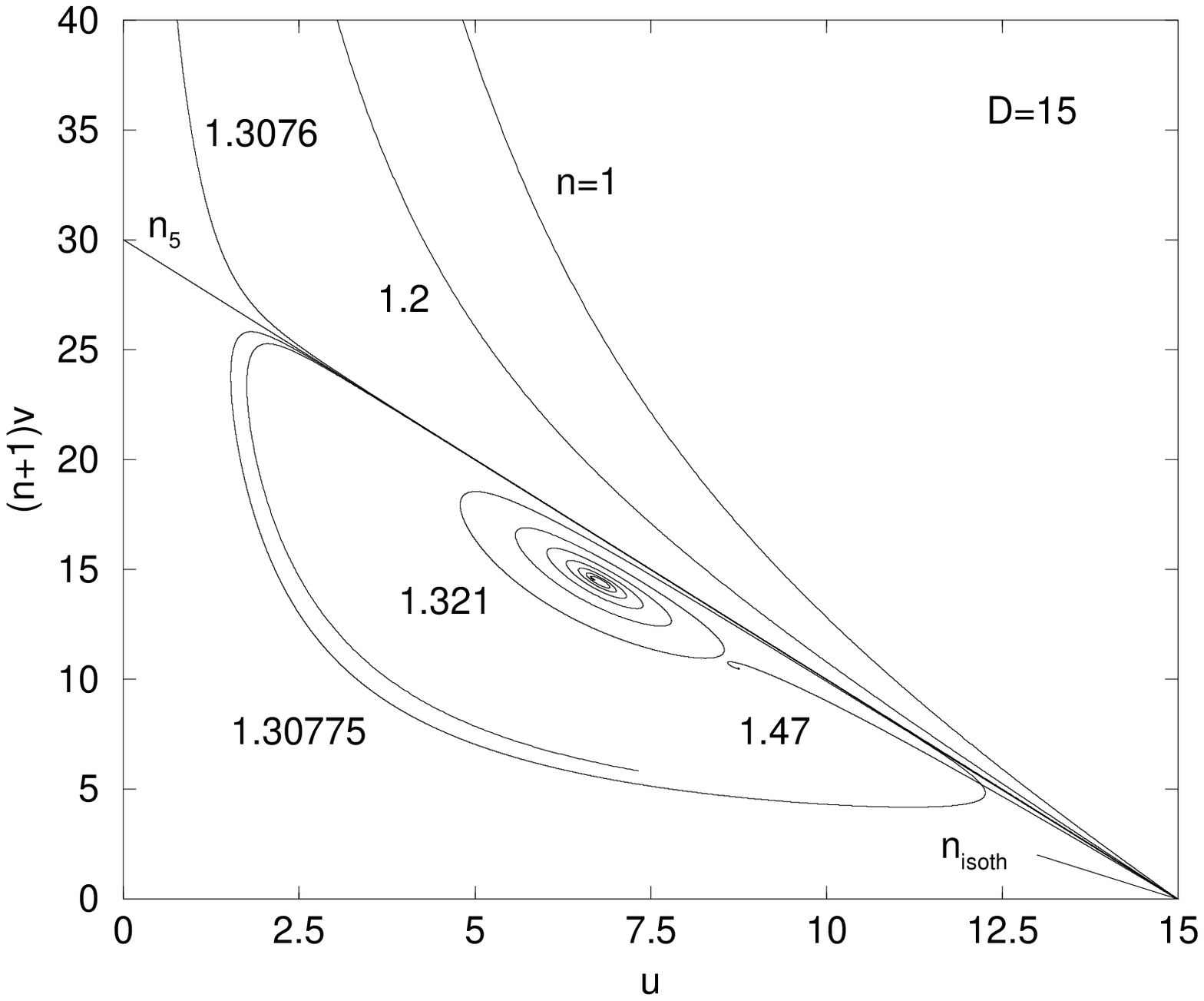,angle=0,height=8.5cm}}
\caption{Phase portrait of the Lane-Emden equation in the $(u,v)$
plane for $D>10$ (specifically $D=15$). For $D=10$,
$n_{-}\rightarrow +\infty$. Therefore, for $n>n_{5}=3/2$, the
phase portrait is always a spiral except for the index $n=+\infty$
for which the spiral is reduced to a point.  } \label{uvD15}
\end{figure}

As is well-known \cite{chandra}, polytropic spheres satisfy a
homology theorem: if $\theta(\xi)$ is a solution of the Lane-Emden
equation, then $A^{2/(n-1)}\theta(A\xi)$ is also a solution, with
$A$ an arbitrary constant. This means that the profile of a
polytropic configuration of index $n$ is always the same
(characterized intrinsically by the function $\theta$), provided
that the central density and the typical radius are rescaled
appropriately. Because of this homology theorem, the second order
differential equation (\ref{emden2}) can be reduced to a {\it
first order} differential equation for the Milne variables
\begin{equation}
u=-{\xi \theta^{n}\over \theta'}\qquad {\rm and}\qquad
v=-{\xi\theta'\over \theta}. \label{milne1}
\end{equation}
Taking the logarithmic derivative of $u$ and $v$ with respect to
$\xi$ and using Eq.~(\ref{emden2}), we get
\begin{equation}
{1\over u}{du\over d\xi}={1\over\xi}(D-nv-u), \label{milne2}
\end{equation}
\begin{equation}
{1\over v}{dv\over d\xi}={1\over\xi}(2-D+u+v). \label{milne3}
\end{equation}
Taking the ratio of the foregoing equations, we obtain
\begin{equation}
{u\over v}{dv\over du}=-{u+v-D+2\over u+nv-D}. \label{milne4}
\end{equation}
The solution curve in the $(u,v)$ plane is plotted in
Figs.~\ref{uvD1}-\ref{uvD15} for different values of $D$ and $n$.
The $(u,v)$ curve is parameterized by $\xi$. It starts, at
$\xi=0$, from the point $(u,v)=(D,0)$ with a slope
$(dv/du)_{0}=-(D+2)/nD$. The points of horizontal tangent are
determined by $u+v-D+2=0$ and the points of vertical tangent by
$u+nv-D=0$. These two lines intersect at
\begin{equation}
u_{s}={(D-2)n-D\over n-1},\qquad v_{s}={2\over n-1},
\label{milne4b}
\end{equation}
which corresponds to the singular solution (\ref{emden4}).

The Milne variables can be expressed in terms of $\xi$ explicitly
for particular values of the polytropic index. For $n=0$, using
Eq. (\ref{emden17}), we have
\begin{equation}
u=D, \qquad v={2\xi^{2}\over 2D-\xi^{2}}. \label{milne5}
\end{equation}
For $n=n_{5}$, using Eq. (\ref{emden21}), we have
\begin{equation}
u={D\over 1+{\xi^{2}\over D(D-2)}},\qquad v={1\over
D}{\xi^{2}\over  1+{\xi^{2}\over D(D-2)}}. \label{milne6}
\end{equation}
Eliminating $\xi$ between these two relations, we get
\begin{equation}
{D\over D-2}v+u=D. \label{milne7}
\end{equation}

More generally, using the asymptotic behavior of $\theta(\xi)$
determined in Sec. \ref{sec_emden}, we can deduce the form of the
solution curve in the $(u,v)$ plane.  For ($2<D<10$, $n>n_{5}$)
and for ($D>10$, $n_{5}<n<n_{-}$), the solution curve spirals
indefinitely around the limit point $(u_{s},v_{s})$.  For ($D>10$,
$n>n_{-}$), the curve reaches the point $(u_{s},v_{s})$ without
spiraling. For $D=10$, $n_{-}\rightarrow +\infty$ and $n_{5}=3/2$.
For $D<2$ and for ($D>2$, $n<n_{5}$), the $(u,v)$ curve is
monotonic and tends to $(u,v)=(0,+\infty)$ as $\xi\rightarrow
\xi_{1}$. More precisely, using Eq. (\ref{emden8}), we have for
$n<+\infty$
\begin{equation}
uv^{n}\sim \omega_{n}^{n-1},\qquad \omega_{n}=-\xi_{1}^{n+1\over
n-1}\theta'_{1} \qquad (\xi\rightarrow \xi_{1}). \label{milne8}
\end{equation}
For $n\rightarrow +\infty$, $\xi_{1}\rightarrow +\infty$ and we
are lead to our previous study \cite{sc}. For $D=2$,
$n_{5}\rightarrow +\infty$.

For $n\rightarrow +\infty$, it is easy to check that the
Lane-Emden function $\theta$ (for polytropes) is related to the
Emden function $\psi$ (for isothermal spheres) by the equivalent
\begin{equation}
\theta^n\sim {1\over n} {\rm e}^{-\psi}.
\label{milne9}
\end{equation}
This suggests to introduce the variables $U=u$ and $V=(n+1)v$ instead
of (\ref{milne1}). For $n\rightarrow +\infty$, $U$ and $V$ tend to the
Milne variables $u$, $v$ defined in the case of isothermal spheres
(see \cite{sc}). We could have introduced these variables since the
beginning but we prefer to respect the notations used by Chandrasekhar
in his classical monograph \cite{chandra}.

\subsection{The thermodynamical parameters}
\label{sec_thermo}

For an incomplete polytrope confined within a box of radius $R$, the solution
of Eq.~(\ref{emden2}) is terminated by the box at the normalized
radius (see Eq. (\ref{emden1}))
\begin{eqnarray}
\alpha=\biggl\lbrack {S_{D} G\rho_{0}^{1-1/n}\over
K(n+1)}\biggr\rbrack^{1/2}R. \label{thermo1}
\end{eqnarray}
We shall now relate the parameter $\alpha$ to the polytropic
constant $K$ (or generalized temperature $T$) and to the energy
$E$. Starting from the relation
\begin{eqnarray}
M=\int_{0}^{R}\rho S_{D} r^{D-1}dr =S_{D}\rho_{0}\biggl\lbrack
{K(1+n)\over S_{D} G
\rho_{0}^{1-1/n}}\biggr\rbrack^{D/2}\int_{0}^{\alpha}
\theta^{n}\xi^{D-1}d\xi,
\label{thermo2}
\end{eqnarray}
and using the Lane-Emden equation (\ref{emden2}), we get
\begin{equation}
M=-S_{D} \rho_{0}\biggl\lbrack {K(1+n)\over
S_{D}G\rho_{0}^{1-1/n}}\biggr\rbrack^{D/2}\alpha^{D-1}\theta'(\alpha).
\label{thermo3}
\end{equation}
Expressing the central density in terms of $\alpha$, using Eq.
(\ref{thermo1}), we obtain after some rearrangements
\begin{equation}
M=-S_{D}\biggl\lbrack {K(1+n)\over S_{D} G}\biggr\rbrack^{n\over
n-1} R^{(D-2)n-D\over n-1}\alpha^{n+1\over n-1}\theta'(\alpha).
\label{thermo4}
\end{equation}
For a complete polytrope with radius $R_{*}<R$, we need to stop the
integration at $\xi=\xi_{1}$. Thus, the equivalent of the foregoing
equation is the
``mass-radius'' relation
\begin{equation}
M^{n-1\over n}R_{*}^{(D-2)(n_3 -n)\over n}={K(1+n)\over G
S_{D}^{1/n}}\omega_{n}^{n-1\over n}, \label{thermo4b}
\end{equation}
where $\omega_n$ is defined by Eq. (\ref{milne8}). For $n=n_{3}$, the
mass is independent on the radius. This mathematical property is
related to the limiting mass of Chandrasekhar for relativistic white
dwarf stars \cite{chandra}. For $n=1$, the radius is independent on
mass. For incomplete polytropes, the parameter
\begin{equation}
\eta\equiv {M\over S_{D}}\biggl\lbrack {S_{D} G\over
K(1+n)}\biggr\rbrack^{n\over n-1}{1\over R^{(D-2)n-D\over n-1} },
\label{thermo5}
\end{equation}
can be considered as a normalized inverse temperature \cite{poly}.
Indeed, for a given mass $M$ and box radius $R$, it simply depends
on the polytropic constant $K$ which is itself related to $\beta$
via Eqs. (\ref{maxent13}) and (\ref{maxent9}).  In addition, for
$n\rightarrow +\infty$, the parameter $\eta$ reduces to the
corresponding one for isothermal spheres ($\eta\sim
\eta_{\infty}/n$, $\eta_{\infty}={\beta GMm\over R^{D-2}}$,
$\beta=1/kT$) \cite{sc}. In terms of this parameter, Eq.
(\ref{thermo4}) can be rewritten
\begin{equation}
\eta=-\alpha^{n+1\over n-1}\theta'(\alpha). \label{thermo6}
\end{equation}
This relation can be expressed in terms of the values of the Milne
variables at the normalized box radius. Writing $u_{0}=u(\alpha)$
and $v_{0}=v(\alpha)$, we get
\begin{equation}
\eta=(u_{0}v_{0}^{n})^{1\over n-1}. \label{thermo7}
\end{equation}
For a given box radius $R$ and a given polytropic constant $K$ (or
generalized temperature $T$), this equation determines the
relation between the mass $M$ and the central density $\rho_0$
(through the parameter $\alpha$). For $D\le 2$ and for ($D>2$,
$n<n_5$), the normalized box radius $\alpha$ is necessarily
restricted by the inequality $\alpha\le\xi_{1}$. For the limiting
value $\alpha=\xi_{1}$, corresponding to a complete polytrope with
radius $R_{*}=R$, we have
\begin{equation}
\eta(\xi_{1})=\omega_{n}, \label{thermo8}
\end{equation}
More generally, for complete polytropes with radius $R_{*}\le R$, we have
\begin{equation}
\eta=\omega_{n}\biggl ({R_{*}\over R}\biggr )^{n(D-2)-D\over n-1}.
\label{complet1a}
\end{equation}
Coming back to incomplete polytropes, we note that for the index
$n=n_{5}$, Eq. (\ref{thermo7}) can be written  explicitly as
\begin{equation}
\eta={\alpha^{D+2\over 2}\over D(1+{\alpha^{2}\over
D(D-2)})^{D/2}}. \label{thermo9}
\end{equation}
For $\alpha\rightarrow +\infty$, we observe that $\eta\sim
(3/\alpha)^{1/2}$ for $D=3$.

The computation of the energy is a little more intricate. We first
recall the expression of the Virial theorem in dimension $D\neq 2$
(see Appendix \ref{sec_vtD}):
\begin{equation}
2K+(D-2)W=DV_{D}R^{D}p(R), \label{thermo10}
\end{equation}
where $V_{D}=S_{D}/D$ is the volume of a hypersphere with unit
radius. For $D=2$,
the expression of the Virial theorem is (see Appendix
\ref{sec_vtD}):
\begin{equation}
2K-{GM^{2}\over 2}=2\pi R^{2}p(R). \label{thermo11}
\end{equation}
The potential energy of a polytrope can be calculated as follows.
Combining the condition of hydrostatic equilibrium
(\ref{eqhydro}) with the equation of state (\ref{maxent12}), we
get
\begin{equation}
(n+1){d\over dr}\biggl ({p\over\rho}\biggr )=-{d\Phi\over dr}.
\label{thermo12}
\end{equation}
This equation integrates to give
\begin{equation}
\Phi=-(n+1)\biggl ({p\over\rho}-{p(R)\over\rho(R)}\biggr
)+\Phi(R). \label{thermo13}
\end{equation}
Inserting this relation in the integral (\ref{maxent3}) defining
the potential energy $W$ and  recalling that the kinetic energy
can be written $K=(D/2)\int p d^D{\bf r}$, we obtain
\begin{equation}
W=-{1\over D}(n+1)K+{1\over 2}(n+1){p(R)\over \rho(R)}M+{1\over
2}M\Phi(R). \label{thermo14}
\end{equation}
For $D\neq 2$, $\Phi(R)=-GM/[(D-2)R^{D-2}]$ and for $D=2$, we take
the convention $\Phi(R)=0$ (see Appendix \ref{sec_gfD}). Eliminating the kinetic energy between
Eqs. (\ref{thermo10}), (\ref{thermo11}) and (\ref{thermo14}), we
obtain for $D\neq 2$:
\begin{equation}
W={-D\over D+2-(D-2)n}\biggl\lbrace
(n+1)V_{D}R^{D}p(R)-(n+1){p(R)\over \rho(R)}M+{GM^{2}\over
(D-2)R^{D-2}}\biggr\rbrace, \label{thermo15}
\end{equation}
and for $D=2$:
\begin{equation}
W=-(n+1){GM^{2}\over 8}-{1\over 2}(n+1)\pi R^{2}p(R)+{1\over
2}(n+1){p(R)\over \rho(R)}M. \label{thermo16}
\end{equation}
For complete polytropes for which $p(R_{*})/\rho(R_{*})=0$, we obtain the
$D$-dimensional generalization of the Betti-Ritter  formula
\cite{chandra}:
\begin{equation}
W={-D\over D+2-(D-2)n}{GM^{2}\over (D-2)R_{*}^{D-2}}\qquad (D\neq 2),
\label{thermo17}
\end{equation}
\begin{equation}
W=-(n+1){GM^{2}\over 8}+{1\over 2}GM^2\ln\biggl ({R_*\over
R}\biggr )\qquad (D=2). \label{thermo18}
\end{equation}
We note that for $D>2$, the potential energy is infinite for
$n=n_5$ (while the mass is finite). Returning to incomplete polytropes,
the total energy $E=K+W$ can be written for $D\neq 2$:
\begin{eqnarray}
E={-1\over  D+2-(D-2)n}\biggl\lbrace {D(4-D)\over
2(D-2)}\biggl\lbrack {GM^{2}\over R^{D-2}}-(n+1)(D-2)M{p(R)\over
\rho(R)}\biggr\rbrack \nonumber\\
-DV_{D}R^{D}(n+1-D)p(R)\biggr\rbrace, \label{thermo19}
\end{eqnarray}
and for $D=2$:
\begin{equation}
E=-(n-1){GM^{2}\over 8}-{1\over 2}(n-1)\pi R^{2}p(R)+{1\over
2}(n+1){p(R)\over \rho(R)}M. \label{thermo20}
\end{equation}
Expressing the pressure $p(R)$ in terms of the Lane-Emden function
$\theta(\alpha)$ using Eqs. (\ref{maxent12}) and (\ref{emden1}), using Eqs.~(\ref{thermo1}) and (\ref{thermo4})
to eliminate the central density $\rho_{0}$ and the polytropic
constant $K$ (or temperature $T$), and introducing the Milne
variables (\ref{milne1}), we finally obtain for $D\neq 2$:
\begin{equation}
\Lambda\equiv -{ER^{D-2}\over GM^{2}}={-1\over (D-2)n-(D+2)}\biggl
\lbrack {D(4-D)\over 2(D-2)}\biggl (1-{D-2\over v_{0}}\biggr
)+{n+1-D\over n+1}{u_{0}\over v_{0}}\biggr\rbrack,
\label{thermo21}
\end{equation}
and for $D=2$:
\begin{equation}
\Lambda={1\over 8}(n-1)+{1\over 4}{n-1\over n+1}{u_{0}\over
v_{0}}-{1\over 2v_{0}}. \label{thermo22}
\end{equation}
For $D\le 2$ and for ($D>2$, $n<n_5$), the normalized box radius
$\alpha$ in necessarily restricted by the inequality
$\alpha\le\xi_{1}$. For the limiting value $\alpha=\xi_{1}$,
corresponding to a complete polytrope with radius $R_{*}=R$, we
have
\begin{equation}
\Lambda(\xi_{1})=\lambda_{n}, \label{thermo23}
\end{equation}
with
\begin{equation}
\lambda_{n}={-D(4-D)\over 2(D-2)[(D-2)n-(D+2)]} \qquad (D\neq 2),
\label{thermo24}
\end{equation}
\begin{equation}
\lambda_{n}= {1\over 8}(n-1)\qquad (D=2). \label{thermo25}
\end{equation}
More generally, for complete polytropes with radius $R_{*}\le R$, the dimensionless energy is
\begin{equation}
\Lambda=\lambda_{n}\biggl ({R\over R_{*}}\biggr )^{D-2}, \qquad (D\neq 2),
\label{complet1b}
\end{equation}
\begin{equation}
\Lambda={1\over 8}(n-1)-{1\over 2}\ln\biggl ({R_{*}\over R}\biggr )\qquad (D=2).
\label{complet4}
\end{equation}
Eliminating $R_{*}$ between Eqs. (\ref{complet1a}), (\ref{complet1b}) and (\ref{complet4}), we obtain
\begin{equation}
\Lambda\eta^{(n-1)(D-2)\over n(D-2)-D}=\lambda_{n}\omega_{n}^{(n-1)(D-2)\over n(D-2)-D}, \qquad (D\neq 2),
\label{complet3}
\end{equation}
\begin{equation}
\Lambda={1\over 8}(n-1)\biggl\lbrack 1+2\ln\biggl ({\eta\over\omega_{n}}\biggr )\biggr\rbrack, \qquad (D=2).
\label{complet5}
\end{equation}
This defines the branch of complete polytropes in the $\Lambda-\eta$ plane.
Coming back to incomplete polytropes, we note finally that, for $D>2$,
Eq. (\ref{thermo21}) is undetermined for $n=n_{5}$. Calculating the
kinetic energy $K=(D/2)\int p d^{D}{\bf r}$ with
Eqs. (\ref{maxent12}), (\ref{emden1}), (\ref{emden21}), and using the
Virial theorem (\ref{thermo10}) to obtain the potential energy, we
find after simplification that
\begin{equation}
\Lambda_{5}={D^{2}\over 4}(4-D)\biggl \lbrack 1+{\alpha^{2}\over
D(D-2)}\biggr \rbrack^{D}{1\over
\alpha^{D+2}}\int_{0}^{\alpha}{\xi^{D-1}\over \bigl \lbrack
1+{\xi^{2}\over D(D-2)}\bigr \rbrack^{D}}d\xi-{D\over
2\alpha^{2}}. \label{thermo26}
\end{equation}
For $D=3$, the integral is explicitly given by
\begin{equation}
\int_{0}^{\alpha}{\xi^{2}\over \bigl (1+{\xi^{2}\over 3}\bigr
)^{3}} d\xi={9\alpha (\alpha^{2}-3)\over
8(\alpha^{2}+3)^{2}}+{3\sqrt{3}\over 8}{\arctan}\biggl
({\alpha\over\sqrt{3}}\biggr ). \label{thermo27}
\end{equation}
We note that for $\alpha\rightarrow +\infty$, the energy diverges
like $\Lambda_{5}\sim (\pi\sqrt{3}/64)\alpha$ for $D=3$.

\subsection{The minimum temperature and minimum energy}
\label{sec_min}

\begin{figure}
\centerline{ \psfig{figure=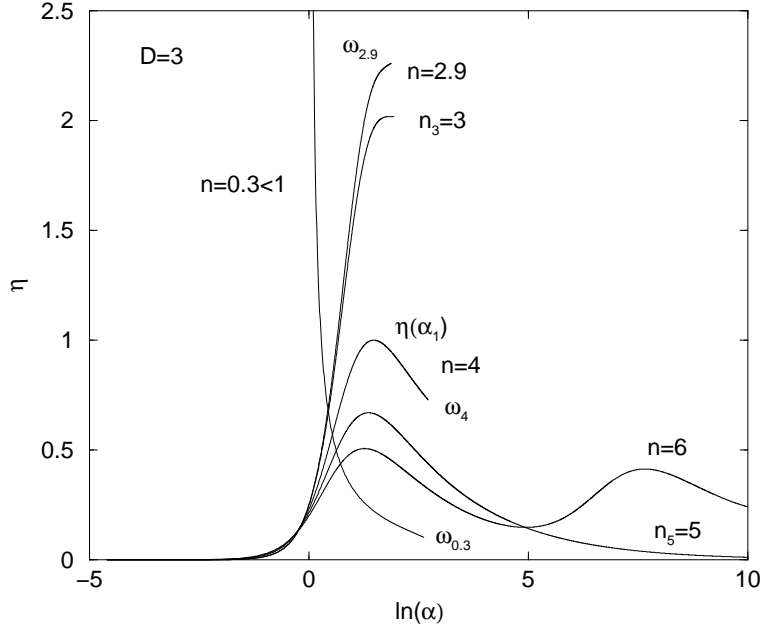,angle=0,height=8.5cm}}
\caption{Mass-central density profiles for polytropic configurations in a
space of dimension $2<D<10$ (specifically $D=3$). A mass peak
appears for the first time for the index $n_{3}$. For
$n>n_{5}$ the profile displays an infinity of peaks. For $n<1$,
$\eta$ is a decreasing function of $\alpha$. For $D\le 2$, the $\eta(\alpha)$
curves are monotonic.} \label{alphaeD3}
\end{figure}

\begin{figure}
\centerline{ \psfig{figure=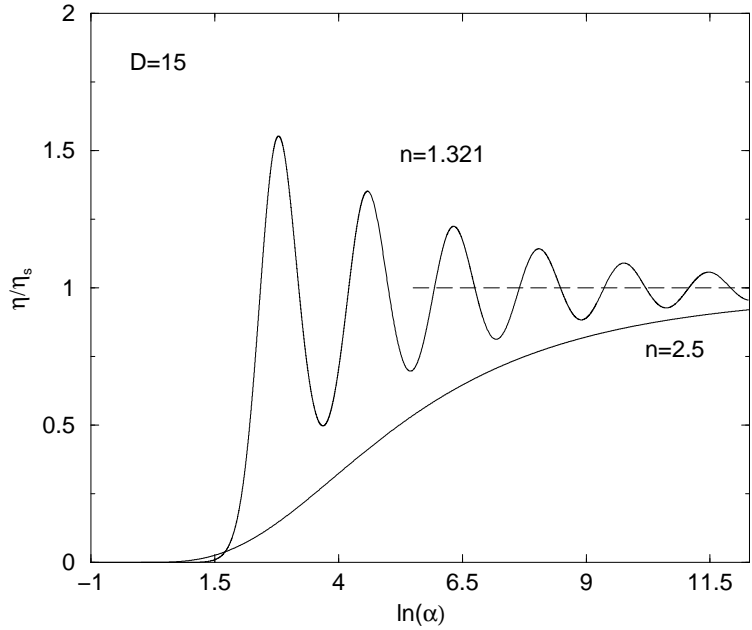,angle=0,height=8.5cm}}
\caption{Mass-central density profiles for polytropic configurations in a
space of dimension $D>10$ (specifically $D=15$). For
$n_{5}<n<n_{-}$ (specifically $n=1.321$) the profile displays an
infinity of peaks. For $n>n_{-}$ (specifically $n=2.5$) the
function $\eta(\alpha)\rightarrow \eta_{s}$ for $\alpha\rightarrow
+\infty$ without oscillating. } \label{alphaeD15}
\end{figure}

The curve $\eta(\alpha)$ presents an extremum at points
$\alpha_{k}$ such that $d\eta/d\alpha(\alpha_{k})=0$. Using
Eqs.~(\ref{thermo7}), (\ref{milne2}) and (\ref{milne3}), we find
that this condition is equivalent to
\begin{equation}
u_{0}={(D-2)n-D\over n-1}=u_{s}. \label{min1}
\end{equation}
Therefore, the points where $\eta$ is extremum are determined by
the intersections between the solution curve in the $(u,v)$ plane
and the straight line defined by Eq. (\ref{min1}). The number of
extrema depends on the value of $D$ and $n$. It can be determined
easily by a graphical construction using Figs.
\ref{uvD1}-\ref{uvD15} (an explicit construction is made in Fig.
\ref{uvparabole3}; see also \cite{poly} for $D=3$). For $D\le 2$,
$u_{s}<0$ for $n>1$ and $u_{s}>D$ for $n<1$ so that there is no
extremum (case A). For $2<D\le 10$, there is no extremum for $n\le
n_{3}$ (case A), there is one maximum for $n_{3}<n\le n_{5}$ (case
$B$) and there is an infinity of extrema for $n>n_{5}$ (case $C$).
They exhibit damped oscillations towards the value $\eta_s$
corresponding to the singular solution (\ref{emden4}).
Asymptotically, the $\alpha_{k}$ follow a geometric progression
$\alpha_k\sim {\rm exp}\lbrace 2k\pi/\sqrt{-\Delta}\rbrace$ (see
\cite{poly} for $D=3$). For $D>10$, there is no extremum for $n\le
n_{3}$ (case $A$), there is one maximum for $n_{3}<n\le n_{5}$
(case $B$), there is an infinity of extrema for $n_{5}<n<n_{-}$
(case $C$) and there is no extremum for $n\ge n_{-}$ (case $D$).
This last case corresponds to an overdamped evolution towards the
value $\eta_{s}$ (in our mechanical analogy of Sec.
\ref{sec_emden}). For $n=n_{5}$, using Eq.  (\ref{thermo9}), we
find that the extremum is located at $\alpha_{1}=\sqrt{D(D+2)}$.  For
incomplete polytropes, the parameter $\eta$ is restricted by the
inequalities (see Figs.
\ref{alphaeD3}-\ref{alphaeD15})
\begin{equation}
\eta\le \omega_{n} \qquad ({\rm case \ A}), \label{min2}
\end{equation}
\begin{equation}
\eta\le \eta(\alpha_{1})\qquad ({\rm cases \ B\ and \ C }),
\label{min3}
\end{equation}
\begin{equation}
\eta\le \eta_{s}\qquad ({\rm case \ D}). \label{min4}
\end{equation}
These inequalities  determine a maximum mass (for given $T$ and
$R$) or a minimum temperature (for given $M$ and $R$) beyond which
the system will either converge toward a complete polytrope with
radius $R_*<R$ (if it is stable) or collapse. This dynamical
evolution will be studied in Sec. \ref{sec_dynamics} for
the nonlinear Smoluchowski-Poisson system.

\begin{figure}
\centerline{ \psfig{figure=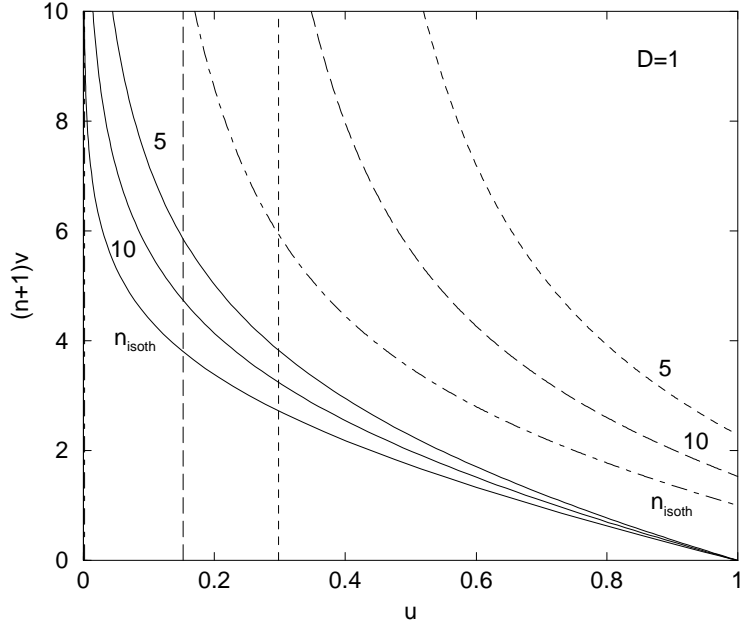,angle=0,height=8.5cm}}
\caption{Graphical construction determining the extrema of
$\Lambda(\alpha)$ for $D<2$ (specifically $D=1$). The solid lines
correspond to the solution curves and the dashed lines to the
curves defined by Eq. (\ref{min5}). The curves are labeled by the
value of the polytropic index $n$. The vertical lines correspond
to the asymptote $u=u_{*}$. For $D<2$, there is no intersection so
that $\Lambda(\alpha)$ has no extremum.    } \label{extremaD1}
\end{figure}

\begin{figure}
\centerline{ \psfig{figure=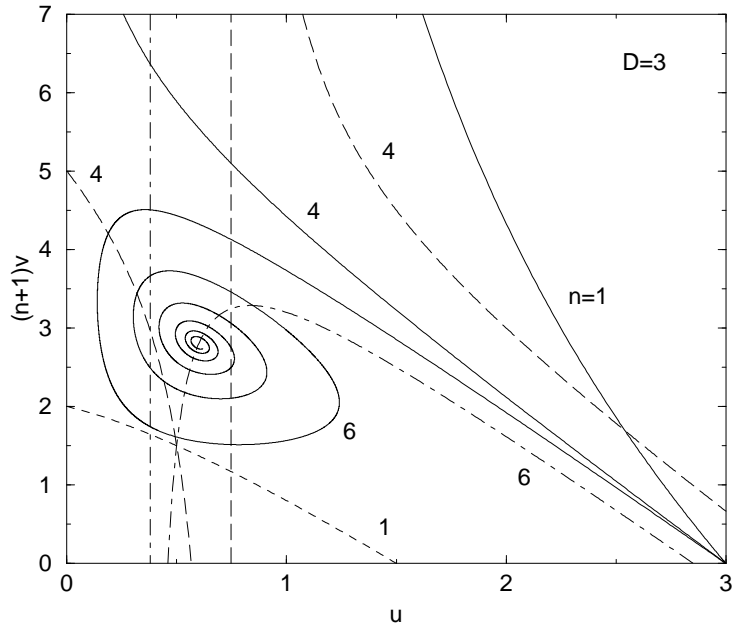,angle=0,height=8.5cm}}
\caption{Same as Fig. \ref{extremaD1} for $2<D<4$ (specifically
$D=3$). The geometrical construction changes for $n=n_{3/2}$, $n=D-1$
and $n=n_{5}$. Typical cases are represented. } \label{extremaD3}
\end{figure}

\begin{figure}
\centerline{ \psfig{figure=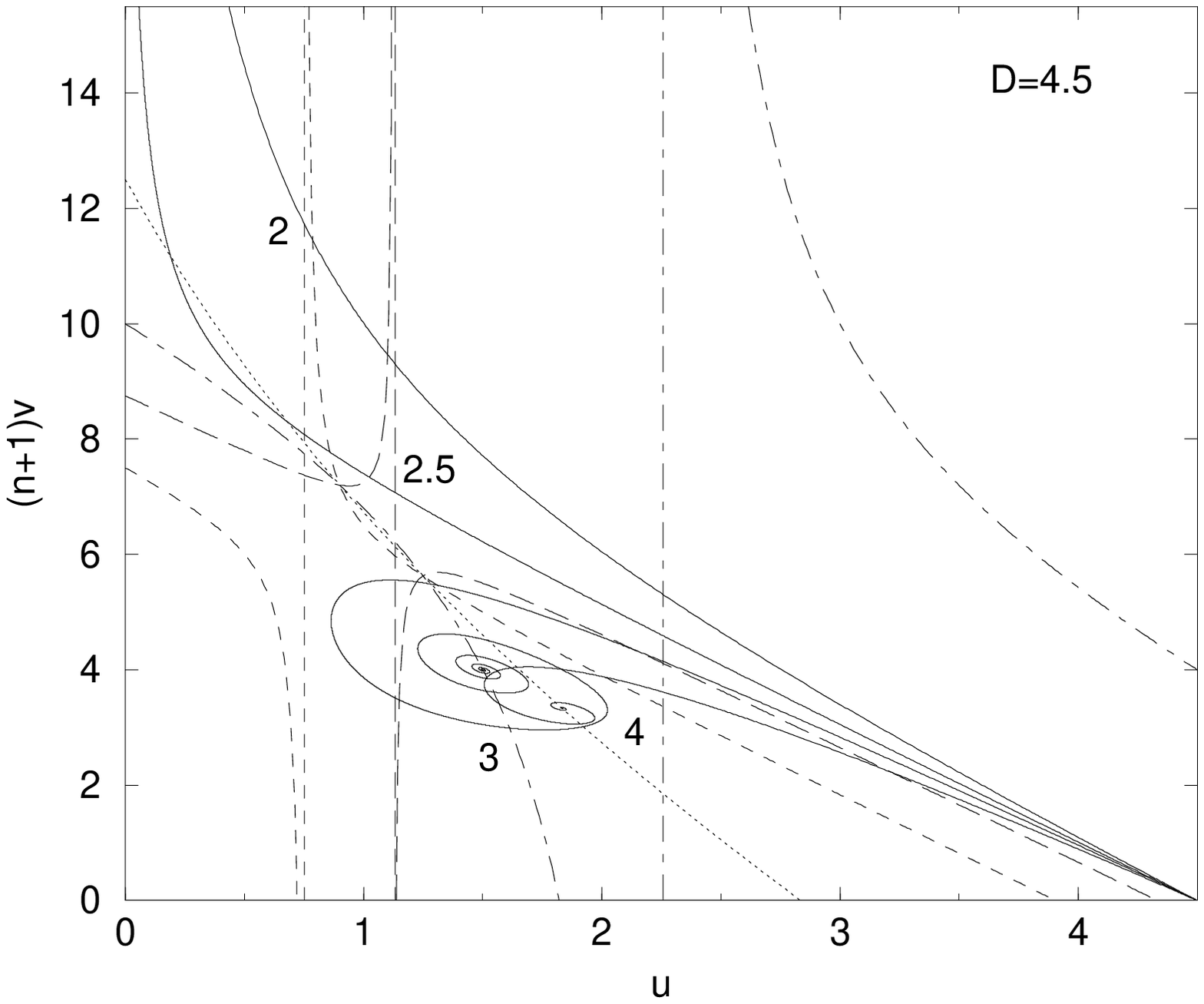,angle=0,height=8.5cm}}
\caption{Same as Fig. \ref{extremaD1} for $4<D<2(1+\sqrt{2})$
(specifically $D=4.5$). The geometrical construction changes for
$n=n_{3/2}$, $n=n_{5}$ and $n=D-1$. Typical cases are represented. The
indices label both the solid curve and the closest broken curve. }
\label{extremaD4.5}
\end{figure}

\begin{figure}
\centerline{ \psfig{figure=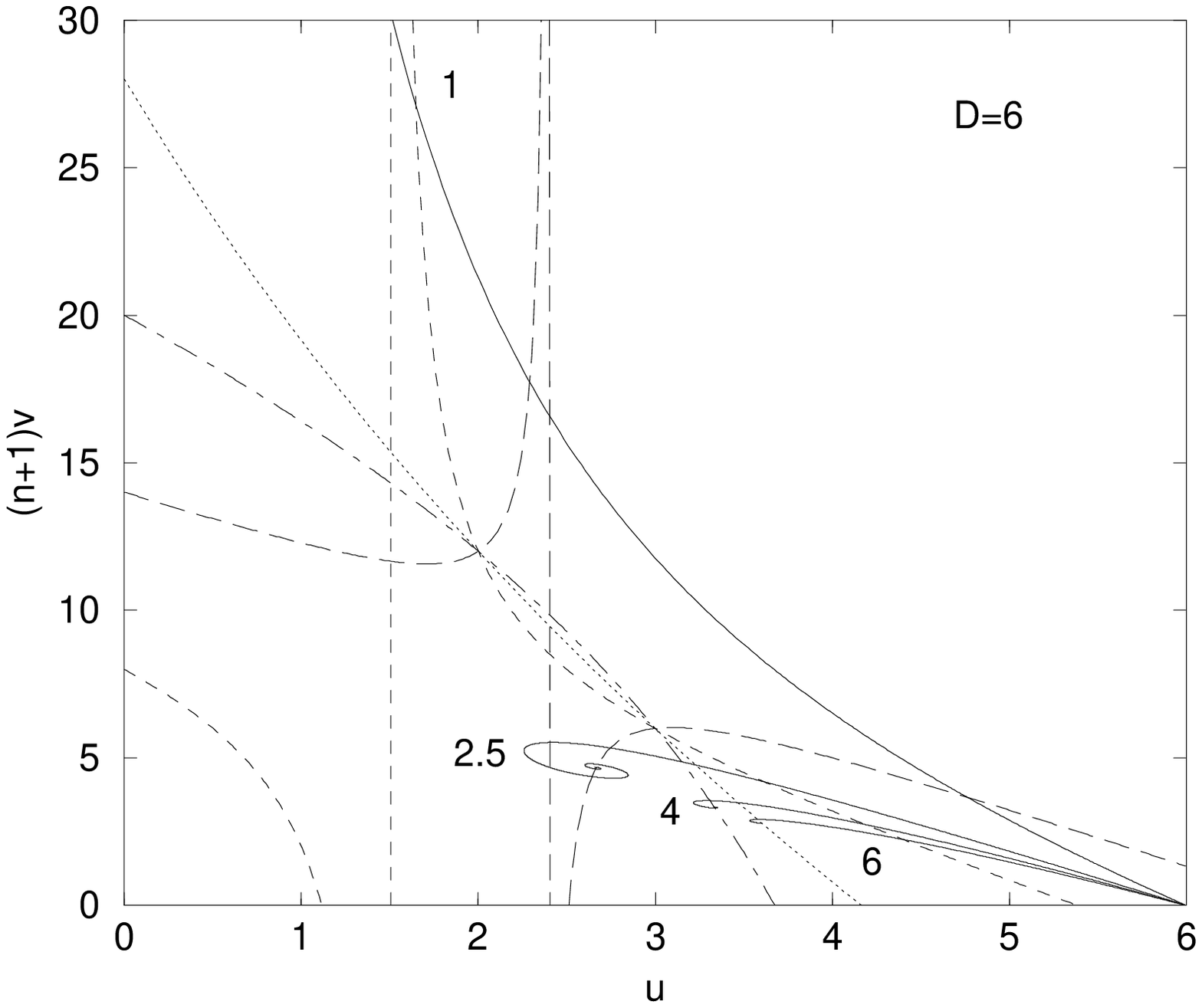,angle=0,height=8.5cm}}
\caption{Same as Fig. \ref{extremaD1} for $2(1+\sqrt{2})<D<10$
(specifically $D=6$). The geometrical construction changes for
$n=n_{5}$, $n=n_{3/2}$ and $n=D-1$. Typical cases are represented. The
indices label both the solid curve and the closest broken curve.
} \label{extremaD6}
\end{figure}

\begin{figure}
\centerline{ \psfig{figure=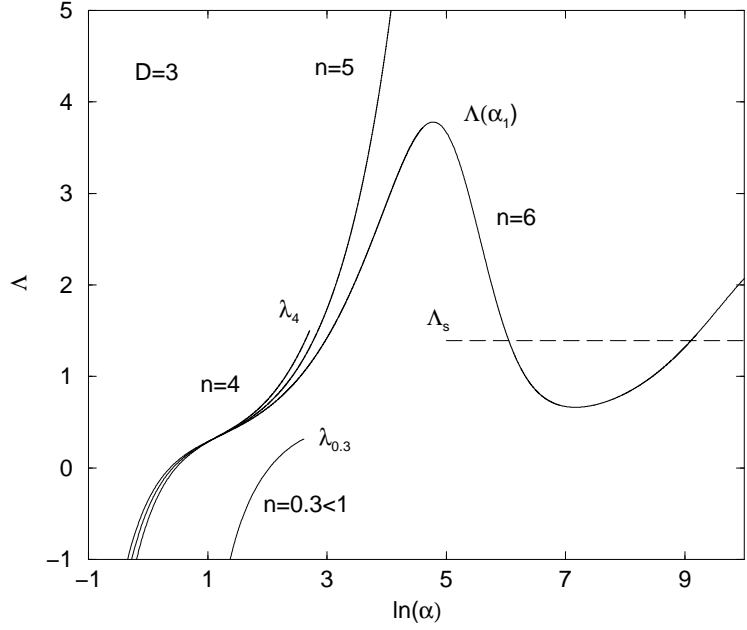,angle=0,height=8.5cm}}
\caption{Evolution of the energy along the series of equilibria
(parameterized by $\alpha$) for $2<D<4$ (specifically $D=3$). For
$n<n_{5}$, the curve has no extremum. For $D\le 2$, the
$\Lambda(\alpha)$ curves are monotonic. } \label{alphalD3}
\end{figure}

\begin{figure}
\centerline{ \psfig{figure=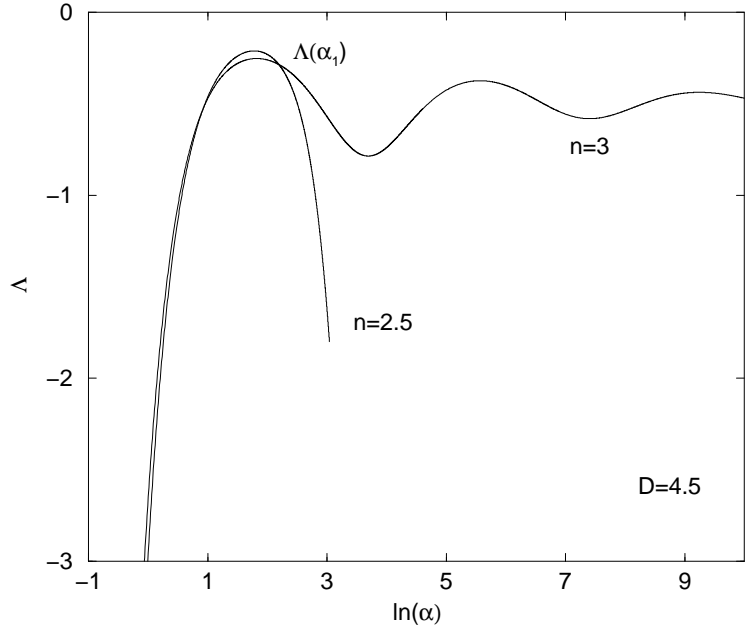,angle=0,height=8.5cm}}
\caption{Evolution of the energy along the series of equilibria
(parameterized by $\alpha$) for $4<D<10$ (specifically $D=4.5$).
For $n<n_{5}$, the curve has one maximum. For $D>10$, the
$\Lambda(\alpha)$ curves are similar to the $\eta(\alpha)$ curves in
Fig. \ref{alphaeD15}.  } \label{alphalD4}
\end{figure}

\begin{figure}
\centerline{ \psfig{figure=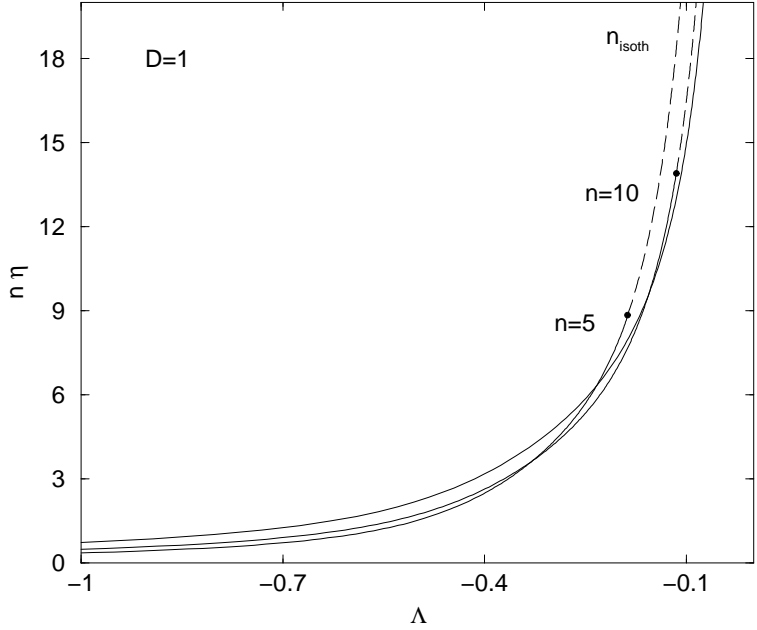,angle=0,height=8.5cm}}
\caption{Generalized caloric curve for $D<2$ (specifically $D=1$).
Note that according to Eq.~(\ref{thermo17}), the potential energy
is necessarily positive for $D<2$, so the region $\Lambda\ge 0$ is
forbidden. We have plotted in dashed line the branch of
complete polytropes with $R_{*}<R$ defined by Eq. (\ref{complet3}). } \label{elD1}
\end{figure}

\begin{figure}
\centerline{ \psfig{figure=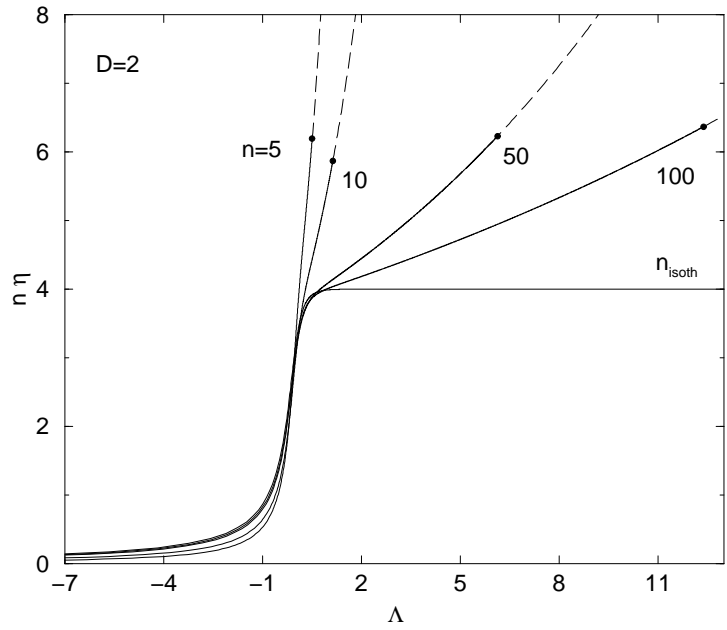,angle=0,height=8.5cm}}
\caption{Generalized caloric curve for $D=2$. We have plotted in 
dashed line the branch of complete polytropes with $R_{*}<R$ defined
by Eq. (\ref{complet5}). }
\label{elD2}
\end{figure}

\begin{figure}
\centerline{ \psfig{figure=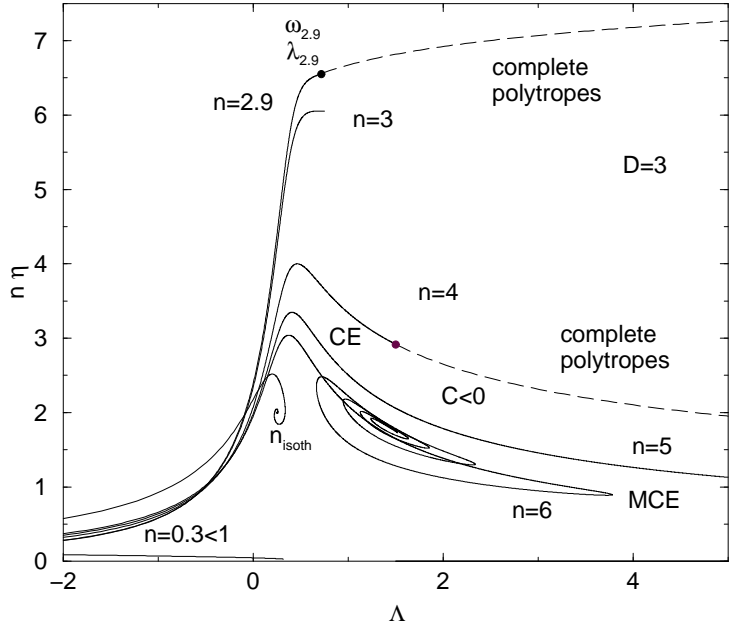,angle=0,height=8.5cm}}
\caption{Generalized caloric curve for $2<D<4$ (specifically
$D=3$). For $n_{3}<n<n_{5}$, the inverse temperature presents a
maximum but not the energy. For $n>n_{3}$, there exists a region of
negative (generalized) specific heats $C=dE/dT<0$ in the
microcanonical ensemble. We have plotted in dashed line the branch of
complete polytropes with $R_{*}<R$ defined by Eq. (\ref{complet3}).}
\label{elD3}
\end{figure}

\begin{figure}
\centerline{ \psfig{figure=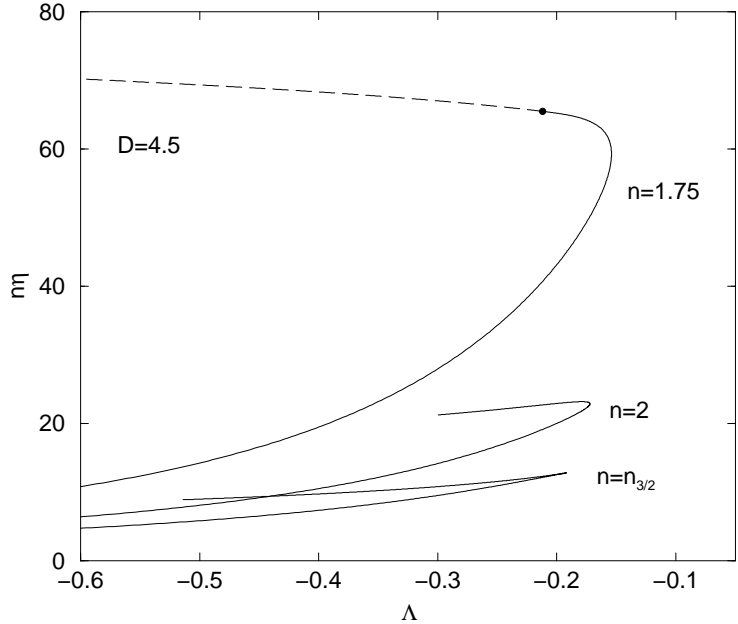,angle=0,height=8.5cm}}
\caption{Generalized caloric curve for $4<D<10$ (specifically
$D=4.5$).  For $n<n_{3}$ (specifically $n_3=1.8$), the energy presents
a minimum but not the inverse temperature. For $n_{3}<n<n_{3/2}$
(specifically $n_{3/2}=2.25$), both the energy and the temperature
present a minimum and the caloric curve $\eta(\Lambda)$ rotates
anti-clockwise. This implies that equilibrium states with positive as
well as {\it negative} specific heats $C=dE/dT$ are stable in the
canonical ensemble. This ``thermodynamical anomaly'' arises because,
as discussed in Sec. \ref{sec_sp}, stellar polytropes with $n<n_{3/2}$
are unphysical (the temperature is negative). For $n=n_{3/2}$ (white
dwarfs), the curve makes an angular point (see Appendix \ref{sec_wd}).}
\label{elD4.5part1}
\end{figure}

\begin{figure}
\centerline{ \psfig{figure=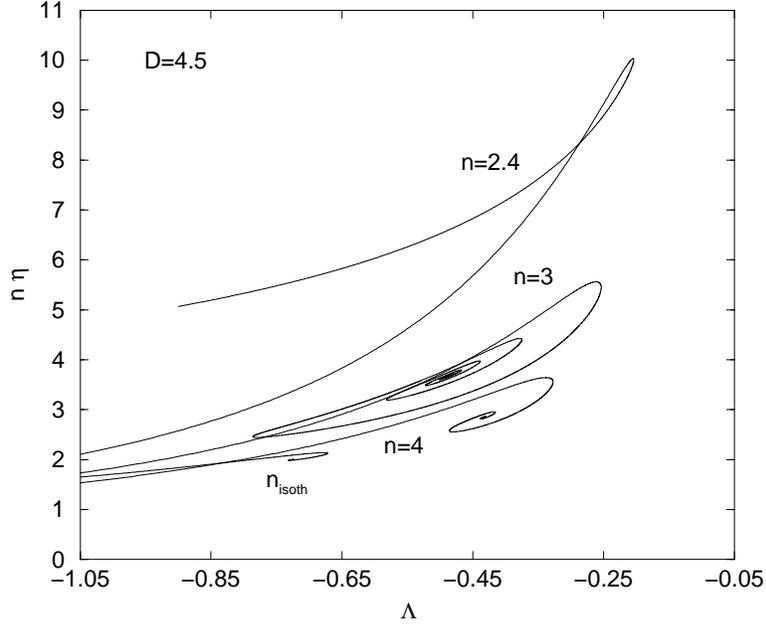,angle=0,height=8.5cm}}
\caption{Continuation of  Fig. \ref{elD4.5part1}.  For $n_{3/2}<n<n_{5}$
(specifically $n_5=2.6$) both the energy and the temperature
present an extremum and the curve rotates clockwise (the curve makes a
``loop''). The region of negative specific heats is now unstable in
the canonical ensemble, as it should. For $n>n_{5}$ (specifically
$n_{5}=2.6$), the energy and temperature present an infinity of
extrema. }
\label{elD4.5part2}
\end{figure}

\begin{figure}
\centerline{ \psfig{figure=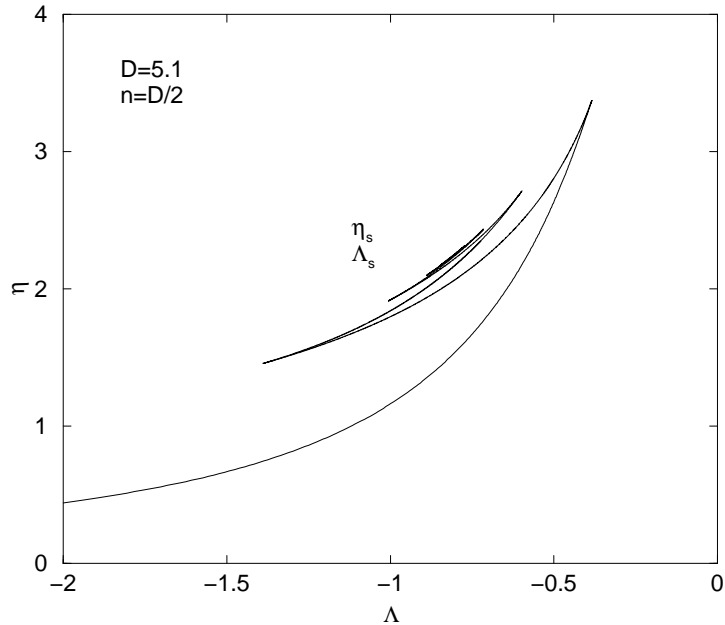,angle=0,height=8.5cm}}
\caption{Generalized caloric curve for $D>2(1+\sqrt{2})$ (specifically
$D=5.1$) and $n=n_{3/2}$. For this particular index, the curve
presents an infinity (because $n_{3/2}>n_{5}$) of angular points towards the singular sphere (see
Appendix \ref{sec_wd}).  } \label{LHpD5.1}
\end{figure}

\begin{figure}
\centerline{ \psfig{figure=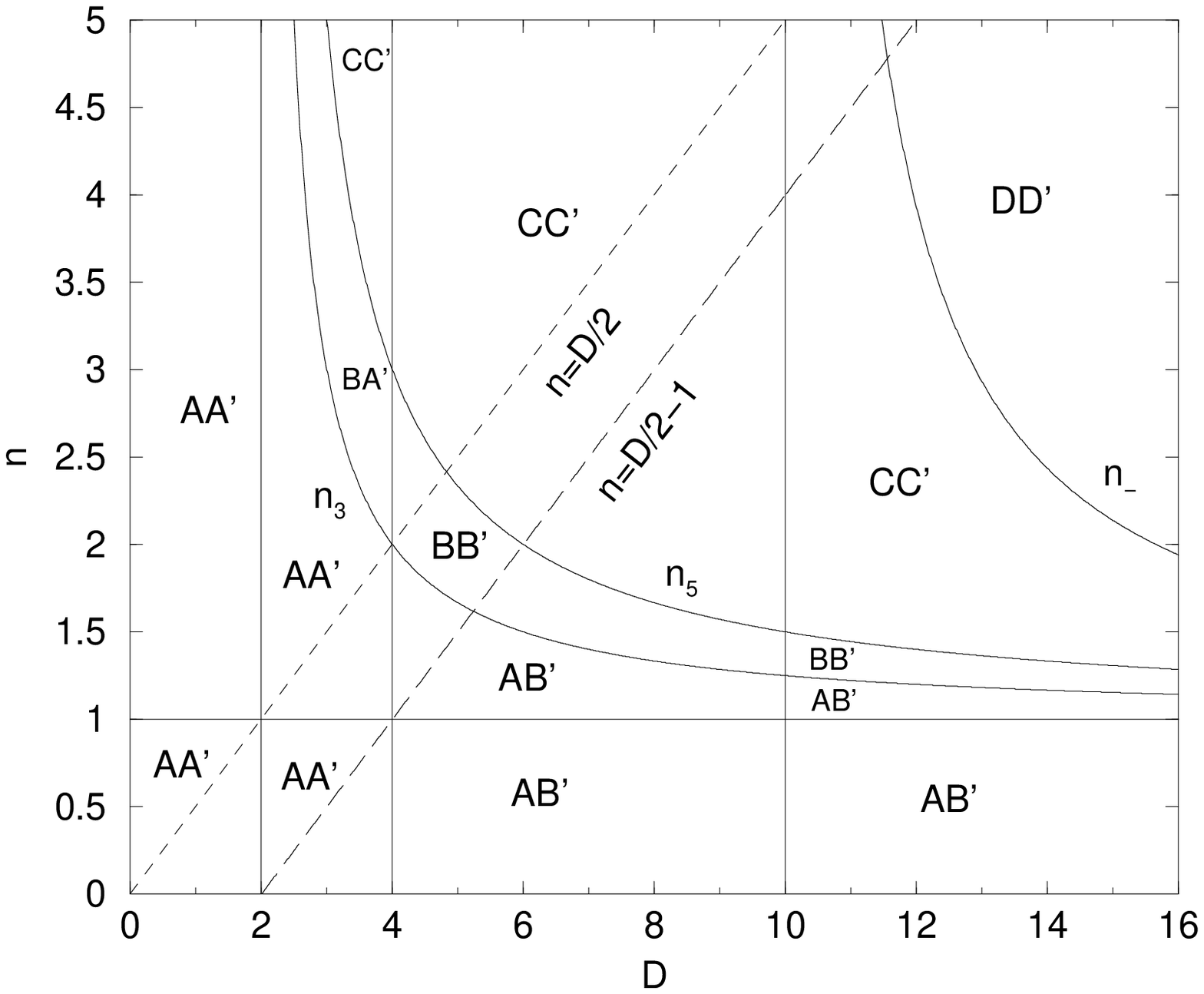,angle=0,height=8.5cm}}
\caption{This figure summarizes the structure of the caloric curve
as a function of the dimension $D$ and the polytropic index $n$.  The
symbols are defined in the text. If we limit ourselves to $n\ge D/2$
(physical stellar polytropes), the region $AB'$ showing a ``thermodynamical
anomaly'' is not accessible. } \label{nD}
\end{figure}

The curve $\Lambda(\alpha)$ presents an extremum at points
$\alpha'_{k}$ such that $d\Lambda/d\alpha(\alpha'_{k})=0$. Using
Eqs.~(\ref{thermo21}), (\ref{milne2}) and (\ref{milne3}), we find
that this condition is equivalent to
\begin{eqnarray}
2(n+1-D)u_{0}^{2}+(n+1)(n+1-D)u_{0}v_{0}+2(D-n-1)(D-1)u_{0}\nonumber\\
+{1\over 2}D(D-4)(n+1)u_{0}+{1\over 2}D(D-4)(n+1)v_{0}+{1\over
2}D(D-4)(2-D)(n+1)=0. \label{min5}
\end{eqnarray}
We can check that the point $(u_{s},v_{s})$ is a solution of this
equation. On the other hand, $v_{0}=D-2$ for $u_{0}=0$ and
$v_{0}\sim -2u_{0}/(n+1)$ for $u_{0}\rightarrow \pm\infty$.
Finally, $v_{0}\rightarrow \infty$ for $u_{0}\rightarrow u_{*}$
where
\begin{eqnarray}
u_{*}={D(4-D)\over 2(n+1-D)}. \label{min6}
\end{eqnarray}
More precisely, for $u_{0}\rightarrow u_{*}$, we have
\begin{eqnarray}
(u_{0}-u_{*})v_{0}\sim {D(D-4)(D-2)\over 2(n+1)(D-n-1)^{2}}
(n-n_{3/2})(n-n_{5})\equiv \Delta_2(n). \label{min7}
\end{eqnarray}
The two roots of $\Delta_2(n)$ are $n=n_{3/2}$ and $n=n_5$.
They coincide at the particular dimension $D=2(1+\sqrt{2})$. We
note also that $u_{*}=0$ for $D=4$. For $D>2$ and $n=D-1$, Eq.
(\ref{min5}) reduces to $u+v=D-2$.

The points where $\Lambda(\alpha)$ is extremum are determined by
the intersections between the solution curve in the $(u,v)$ plane
and the curve defined by Eq. (\ref{min5}).  The number of extrema
can thus be determined by a graphical construction using Figs.
\ref{uvD1}-\ref{uvD15}. This graphical construction depends on the
values of $D$ and $n$ and the different cases are shown in Figs.
\ref{extremaD1}-\ref{extremaD6}.  For $D<2$, there is no extremum
(case $A'$). For $2<D<4$, there is no extremum for $n<n_{5}$ (case
$A'$) and there is an infinity of extrema for $n>n_{5}$ (case
$C'$). They exhibit damped oscillations toward the value
$\Lambda_{s}$ corresponding to the singular solution
(\ref{emden4}). For $4<D<10$, there is one maximum for $n<n_{5}$
(case $B'$) and there is an infinity of extrema for $n>n_{5}$
(case $C'$).  For $D>10$, there is one maximum for $n<n_{5}$ (case
$B'$), there is an infinity of extrema for $n_{5}<n<n_{-}$ (case
$C'$), and there is no extremum for $n>n_{-}$ (case $D'$). This
last case corresponds to an overdamped evolution towards the value
$\Lambda_{s}$ (in our mechanical analogy of Sec.
\ref{sec_emden}). The appearance of a maximum for $n<n_5$ when $D>4$
was a surprise in view of  preceding analysis for $D=3$
\cite{taruya1,poly,grand}. The parameter $\Lambda$ is restricted by the
inequalities (see Figs. \ref{alphalD3} and \ref{alphalD4})
\begin{equation}
\Lambda\le \lambda_{n} \qquad ({\rm case \ A'}), \label{min8}
\end{equation}
\begin{equation}
\Lambda<\Lambda(\alpha_{1})\qquad ({\rm cases \ B'\ and \ C' }),
\label{min9}
\end{equation}
\begin{equation}
\Lambda\le \Lambda_{s}\qquad ({\rm case \ D'}). \label{min10}
\end{equation}
These inequalities  determine a minimum energy (for given $M$ and
$R$)  below which the system will either converge toward a
complete polytrope with radius $R_*<R$ (if it is stable) or
collapse.

In Figs.~\ref{elD1}-\ref{LHpD5.1}, we have plotted the generalized
caloric curve $\Lambda-\eta$, giving the inverse temperature as a
function of the energy, for different dimensions $D$ and
polytropic index $n$. This extends the results of our previous
analysis in $D=3$ \cite{poly}. The number of turning points as a
function of $D$ and $n$ is recapitulated in Fig. \ref{nD}.

\subsection{The generalized thermodynamical stability}
\label{sec_stab}

We now come to the generalized thermodynamical stability problem.  We
shall say that a polytrope is stable if it corresponds to a maximum of
Tsallis entropy (free energy) at fixed mass and energy (temperature) in
the microcanonical (canonical) ensemble.  The stability analysis has
already been performed for $D=3$
\cite{taruya1,poly,taruya2,grand} and is extended here to a space of
arbitrary dimension. This stability analysis can be used to settle
either (i) the {\it dynamical} stability of stellar and gaseous
polytropes (see Sec. \ref{sec_dyn}) or (ii) the {\it generalized
thermodynamical} stability of self-gravitating Langevin particles (see
Sec. \ref{sec_dynamics}).

We start by the canonical ensemble which is simpler in a first
approach.  A polytropic distribution is a local {\it minimum} of
free energy at fixed mass and temperature if, and only if, the
second order variations (see Appendix \ref{sec_so})
\begin{equation}
\delta^{2}F={n+1\over 2n}\int
{p\over\rho^{2}}(\delta\rho)^{2} d^{D}{\bf r}+{1\over 2}\int
\delta\rho\delta\Phi d^{D}{\bf r}, \label{stab1}
\end{equation}
are positive for any perturbation $\delta\rho$ that conserves mass, i.e.
\begin{equation}
\int\delta\rho \ d^{D}{\bf r}=0. \label{stab2}
\end{equation}
This is the condition of (generalized) thermodynamical stability
in the canonical ensemble. Introducing the function $q(r)$ by the
relation
\begin{equation}
\delta\rho={1\over S_{D}r^{D-1}}{dq\over dr}, \label{stab3}
\end{equation}
and integrating by parts,  we can put the second order variations
of free energy in the quadratic form
\begin{equation}
\delta^{2}F=-{1\over 2}\int_{0}^{R}dr q\biggl\lbrack K\gamma
{d\over dr}\biggl ({\rho^{\gamma-2}\over S_{D} r^{D-1}}{d\over
dr}\biggr )+{G\over r^{D-1}}\biggr \rbrack q. \label{stab4}
\end{equation}
The second order variations of free energy can be negative
(implying instability) only if the differential operator which
occurs in the integral has positive eigenvalues. We need therefore
to consider the eigenvalue problem
\begin{equation}
\biggl\lbrack K\gamma {d\over dr}\biggl ({\rho^{\gamma-2}\over
S_{D} r^{D-1}} {d\over dr}\biggr )+{G\over r^{D-1}}\biggr \rbrack
q_{\lambda}(r)=\lambda q_{\lambda}(r). \label{stab5}
\end{equation}
with $q_{\lambda}(0)=q_{\lambda}(R)=0$ in order to satisfy the
conservation of mass. If all the eigenvalues $\lambda$ are
negative, the polytrope is a {minimum} of free energy. If at least
one eigenvalue is positive, the polytrope is an unstable saddle
point. The point of marginal stability in the series of equilibria
is determined by the condition that the largest eigenvalue is
equal to zero ($\lambda=0$). We thus have to solve the
differential equation
\begin{equation}
K\gamma {d\over dr}\biggl ({\rho^{\gamma-2}\over S_{D}
r^{D-1}}{dF\over dr}\biggr )+{GF\over r^{D-1}}=0, \label{stab6}
\end{equation}
with $F(0)=F(R)=0$. The same eigenvalue equation is obtained by
studying the linear stability of the Euler-Jeans equation
\cite{poly,grand}. Introducing the dimensionless variables defined
previously, we can rewrite this equation in the form
\begin{equation}
{d\over d\xi}\biggl ({\theta^{1-n}\over \xi^{D-1}}{dF\over
d\xi}\biggr )+{nF\over \xi^{D-1}}=0, \label{stab7}
\end{equation}
with $F(0)=F(\alpha)=0$. If
\begin{equation}
{\cal L}\equiv {d\over d\xi}\biggl ({\theta^{1-n}\over
\xi^{D-1}}{d\over d\xi}\biggr )+{n\over \xi^{D-1}} \label{stab7a}
\end{equation}
denotes the differential operator occurring in Eq.~(\ref{stab7}),
we can check by using the Emden Eq.~(\ref{emden2}) that
\begin{equation}
{\cal L}(\xi^{D-1}\theta')=(n-1)\theta',\qquad {\cal
L}(\xi^{D}\theta^{n})=\lbrack (2-D)n+D\rbrack \theta'.
\label{stab7b}
\end{equation}
Therefore, the general solution of Eq.~(\ref{stab7}) satisfying
the boundary conditions at $\xi=0$ is
\begin{equation}
F(\xi)=c_{1}\biggl \lbrack \xi^{D}\theta^{n}+{(D-2)n-D\over
n-1}\xi^{D-1}\theta'\biggr \rbrack. \label{stab8}
\end{equation}
Using Eq.~(\ref{stab8}) and introducing the Milne variables
(\ref{milne1}), the condition $F(\alpha)=0$ can be written
\begin{equation}
u_{0}={(D-2)n-D\over n-1}=u_{s}. \label{stab8a}
\end{equation}
This relation determines the points at which a new eigenvalue becomes
positive ($\lambda=0^{+}$). Comparing with Eq.~(\ref{min1}), we see
that a mode of stability is lost each time that $\eta$ is extremum in
the series of equilibria, in agreement with the turning point
criterion of Katz \cite{katz78} in the canonical ensemble. When the
curve $\eta(\alpha)$ is monotonic (cases $A$ and $D$), the system is
always stable because it is stable at low density contrasts
$(\alpha\rightarrow 0)$ and no change of stability occurs
afterward. When the curve $\eta(\alpha)$ presents extrema (cases $B$
and $C$), the series of equilibria becomes unstable at the point of
minimum temperature (or maximum mass) $\alpha_{1}$. In
Fig. \ref{elD3}, this corresponds to a point of infinite specific heat
$C=dE/dT\rightarrow \infty$, just before entering the region of
negative specific heats $C<0$.  When the curve $\eta(\alpha)$ presents
several extrema (case $C$), secondary modes of instability appear at
values $\alpha_{2}$, $\alpha_{3}$,... (see \cite{poly} in $D=3$). We
note that complete polytropes (with $n<n_{5}$ if $D>2$) are stable in
the canonical ensemble if $D\le 2$ and if ($D>2$, $n\le n_3$). They
are unstable otherwise. In the {\it thermodynamical analogy} developed
in \cite{poly,grand}, this is a condition of nonlinear dynamical
stability for gaseous polytropes with respect to the Jeans-Euler
equations. In particular, the self-gravitating Fermi gas at zero
temperature (a classical white dwarf star) is dynamically stable if
$n_{3/2}<n_{3}$, i.e. $D<4$, and unstable otherwise.

\begin{figure}
\centerline{ \psfig{figure=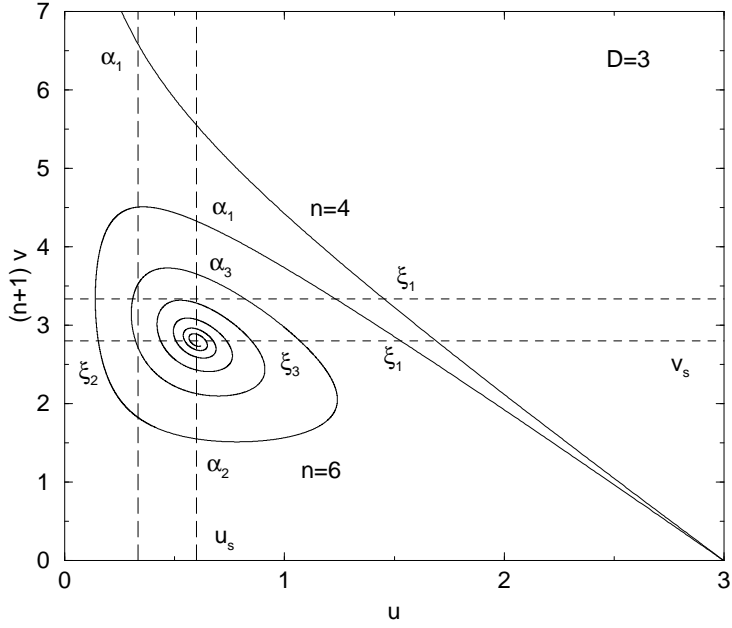,angle=0,height=8.5cm}}
\caption{Location of the turning points of temperature in the
$(u,v)$ plane for systems with dimension $2<D<10$ (specifically
$D=3$). The line $u=u_{s}$ determines the extrema of $\eta$ and
the line $v=v_{s}$ determines the nodes of the density profiles
that trigger the instabilities in the canonical ensemble.}
\label{uvparabole3}
\end{figure}

According to Eq.~(\ref{stab3}), the perturbation profile that
triggers a mode of instability at the critical point $\lambda=0$
is given by
\begin{equation}
{\delta\rho\over\rho_{0}}={1\over S_{D}\xi^{D-1}}{dF\over d\xi},
\label{stab9}
\end{equation}
where $F(\xi)$ is given by Eq.~(\ref{stab8}). Introducing the
Milne variables (\ref{milne1}), we get
\begin{equation}
{\delta\rho\over\rho}={nc_{1}\over S_{D}}(v_{s}-v). \label{stab9b}
\end{equation}
The density perturbation $\delta\rho$ becomes zero at point(s)
$\xi_{i}$ such that $v(\xi_{i})=v_s$. The number of nodes is therefore
given by the number of intersections between the solution curve in the
$(u,v)$ plane and the line $v=v_{s}$. It can be determined by
straightforward graphical constructions in the Milne plane, using
Figs. \ref{uvD1}-\ref{uvD15} (see, e.g., Fig. \ref{uvparabole3}). When
the solution curve is monotonic (case $B$), the density perturbation
profile has only one node. In particular, for $n=n_{5}$, the
perturbation profile at the point of marginal stability is given by
\begin{equation}
{\delta\rho\over\rho}={(D+2)c_{1}\over 2S_{D}}{D(D-2)-\xi^{2}\over
D(D-2)+\xi^{2}}. \label{stab9c}
\end{equation}
It vanishes for $\xi_{(1)}=\sqrt{D(D-2)}$. When the solution curve
forms a spiral (case $C$), the density perturbation $\delta\rho$
corresponding to the $k$-th mode of instability has $k$ zeros
$\xi_{1}, \xi_{2},..., \xi_{k}<\alpha_{k}$. In particular, the first
mode of instability has only {\it one} node. For high modes of
instability, the zeros asymptotically follow a geometric progression
with ratio $1:{\rm exp}\lbrace 2\pi/\sqrt{-\Delta}\rbrace$ (see
Ref. \cite{poly} for $D=3$).

In the microcanonical ensemble, a polytrope is a {\it maximum} of
entropy at fixed mass and energy  if, and only if, the second
order variations (see Appendix \ref{sec_so})
\begin{eqnarray}
{\delta^{2}S}=\beta\Biggl\lbrace-{1\over 2}\gamma\int p
{(\delta\rho)^{2}\over \rho^{2}}d^{D}{\bf r}-{1\over 2}
\int \delta\rho\delta\Phi d^{D}{\bf r}\nonumber\\
-{2n\over D(2n-D)}{1\over\int p d^{D}{\bf r}}\biggl\lbrack \int
\biggl (\Phi+{D\over 2}\gamma {p\over\rho}\biggr )\delta \rho
d^{D}{\bf r}\biggr \rbrack^{2}\Biggr\rbrace, \label{stab10}
\end{eqnarray}
are negative for any variation $\delta\rho$ that conserves mass to
first order (the conservation of energy has already been taken
into account in obtaining Eq.~(\ref{stab10})). Now, using Eq.
(\ref{stab3}) and integrating by parts, the second variations of
entropy can be put in a quadratic form
\begin{equation}
\delta^{2}S=\int_{0}^{R}\int_{0}^{R}dr dr' q(r)K(r,r')q(r'),
\label{stab11}
\end{equation}
with
\begin{eqnarray}
K(r,r')=-{2n\over D(2n-D)}{1\over \int p d^{D}{\bf r}}\biggl
(\Phi+{D\over 2}K\gamma\rho^{\gamma-1}\biggr )'(r)
\biggl (\Phi+{D\over 2}K\gamma\rho^{\gamma-1}\biggr )'(r')\nonumber\\
+{1\over 2}\delta(r-r')\biggl \lbrack K\gamma {d\over dr}\biggl
({\rho^{\gamma-2}\over S_{D} r^{D-1}}{d\over dr}\biggr )+{G\over
r^{D-1}}\biggr\rbrack. \label{stab12}
\end{eqnarray}
The problem of stability can therefore be reduced to the study of
the eigenvalue equation
\begin{equation}
\int_{0}^{R}dr' K(r,r')q_{\lambda}(r')=\lambda q_{\lambda}(r),
\label{stab13}
\end{equation}
with $q_{\lambda}(0)=q_{\lambda}(R)=0$. The point of marginal
stability ($\lambda=0$) will be determined by solving the
differential equation
\begin{equation}
K\gamma {d\over dr}\biggl ({\rho^{\gamma-2}\over S_{D}
r^{D-1}}{dF\over dr}\biggr )+{GF\over r^{D-1}}={K\gamma^{2}\over D
n S_{D}}(2n-D)V\rho^{\gamma-2}{d\rho\over dr}, \label{stab14}
\end{equation}
with
\begin{equation}
V={\int_{0}^{R}\rho^{\gamma-2}{d\rho\over dr} Fdr\over
\int_{0}^{R}\rho^{\gamma} r^{D-1}dr}. \label{stab15a}
\end{equation}
In arriving at this expression, we have used the relation
\begin{equation}
\biggl (\Phi+{D\over 2}K\gamma\rho^{\gamma-1}\biggr
)'={K\gamma\over 2n}(D-2n)\rho^{\gamma-2}{d\rho\over dr},
\label{stab15}
\end{equation}
which results from the condition of hydrostatic equilibrium
(\ref{eqhydro}) with the polytropic equation of state
(\ref{maxent12}). Introducing the dimensionless variables defined
previously, Eqs. (\ref{stab14}) and (\ref{stab15a})  can be
rewritten
\begin{equation}
{d\over d\xi}\biggl ({\theta^{1-n}\over \xi^{D-1}}{dF\over
d\xi}\biggr )+{nF\over \xi^{D-1}}=\chi\theta', \label{stab16}
\end{equation}
with
\begin{equation}
\chi={1\over D}(n+1)(2n-D){\int_{0}^{\alpha}\theta' Fd\xi\over
\int_{0}^{\alpha}\theta^{n+1} \xi^{D-1}d\xi}, \label{stab17}
\end{equation}
and $F(0)=F(\alpha)=0$. Using the identities (\ref{stab7b}), we
can check that the general solution of Eq.~(\ref{stab16})
satisfying the boundary conditions for $\xi=0$ and $\xi=\alpha$ is
\begin{equation}
F(\xi)={\chi\over
(n-1)u_{0}+D-(D-2)n}(\xi^{D}\theta^{n}+u_{0}\xi^{D-1}\theta').
\label{stab18}
\end{equation}
The point of marginal stability is then obtained by substituting
the solution (\ref{stab18}) in Eq.~(\ref{stab17}). Using the
identities (see Appendix \ref{sec_id})
\begin{equation}
\int_{0}^{\alpha}\xi^{D-1}(\theta')^{2}d\xi=
{\alpha^{D}\theta'(\alpha)^{2}\over
D+2-(D-2)n}\biggl (n+1+2{u_{0}\over v_{0}}-{2D\over v_{0}}\biggr
), \label{stab19}
\end{equation}
\begin{equation}
\int_{0}^{\alpha}\theta^{n+1}\xi^{D-1}d\xi={\alpha^{D}
\theta'(\alpha)^{2}\over
D+2-(D-2)n}\biggl (n+1+2{u_{0}\over v_{0}}-{(D-2)(n+1)\over
v_{0}}\biggr ), \label{stab20}
\end{equation}
\begin{equation}
\int_{0}^{\alpha}\theta^{n}\theta'\xi^{D}d\xi={\alpha^{D}
\theta'(\alpha)^{2}\over
D+2-(D-2)n}\biggl (-D-(D-2){u_{0}\over v_{0}}+{D(D-2)\over
v_{0}}\biggr ), \label{stab20b}
\end{equation}
which result from simple integrations by parts and from the properties
of the Lane-Emden equation (\ref{emden2}), it is found that the point
of marginal stability is determined by the condition
(\ref{min5}). Therefore, the series of equilibria becomes unstable at
the point of minimum energy in agreement with the turning point
criterion of Katz \cite{katz78} in the microcanonical ensemble.  When
the curve $\Lambda(\alpha)$ is monotonic (cases $A'$ and $D'$), the
system is always stable. When the curve $\Lambda(\alpha)$ presents
extrema (cases $B'$ and $C'$), the series of equilibria becomes
unstable at the point of minimum energy $\alpha'_{1}$. In
Fig. \ref{elD3}, this corresponds to the point where the specific heat
$C=dE/dT=0$, passing from negative to positive values.  Note that the
branch of negative specific heats between $CE$ and $MCE$ is stable in
the microcanonical ensemble although it is unstable in the canonical
ensemble.  When the curve $\Lambda(\alpha)$ presents several extrema
(case $C'$), secondary modes of instability appear at values
$\alpha'_{2}$, $\alpha'_{3}$,... We note that complete polytropes
(with $n<n_{5}$ if $D>2$) are stable in the microcanonical ensemble if
$D<4$ and unstable if $D\ge 4$. Owing to the {\it thermodynamical
analogy}, this is a condition of nonlinear dynamical stability for
stellar polytropes with respect to the Vlasov equation
\cite{grand}. The difference between the dynamical stability of gaseous
polytropes ($n\le n_{3}$) and stellar polytropes ($n\le n_{5}$) was
related in \cite{poly,grand} to a situation of ensemble inequivalence
(and the existence of a negative specific heat region) in
thermodynamics. Since the caloric curve is monotonic in $D=2$, we also
conclude that polytropic vortices
\cite{Ctsallis} are always nonlinearly dynamically stable with
respect to the 2D Euler equation.

According to Eqs. (\ref{stab9}) and (\ref{stab18}), the
perturbation profile that triggers a mode of instability at the
critical point $\lambda=0$ is given by
\begin{equation}
{\delta\rho\over\rho}={\chi\over S_{D}}{1\over
(n-1)u_{0}+D-(D-2)n}(D-nv-u_{0}), \label{stab21}
\end{equation}
where we have used the Emden equation (\ref{emden2}) and
introduced the Milne variables (\ref{milne1}). The number of nodes
in the perturbation profile can be determined by a graphical
construction similar to the one described in Refs.~\cite{pad,sc}
for $n=\infty$ (isothermal case).

\section{Self-gravitating Langevin particles}
\label{sec_dynamics}

\subsection{The nonlinear Smoluchowski-Poisson system}
\label{sec_brown}

Let us consider a system of $N$ self-gravitating Brownian
particles described by the stochastic equations ($i=1,...,N$)
\begin{equation}
{d{\bf r}_i\over dt}={\bf v}_i, \qquad {d{\bf v}_i\over
dt}=-\nabla\Phi_i-\xi {\bf v}_i+\sqrt{2T\xi}\ {\bf R}_i(t),
\label{Nbrown2}
\end{equation}
where $\Phi_{i}=\Phi({\bf r}_{i},t)$ is the self-consistent
gravitational potential created by the particles, $-\xi {\bf v}_{i}$
is a friction force originating from the presence of an inert medium
and ${\bf R}_{i}(t)$ is a white noise satisfying $\langle {\bf
R}_{i}(t)\rangle=0$ and $\langle
{R}_{i,a}(t){R}_{j,b}(t')\rangle=\delta_{ij}\delta_{ab}\delta(t-t')$. To
simplify the problem, we consider the high friction limit
$\xi\rightarrow +\infty$, where $\xi$ is the friction coefficient
\cite{crs}. This regime is achieved for times $t\gg \xi^{-1}$. In the
mean-field approximation, the evolution of
the density of particles is governed by the Smoluchowski equation \cite{risken}
\begin{equation}
{\partial\rho\over\partial t}=\nabla\biggl \lbrack \nabla
(D\rho)+{1\over \xi}\rho\nabla\Phi\biggr \rbrack, \label{brown1}
\end{equation}
coupled to the Newton-Poisson equation (\ref{maxent4}). In the usual
case \cite{crs,sc,sc2}, the diffusion coefficient $D$ is constant and the
condition that the Boltzmann distribution $\rho\sim {\rm e}^{-\Phi/T}$
is a stationary solution of Eq.  (\ref{brown1}) is ensured by the Einstein
relation $\xi D=T$.  It can be shown \cite{crs} that the
SP system decreases the free energy $F=E-TS$ constructed with the
Boltzmann entropy. Hence, the equilibrium state minimizes
\begin{equation}
F={1\over 2}\int\rho\Phi d^{D}{\bf r}+T\int\rho\ln\rho d^{D}{\bf r}, \label{brown1add}
\end{equation}
at fixed $M$ and $T$.

We want to consider here a more general situation in which the
diffusion coefficient $D$ depends on the density $\rho$ while the
drift coefficient $\xi$ is still constant. In the absence of drift,
this would lead to a situation of {\it anomalous diffusion}. In the
presence of drift, a notion of {\it generalized thermodynamics}
emerges \cite{Ctsallis}. Indeed, writing the diffusion coefficient in
the form $D(\rho)={1\over \xi}p(\rho)/\rho$, we obtain the generalized
Smoluchowski equation
\begin{equation}
{\partial\rho\over\partial t}=\nabla\biggl \lbrack
{1\over\xi}(\nabla p+\rho\nabla\Phi)\biggr \rbrack.
\label{brown2a}
\end{equation}
In Ref. \cite{Ctsallis}, it is shown that generalized Smoluchowski
equations of this type satisfy a form of canonical $H$-theorem. The Lyapunov
functional decreasing monotonically with time is
\begin{equation}
F=\int\rho\int_{0}^{\rho}{p(\rho')\over \rho^{'2}}d\rho'd^{D}{\bf
r}+{1\over 2}\int\rho\Phi d^{D}{\bf r}. \label{brown2b}
\end{equation}
This can be interpreted as a free energy associated with a generalized
entropy functional (see \cite{Ctsallis} for more details). The
equilibrium state minimizes $F$ at fixed $M$. In the present context,
this is a condition of generalized thermodynamical stability in
canonical ensemble.

The generalized Smoluchowski equation (\ref{brown2a}) can be
obtained by combining the {ordinary} Fokker-Planck equation
with a Langevin equation of the form \cite{Ctsallis}:
\begin{equation}
{d{\bf r}\over dt}=-{1\over\xi}\nabla\Phi+\sqrt{2p(\rho)\over \xi
\rho}{\bf R}(t), \label{brown2c}
\end{equation}
where ${\bf R}(t)$ is a white noise. When $p(\rho)$ is a power-law,
Eq. (\ref{brown2c}) reduces to the stochastic equations studied by
Borland \cite{borland} in connexion with Tsallis thermodynamics. Since
the function in front of ${\bf R}(t)$ depends on ${\bf r}$, the last
term in Eq. (\ref{brown2c}) can be interpreted as a multiplicative
noise. Note that  the noise depends on ${\bf r}$ through
the density $\rho({\bf r})$.  Kaniadakis
\cite{kaniadakis} has also introduced {generalized}
Fokker-Planck equation arising from a modified form of transition
probabilities. In these works, the Langevin particles evolve in an
external potential. The case of Langevin particles in {\it
interaction} was considered by Chavanis
\cite{Ctsallis}. He introduced a generalized Fokker-Planck
equation (see in particular Eq. (81) of \cite{Ctsallis}) valid for an
arbitrary equation of state $p=p(\rho)$, or diffusion coefficient
$D(\rho)$, and for an arbitrary binary potential of interaction
$u({\bf r}-{\bf r}')$. This equation  has been studied recently in
\cite{csr,crs,sc,sc2} for an isothermal equation of state $p=\rho T$
(constant $D$) and a gravitational interaction. This study has been extended
in \cite{ribot} to a Fermi-Dirac equation of state. In this 
paper, we consider the case where the function $D(\rho)$ is a power
law and write $\xi D(\rho)=K \rho^{\gamma-1}$. This corresponds to a
polytropic equation of state $p=K\rho^\gamma$.  Then,
Eq. (\ref{brown1}) can be rewritten
\begin{equation}
{\partial\rho\over\partial t}=\nabla\biggl \lbrack
{1\over\xi}(K\nabla\rho^{\gamma}+\rho\nabla\Phi)\biggr \rbrack.
\label{brown2}
\end{equation}
For the nonlinear Smoluchowski-Poisson system, the
Lyapunov functional decreasing monotonically with time is \cite{Ctsallis}
\begin{equation}
F={K\over \gamma-1}\int (\rho^{\gamma}-\rho)d^{D}{\bf r}+{1\over
2}\int\rho\Phi d^{D}{\bf r}. \label{brown3}
\end{equation}
It can be interpreted as a free energy associated
with Tsallis entropy. In this context, the
polytropic index $\gamma$ plays the role of the $q$-parameter and
the polytropic constant $K$ plays the role of the temperature (see
Ref. \cite{Ctsallis} for details and subtleties). Therefore,
keeping $K$ fixed corresponds to a canonical situation
\cite{poly}. For $\gamma\rightarrow 1$, we recover the Boltzmann
free energy studied in \cite{crs,sc}. For $\gamma=5/3$, i.e.
$n=3/2$, Eq. (\ref{brown2}) describes self-gravitating Brownian fermions at
$T=0$ (in $D=3$) \cite{ribot}.

The nonlinear Smoluchowski equation can be obtained from a
variational principle, called Maximum Entropy Production
Principle, by maximizing the rate of free energy $\dot F$ for a
fixed total mass $M$ \cite{csr,Ctsallis}. It is straightforward to check
that the rate of free energy dissipation can be put in the form
\begin{equation}
\dot F=-\int {1\over
\rho\xi}(K\nabla\rho^{\gamma}+\rho\nabla\Phi)^{2}d^{D}{\bf r}\le
0. \label{brown4}
\end{equation}
For a stationary solution, $\dot F=0$, and we obtain a polytropic
distribution  which is a critical point of $F$ at fixed $M$.
Considering a small perturbation around equilibrium, we can
establish the identity \cite{Ctsallis}:
\begin{equation}
\delta^{2}\dot F=2\lambda\delta^{2}F\le 0, \label{brown5}
\end{equation}
where $\lambda$ is the growth rate of the perturbation defined
such that $\delta\rho\sim {\rm e}^{\lambda t}$. This relation
shows that a stationary solution of the nonlinear
Smoluchowski-Poisson (NSP) system is dynamically stable for small
perturbations ($\lambda<0$) if and only if it is a local {\it
minimum} of free energy ($\delta^{2}F>0$). In addition, it is
shown in Refs. \cite{crs,Ctsallis} that the eigenvalue problem
determining the growth rate $\lambda$ of the perturbation is
similar to the eigenvalue problem (\ref{stab6}) associated with
the second order variations of free energy (they coincide for
marginal stability). This shows the equivalence between dynamical
and generalized thermodynamical stability for self-gravitating
Langevin particles exhibiting anomalous diffusion (this result was
obtained independently by Shiino \cite{shiino} in the specific
context of Tsallis thermodynamics). In fact, our formalism is
valid for more general functionals than the Boltzmann or the
Tsallis entropies \cite{Ctsallis}. These functionals
(\ref{brown2b}) arise when the diffusion coefficient is of the
general form $D(\rho)$, not necessarily a power law. We note that
the NSP system satisfies a form of Virial theorem \cite{Ctsallis}.
For $D\neq 2$, it reads
\begin{equation}
{1\over 2}\xi{dI\over dt}=2K+(D-2)W-Dp_{b}V, \label{brown6}
\end{equation}
where
\begin{equation}
I=\int \rho r^{2}d^{D}{\bf r}, \label{brown7}
\end{equation}
is the moment of inertia and $p_{b}$ the pressure on the box
(assumed uniform). For $D=2$, the term $(D-2)W$ is replaced by
$-GM^{2}/2$. For a stationary solution $dI/dt=0$, we recover the
Virial theorem (\ref{thermo10}).

Our model of self-gravitating Brownian (or Langevin) particles has no
clear astrophysical applications, so it must be regarded essentially
as a {\it toy model} of gravitational dynamics. It may find
application for the formation of planetesimals in the solar nebula
since the dust particles experience a friction with the gas and a
noise due to small-scale turbulence \cite{planetes}.  However, even in
this context, the model has to be refined so as to take into account
the attraction of the Sun and the rotation of the disk. Anyway, the
self-gravitating Brownian (or Langevin) gas model is well-posed
mathematically, and it possesses a rigorous thermodynamical structure
corresponding to the {\it canonical} ensemble. Therefore, it can be
used as a simple model to illustrate some aspects of the 
thermodynamics of self-gravitating systems. Since it minimizes
$F={\cal W}[\rho]$ at fixed mass, it can also be used as a powerful
numerical algorithm to construct nonlinearly dynamically stable
stationary solutions of the Euler-Jeans equations (see
Sec. \ref{sec_bs}). Coincidentally, the SP system also provides a simple model
for the chemotactic aggregation of bacterial populations
\cite{murray}. The name chemotaxis refers to the motion of organisms
induced by chemical signals. In some cases, the biological organisms
secrete a substance that has an attractive effect on the organisms
themselves. This is the case for the bacteria {\it Escherichia
coli}. In the simplest model, the bacteria have a diffusive motion and
they also move systematically along the gradient of concentration of
the chemical they secrete. Since the density $\Phi({\bf r},t)$ of the
secreted substance is induced by the particles themselves, the drift
is directed toward the region of higher density. This attraction
triggers a self-accelerating process until a point at which
aggregation takes place. If we assume in a first step that $\Phi({\bf
r},t)$ is related to the bacterial density $\rho({\bf r},t)$ by a
Poisson equation, this phenomenon can be modeled by the SP
system. Now, it has been observed in many occasions in biology that
the diffusion of the particles is anomalous \cite{murray}. This is a
physical motivation to study the NSP system in which the diffusion
coefficient is a power law of the density. In the following section,
we show that the NSP system admits self-similar solutions describing
the collapse of the self-gravitating Langevin gas or of the bacterial
population. These theoretical results are confirmed in
Sec. \ref{sec_num} where we numerically solve the NSP system.

\subsection{Self-similar solutions of the nonlinear
Smoluchowski-Poisson system} \label{sec_dim}

From now on, we set $M=R=G=\xi=1$ without loss of generality. The
equations of the problem become
\begin{equation}
{\partial\rho\over\partial t}=\nabla
(K\nabla\rho^{\gamma}+\rho\nabla\Phi), \label{dim1}
\end{equation}
\begin{equation}
\Delta\Phi=S_{D}\rho, \label{dim2}
\end{equation}
with boundary conditions
\begin{equation}
{\partial\Phi\over\partial r}(0,t)=0, \qquad \Phi(1)={1\over 2-D},
\qquad K{\partial \rho^{\gamma}\over\partial r}(1)+\rho(1)=0,
\label{dim3}
\end{equation}
for $D\neq 2$. For $D=2$, we take $\Phi(1)=0$ on the boundary. We
restrict ourselves to spherically symmetric solutions. Integrating
Eq.~(\ref{dim2}) once, we can rewrite the NSP system in the form
of a single integrodifferential equation
\begin{equation}
{\partial\rho\over\partial t}={1\over
r^{D-1}}{\partial\over\partial r}\biggl\lbrace r^{D-1}\biggl
\lbrack (S_{D}\rho)^{1/n}\Theta{\partial\rho\over\partial
r}+{\rho\over r^{D-1}}\int_{0}^{r}\rho(r')S_{D}r^{'D-1}dr'\biggr
\rbrack\biggr \rbrace, \label{dim4}
\end{equation}
where we have set
\begin{equation}
\Theta\equiv {K(1+n)\over nS_{D}^{1/n}}={1\over n\eta^{1-1/n}}.
\label{dim5}
\end{equation}
The quantity $\Theta$ can be seen as a sort of generalized
temperature (sometimes called a polytropic temperature
\cite{poly}) and it reduces to the ordinary temperature $T$ for
$n\rightarrow +\infty$. We note that the proper description of a gas
of Langevin particles in interactions is the canonical ensemble where
$\Theta$ is fixed. However, we can formally set up a microcanonical
description of self-gravitating Langevin particles by letting the
temperature $\Theta(t)$ depend on time so as to conserve the total
energy:
\begin{equation}
E={D\over 2}{nS_{D}^{1/n}\over 1+n}\int\Theta(t) \rho^{1+{1\over
n}}d^{D}{\bf r}+{1\over 2}\int\rho\Phi d^{D}{\bf r}. \label{dim5a}
\end{equation}
The two situations have been considered in the case of
self-gravitating Brownian particles ($n\rightarrow +\infty$) in
\cite{csr,crs,sc,sc2}.

The NSP system is also equivalent to a single differential
equation
\begin{equation}
\frac{\partial M}{\partial t}=\Theta \biggl ({1\over
r^{D-1}}{\partial M\over\partial r}\biggr )^{1/n}\biggl\lbrack
{\partial^{2}M\over\partial r^{2}}-{D-1\over r}{\partial
M\over\partial r}\biggr\rbrack +{M\over r^{D-1}}{\partial
M\over\partial r}, \label{dim6}
\end{equation}
for the quantity
\begin{equation}
M(r,t)=\int_{0}^{r}\rho(r',t)S_{D}r^{'D-1}dr', \label{dim7}
\end{equation}
which represents the mass contained within the sphere of radius
$r$. The appropriate boundary conditions are
\begin{equation}
M(0,t)=0,\qquad M(1,t)=1. \label{dim8}
\end{equation}
It is also convenient to introduce the function
$s(r,t)=M(r,t)/r^{D}$ satisfying
\begin{equation}
{\partial s\over\partial t}=\Theta\biggl (r{\partial
s\over\partial r} +D s\biggr )^{1/n}\biggl
({\partial^{2}s\over\partial r^{2}}+{D+1\over r}{\partial
s\over\partial r}\biggr )+\biggl (r{\partial s\over\partial
r}+Ds\biggr )s. \label{dim9}
\end{equation}
For $n\rightarrow +\infty$, these equations reduce to those
studied in Refs. \cite{crs,sc,sc2} in the isothermal case. We look for
self-similar solutions of the form
\begin{equation}
\rho(r,t)=\rho_{0}(t)f\biggl ({r\over r_{0}(t)}\biggr ), \qquad
r_{0}=\biggl ({\Theta\over \rho_{0}^{1-1/n}}\biggr )^{1/2}.
\label{dim10}
\end{equation}
The radius $r_{0}$ defined by the foregoing equation provides a
typical value of the core radius of an incomplete polytrope (with
$n>n_{5}$). It reduces to the King's radius \cite{bt} as
$n\rightarrow +\infty$. In terms of the mass profile, we have
\begin{equation}
M(r,t)=M_{0}(t)g\biggl ({r\over r_{0}(t)}\biggr ), \qquad {\rm
with}\qquad M_{0}(t)=\rho_{0}r_{0}^{D}, \label{dim11}
\end{equation}
and
\begin{equation}
g(x)=\int_{0}^{x}f(x')S_{D}x^{'D-1}dx'. \label{dim12}
\end{equation}
In terms of the function $s$, we have
\begin{equation}
s(r,t)=\rho_{0}(t)S\biggl ({r\over r_{0}(t)}\biggr ), \qquad {\rm
with}\qquad S(x)={g(x)\over x^{D}}. \label{dim13}
\end{equation}

Substituting the {\it ansatz} (\ref{dim13}) into Eq.~(\ref{dim9}),
and using the definition of $r_0$ in Eq. (\ref{dim10}), we find
that
\begin{equation}
{d\rho_{0}\over dt}S-{\rho_{0}\over r_{0}}{dr_{0}\over dt}x
S'=\rho_{0}^{2}(xS'+DS)^{1/n}\biggl (S''+{D+1\over x}S'\biggr
)+\rho_{0}^{2}(xS'+DS)S, \label{dim14}
\end{equation}
where we have set $x=r/r_{0}$. We now assume that there exists
$\alpha$ such that
\begin{equation}
\rho_0\sim r_0^{-\alpha}. \label{scaa}
\end{equation}
Inserting this relation into  Eq.~(\ref{dim14}), we find
\begin{equation}
{d\rho_{0}\over dt}\biggl (S+{1\over\alpha}xS'\biggr
)=\rho_{0}^{2}\biggl\lbrack (xS'+DS)^{1/n}\biggl (S''+{D+1\over
x}S'\biggr )+(xS'+DS)S\biggr\rbrack, \label{dim17}
\end{equation}
which implies that $(1/\rho_{0}^{2})(d\rho_{0}/dt)$ is a constant
that we arbitrarily set equal to $\alpha$. This leads to
\begin{equation}
\rho_{0}(t)={1\over\alpha}(t_{coll}-t)^{-1}, \label{dim18}
\end{equation}
so that the central density becomes infinite in a finite time
$t_{coll}$. The scaling equation now reads
\begin{equation}
\alpha S+xS'=(xS'+DS)^{1/n} \biggl (S''+{D+1\over x}S'\biggr
)+(xS'+DS)S. \label{dim19}
\end{equation}
For $x\rightarrow +\infty$, we have asymptotically
\begin{equation}
S(x)\sim x^{-\alpha}, \qquad g(x)\sim x^{D-\alpha}, \qquad
f(x)\sim x^{-\alpha}. \label{dim20}
\end{equation}

In the canonical ensemble where $\Theta$ is constant,
Eq.~(\ref{dim10}) and Eq.~(\ref{scaa}) lead to $\alpha=\alpha_n$,
with
\begin{equation}
\alpha_n=\frac{2n}{n-1}.
\end{equation}
Note that for $n\to\infty$, we recover the result of \cite{sc},
$\alpha_{\infty}=2$.  Equation (\ref{dim20}) implies that for large
$x$, $\rho\sim (D-\alpha)S>0$, which enforces $\alpha < D$ (this
also guarantees that the mass of the power-law profile $\rho=C
r^{-\alpha}$ at $t=t_{coll}$ is finite). The limit value
$\alpha_{n}=D$ corresponds to $n=n_{3}$. Therefore, there is no
scaling solution for $n<n_{3}$. This is consistent with our
finding that the collapse occurs only for $n>n_3$. For $n<n_3$ and
$\eta>\omega_{n}$, the system converges toward a complete
polytrope with radius $R_*<1$  which is stable (see Sec.
\ref{sec_stab} and Fig. \ref{elD3}). For $n_{3}<n<n_{5}$ 
and $\eta>\eta(\alpha_{1})$, we can formally construct a complete
polytrope with radius $R_*<1$, but this structure is unstable
(Sec. \ref{sec_stab}) so that the system undergoes gravitational
collapse.

In the microcanonical ensemble, the value of $\alpha\geq \alpha_n$
cannot be obtained by dimensional analysis. It will be selected by
the dynamics. In the case $n\to+\infty$, we have found in
\cite{sc} that Eq.~(\ref{dim19}) for the scaling profile has
physical solutions only for $2\leq\alpha\leq \alpha_{\rm max}(D)$
(with $\alpha_{\rm max}(D=3)=2.209733...$).  For arbitrary $n$,
such a $\alpha_{\rm max}(D,n)\ge \alpha_n$ also exists (see Sec.
\ref{sec_num}). It is easy to see that the maximum value for
$\alpha$ leads to the maximum divergence of temperature and
entropy. Therefore, it is natural to expect that the value
$\alpha_{\rm max}$ will be selected by the dynamics except if some
kinetic constraints forbid this natural evolution (see below). In
fact, as already noted in \cite{crs}, a value of
$\alpha>\alpha_{n}$ poses problem with respect to the conservation
of energy. We recall (and generalize) the argument below.
According to Eqs. (\ref{dim10}), (\ref{scaa}) and (\ref{dim18}), during 
collapse, the  temperature behaves as
\begin{equation}
\Theta\sim\rho_0^{1-1/n-2/\alpha}\sim
(t_{coll}-t)^{-(2/\alpha_n-2/\alpha)},
\end{equation}
and the kinetic energy (\ref{maxent16}) behaves as
\begin{equation}
K\sim\Theta\int_0^1\rho^\gamma(r,t)r^{D-1}\,dr\sim\Theta (\rho_0
r_0^\alpha)^{\gamma}{\times} \int_{r_0}^1 r^{D-1-\gamma\alpha}\,dr
\sim\Theta\int_{r_0}^1 r^{D-1-\gamma\alpha}\,dr.\label{EK}
\end{equation}
First consider the case $n>n_5$. If $\alpha<D/\gamma$ (which is
the case in practice since $\alpha_{n}<D/\gamma$ implies $n>n_5$),
the integral is finite and the kinetic energy behaves as $K\sim
\Theta$. Therefore, it {\it diverges} at $t_{coll}$ for any
$\alpha>\alpha_{n}$. On the other hand, the {scaling contribution}
to the potential energy behaves as
\begin{equation}
W\sim\int_0^1\frac{M^2(r,t)}{r^{D-1}}\,dr\sim
(\rho_0r_0^\alpha)^2{\times} \int_{r_0}^1 r^{D+1-2\alpha}\,dr\sim
\int_{r_0}^1 r^{D+1-2\alpha}\,dr. \label{Ep}
\end{equation}
If $\alpha<(D+2)/2$ (which is the case in practice since
$\alpha_{n}<(D+2)/2$ implies $n>n_5$), the scaling part of $W$
remains {\it finite} at $t_{coll}$. Energy conservation would then
imply that $\alpha=\alpha_{n}$. In a first series of numerical
experiments reaching moderately high values of the central density
\cite{crs}, we measured (by different methods) a scaling exponent
$\alpha\simeq 2.2>\alpha_{\infty}=2$ (for the isothermal case
$n=\infty$). Combined with the fact that the Smoluchowski-Poisson
system must lead  to a diverging entropy, we argued that
$\alpha_{\rm max}$ is selected by the dynamics (while being
careful not to rigorously reject the possibility that
$\alpha=\alpha_{\infty}=2$). Then, in order to account for energy
conservation, we proposed a heuristic scenario showing how
subscaling contributions could lead to the divergence of potential
energy. In fact, the numerical simulations were not really
conclusive in showing the divergence of temperature, as the
expected exponent is very small ($2/\alpha_n-2/\alpha_{\rm
max}=0.09491...$, for $D=3$ and $n\to+\infty$).  Recently, we have
conducted a new series of numerical simulations allowing to
achieve much higher values of density (see Sec. \ref{sec_num}).
These simulations tend to favor a value of $\alpha=\alpha_{n}$
leading to a finite value of the temperature at $t_{coll}$.
However, the convergence to the value $\alpha_{n}$ is obtained for
values of $t$ very close to $t_{coll}$ and, at intermediate times
for which the temperature has not reached its asymptotic value,
the numerical scaling function tends to display an {\it effective
exponent} between $\alpha_{n}$ and $\alpha_{\rm max}$.  This situation is
reminiscent of the $(D=2, n=+\infty)$ case studied in \cite{sc},
although the situation is not exactly equivalent. Numerical
simulations of the $(D=3,n=+\infty)$ case have been conducted
independently by \cite{guerra} and favor also a value of
$\alpha=\alpha_{n}$. Note that there is no rigorous result
proving that $\alpha=\alpha_{n}$ in the microcanonical
situation, so this point remains an open mathematical problem.

For $n<n_5$, the kinetic and potential energies diverge in a
consistent way as
\begin{eqnarray}
K &\sim &\Theta\int_{r_0}^1 r^{D-1-\gamma\alpha}\,dr\sim \Theta
r_0^{D-\gamma\alpha}\sim \rho_0^{2-(D+2)/\alpha},
\\
W &\sim &\int_{r_0}^1 r^{D+1-2\alpha}\,dr\sim r_0^{D+2-2\alpha}
\sim \rho_0^{2-(D+2)/\alpha}.
\end{eqnarray}
However, in the microcanonical ensemble, the system is expected to
reach a self-confined polytrope for $n<n_5$ and $\Lambda>\lambda_{n}$
since it is stable (see Sec. \ref{sec_stab} and
Fig. \ref{elD3}). Probably, the choice of evolution will depend on a
notion of {\it basin of attraction} as in \cite{crs}.

We now focus on self-gravitating Brownian particles ($n=\infty$). In case
of collapse, the previous discussion shows that the system has the
{\it desire} to achieve a value of $\alpha>2$ leading to a
divergence of temperature and entropy. This is indeed a natural
evolution in a thermodynamical sense. This is also consistent with the
notion of {\it gravothermal catastrophe} introduced in the context of
globular clusters
\cite{lbw,larson,cohn,lbe}. However, the energy constraint (\ref{dim5a})
seems to prevent this natural evolution (the divergence of entropy
occurs in the post-collapse regime
\cite{sc2}). This is related to the assumption that the
temperature is {\it uniform} although this assumption clearly breaks
down during the late stage of the collapse. Therefore, we expect that
if the temperature is not constrained to remain uniform, the system
will select a value of $\alpha>2$ as in other models of microcanonical
gravitational collapse \cite{larson,cohn,lbe}. Below, we give a heuristic
hint so as how this can happen. We consider the Smoluchowski equation
\begin{equation}
{\partial\rho\over\partial t}=\nabla \lbrack \nabla
(T\rho)+\rho\nabla\Phi \rbrack, \label{Tvar}
\end{equation}
where $T=T({\bf r},t)$ is now position dependant. In any model
where the temperature $T({\bf r},t)$ satisfies a {\it local}
conservation of energy, we expect the following scaling
\begin{equation}
T(r,t)=T_0(t)\theta \biggl ({r\over r_0(t)}\biggr ), \quad{\rm
with}\quad \theta (x)\sim x^{2-\alpha},\quad{\rm when}\quad x\to
+\infty .\label{scaT}
\end{equation}
Such a scaling is indeed observed in the globular cluster model of
\cite{lbe}. The decay exponent is obtained by using the definition
of $T_0\sim \rho_0r_0^2$ and the fact that the temperature at
distances $r\gg r_0$ should be of order unity. The precise density
and temperature profiles and the value of $\alpha$ depend on the
model considered for the energy transport equation. It is not our
purpose to discuss a precise model in the present paper and we
postpone this study for a future work \cite{sc3}. However, we can
give an analytical argument showing why $\alpha>2$ should now be
selected in a unique way.  Equation (\ref{scaT}) shows that
temperature scales  as the potential $\Phi$ since
\begin{equation}
\Phi(r,t)=\Phi(1)-\int_r^1 s(r',t)r'\,dr'\approx
\Phi(1)-\rho_0r_0^2\int_{r/r_0}^{1/r_0} S(x) x \,dx \approx
-T_0\int_{r/r_0}^{+\infty} S(x) x \,dx. \label{scaphi1}
\end{equation}
The scaling function for $\Phi(r,t)$ is then
\begin{equation}
\phi(x)=-\int_{x}^{+\infty} S(x')x'\,dx'\sim
x^{2-\alpha},\quad{\rm when}\quad x\to +\infty.\label{scaphi2}
\end{equation}
Equations (\ref{scaT}), (\ref{scaphi1}) and (\ref{scaphi2}) imply that
both the kinetic and potential energies remain bounded for all
times $t\leq t_{coll}$, at least for $\alpha<(D+2)/2$, even if the
central temperature $T_{0}(t)$ diverges. Indeed, the temperature
increases in the core but the core mass goes to zero so that the
kinetic energy of the core $\sim M_{0}T_{0}$ goes to zero. On the
other hand, the temperature remains of order unity in the halo
leading to a finite kinetic energy in the halo. This ``core-halo''
structure for the temperature is more satisfactory than a model in
which the temperature is uniform everywhere, even in the collapse
phase.  Before introducing a precise equation for $T(r,t)$
\cite{sc3}, we make the reasonable claim that the temperature and
the potential energy are simply proportional in the core region as
they exhibit a similar scaling relation. Defining $T_0(t)$ such
that $\theta(0)=1$, we end up with the hypothesis
\begin{equation}
\theta
(x)=\frac{\phi(x)}{\phi(0)}=\frac{\lambda}{D}\int_{x}^{+\infty}
S(x')x'\,dx',\label{tphi}
\end{equation}
with
\begin{equation}
\lambda=-\frac{D}{\phi(0)}=\frac{D}{\int_{0}^{+\infty}
S(x')x'\,dx'}.\label{lambda}
\end{equation}
Using Eq.~(\ref{tphi}), we find that the scaling equation for
$S(x)$ is now
\begin{equation}
\alpha S+xS'= \frac{\lambda}{D}{\phi(x)}\biggl(S''+{D+1\over
x}S'\biggr )+\left(1-\frac{\lambda}{D}\right)S(xS'+DS).
\label{scavar}
\end{equation}
For a given $\lambda$, this equation is an eigenvalue problem in
$\alpha$. In the limit of large dimension and proceeding exactly along
the line of \cite{sc}, it can be seen that $S\sim O(D^{-1})$ and
$\lambda\sim O(D^{0})$. Using the method of \cite{sc}, $S(x)$ and
$\alpha$ can be computed easily up to order $O(D^{-1})$, as a function
of $\lambda$ and $z=D S(0)/2$. Now imposing the constraint of
Eq.~(\ref{lambda}), this selects a {\it unique} value for
$\alpha$. After straightforward calculations, we find a simple
parametrization of $\lambda$ and $\alpha$ as a function of $z=D
S(0)/2>2$~:
\begin{eqnarray}
\alpha-2&=&\frac 4D\left(1-2z^{-1}\right)+O(D^{-2}),\\
\lambda
&=&2\left(1-z^{-1}\right)\left(1-2z^{-1}\right)+O(D^{-1})\label{lz},
\end{eqnarray}
which can be recast in the form
\begin{eqnarray}
\alpha-2&=&\frac 2D\left(\sqrt{1+4\lambda}-1\right)+O(D^{-2}),\\
S(0)&=&\frac 8D\left(3-\sqrt{1+4\lambda}\right)^{-1}+O(D^{-2}).
\end{eqnarray}
Equation (\ref{lz}) implies that for $\lambda\geq 2$ (up to order
$O(D^{-1})$), there is no solution to the scaling equation, and
that for $\lambda<2$, there is a {unique} $\alpha$ corresponding to
a physical solution. In general, the actual value of $\lambda$
will be selected dynamically by an additional evolution equation
for the temperature profile. However, assuming $\lambda\simeq 1$
is natural (although this point needs to be confirmed), since it
corresponds naively to a local energy conservation condition
\begin{equation}
\frac{D}{2}T(r,t)\sim-\frac{1}{2}\Phi(r,t).\label{conse}
\end{equation}
In that case, we obtain
\begin{equation}
\alpha-2=\frac 2D\left(\sqrt{5}-1\right)+O(D^{-2}).
\end{equation}
The above argument is a strong indication that if the uniform
temperature constraint is abandoned, a non trivial value for
$\alpha>2$ will be selected. We conjecture that this eigenvalue
will be close to $\alpha_{\rm max}\simeq 2.21$ as found in other
models of microcanonical gravitational collapse with a non-uniform
temperature \cite{larson,cohn,lbe}.

\section{Numerical simulations}
\label{sec_num}

In this section, we illustrate numerically some of the theoretical
results presented in the previous sections, but we restrain
ourselves to $D=3$.

\begin{figure}
\centerline{ \psfig{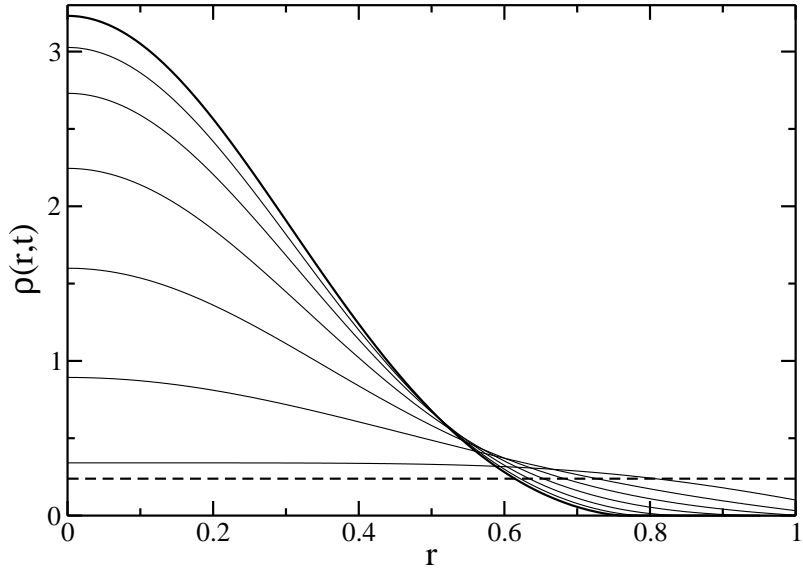}}
\vskip 0.3cm \caption{Evolution of the density profile $\rho(r,t)$
for $n=3/2$ and $\Theta=0.1$ (corresponding to $\eta>\omega_n$).
The profile converges to a complete polytrope strictly confined
inside the box (thick line). The dashed line is the initial
uniform density profile.} \label{cano32}
\end{figure}

\begin{figure}
\centerline{ \psfig{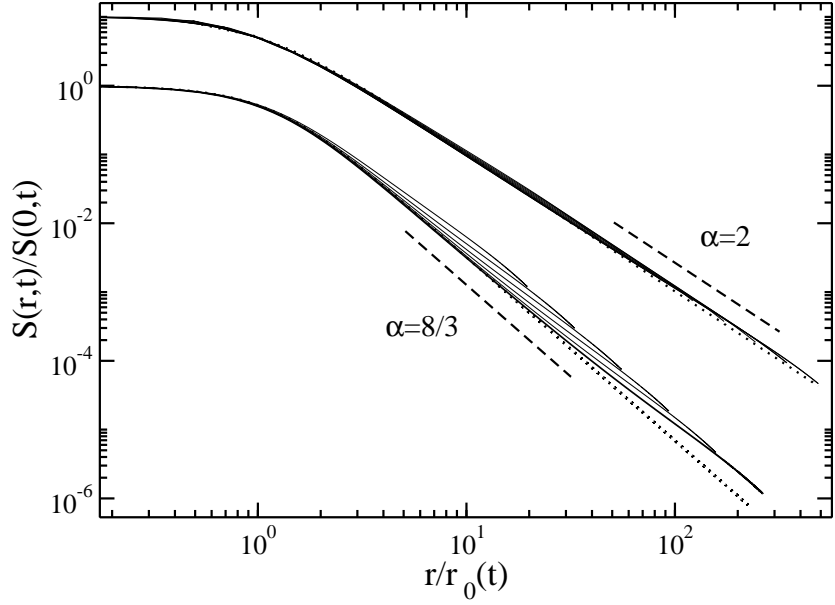}}
\caption{For $n=4$ ($\alpha_4=8/3$) and $\Theta=0.1$ (canonical
description), we plot $S(r,t)/S(0,t)$ as a function of $r/r_0(t)$,
where $r_0(t)$ is defined by Eq.~(\ref{dim10}), for different
times corresponding to central densities in the range $2.10^2\sim
4.10^5$ (bottom data collapse). This is compared to the scaling
function obtained by solving Eq.~(\ref{dim19}) numerically (dotted
line). The same is plotted in the case $n=\infty$
($\alpha_\infty=2$), for which the scaling profile is known
analytically \cite{sc}~: $S(x)/S(0)=(1+x^2)^{-1}$ (upper data
collapse). The two curves have been shifted for clarity. In the
$n=\infty$ case, the asymptotic scaling profile (dotted line) is
almost indistinguishable from the data collapse. Dashed lines have
respective slopes $-8/3$ and $-2$.  } \label{cano4}
\end{figure}

\begin{figure}
\centerline{ \psfig{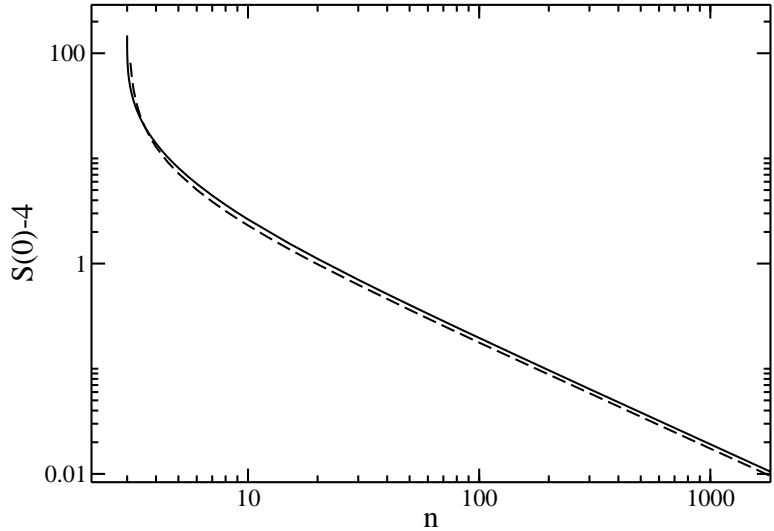}}
\caption{We plot $S(0)$ as a function of $n$ (full line), and
compare it to a simple theory explained in the text, which is
inspired by the large $D$ perturbation introduced in \cite{sc}
(dashed line). Note that $S(0)\simeq 4+{C/n}+O(n^{-2})$ for large $n$
with $C\simeq 19$.} \label{s0n}
\end{figure}

We first consider the dynamics in the canonical ensemble (fixed
$\Theta$).  In Fig.~\ref{cano32}, we show the different steps of the
formation of a self-confined polytrope of index $n=3/2<n_3=3$ similar
to a classical ``white dwarf'' star. In this range of $n$, the system
always converges to an equilibrium state (see Fig. \ref{elD3}). If
$\eta<\omega_{n}$ the equilibrium state is confined by the box
(incomplete polytrope) while for $\eta>\omega_{n}$ the density
vanishes at $R_{*}<R$ (complete polytrope). In Fig.~\ref{cano4}, we
illustrate the collapse dynamics at low temperatures for
$n=4\in[n_3,n_5]$ and $n=\infty>n_5=5$. This is compared to the
predicted scaling profiles. The convergence to scaling is slower for
$n=4$ ($\alpha=8/3$) than for $n=\infty$ ($\alpha=2$). This is
expected since, in the former case, $r_0\sim\rho_0^{-1/\alpha}$
decreases more slowly as $\alpha$ is bigger. Thus, the scaling regime
$r_0\ll 1$ is reached in a slower way.  For instance, for comparable
final densities of order $10^6$, and for the considered temperatures,
we find that the minimum $r_0$ obtained for $n=4$ is roughly 4 times
bigger than in the $n=\infty$ simulations. For $n=\infty$ and in the
large $D$ limit, we have shown in \cite{sc} that the scaling function
$S(x)$ takes the form
\begin{equation}
S(x)=\frac{\alpha}{D}\left[1+\left(1-\frac{\alpha}{2z}\right)
\left(\frac{x^2}{x_0^2}-1\right)
 \left(\frac{x^2}{x_1^2}+1\right)^{\frac{\alpha}{2}-1}\right]^{-1},
\label{snextexp}
\end{equation}
where $x_0$ is such that $S(x_0)=\alpha/D$. The quantities
$z=DS(0)/2$ and $\alpha(z)$ have been exactly calculated in this
limit. In the present case, and for a given $n$ (yielding
$\alpha_n=2n/(n-1))$, we compute $S(0)$, by assuming the above
functional form. The parameter $x_0$, $x_1$ and $S(0)$ are
numerically calculated by imposing the exact value for $S''(0)$
extracted from Eq.~(\ref{dim19}), as well as the two conditions
that $x_0$ must satisfy (see \cite{sc} for more details). The
comparison of this approximate theory with actual numerical data
is satisfactory (see Fig.~\ref{s0n}).  Note that we have been
unable  to develop a large $n$ perturbation theory in the same
spirit as the large $D$ expansion scheme derived in \cite{sc}.

\vskip 0.3cm

\begin{figure}
\centerline{\psfig{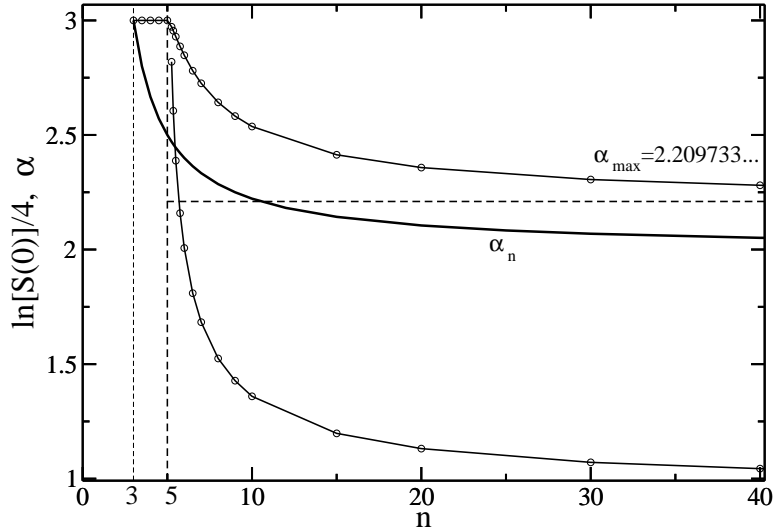}} \vskip
0.3cm \caption{We plot $\alpha_{\rm max}(D=3,n)$ as a function of
$n$ (top plot), as well as the associated value of $\frac 14\ln
S(0)$ (bottom plot). The horizontal dashed line represents the
asymptotic value of $\alpha_{\rm max}$ for $n\to+\infty$, and we
find $\alpha_{\rm max}(D=3,n)\approx\alpha_{\rm
max}(D=3,n=\infty)+C_3/n+ O(n^{-2})$ for large $n$, with $C_3\sim
2.7$. For $n\le n_{5}=5$, $\alpha_{\rm max}=D^{-}=3^{-}$ (strictly
speaking, the scaling solution associated to $\alpha=D=3$ does not
exist below $n_5=5$). We observe that $\ln S(0)\sim (n-5)^{-1/4}$
provides an excellent fit of $S(0)$ for $n\in [5,10]$.  We have
also plotted $\alpha_n=2n/(n-1)$ (thick line). The scaling
equation (\ref{dim19}) admits solutions for $\alpha_{n}\le \alpha\le
\alpha_{\rm max}$. The two curves intersect when $\alpha_{n}=D$ which
corresponds to $n=n_{3}$. There is no scaling collapse solution
for $n\leq n_3$.} \label{amaxn}\vskip 0.9cm
\end{figure}

\begin{figure}
\centerline{\psfig{figure=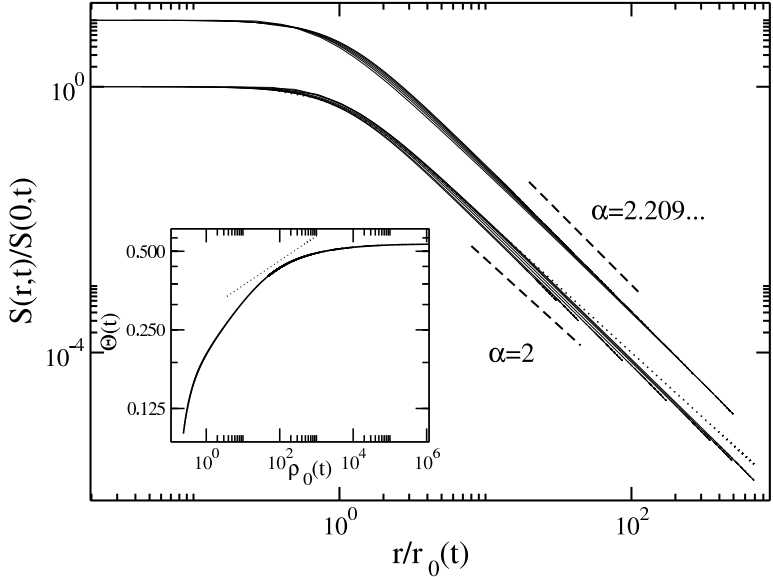,angle=0,height=9cm}}
\caption{For $n=+\infty$ and $E=-0.45$ (microcanonical
description), we plot $S(r,t)/S(0,t)$ as a function of $r/r_0(t)$
where $r_0(t)\sim\rho_0(t)^{1/\alpha}$, for times corresponding to
central densities in the range $2.10^2\sim 4.10^5$ (for
comparison, our previous simulations \cite{crs} did not exceed
$\rho_{0}\sim 1000$). We try both values $\alpha_{\infty}=2$
(bottom plot) and $\alpha=\alpha_{\rm max}=2.209733...$ (top
plot), and compare both data collapses to the associated scaling
function (dotted lines). The two curves have been shifted for
clarity.  The scaling associated to $\alpha_{\rm max}$ is clearly
more convincing than that for $\alpha=2$, especially at large
distances. However, our simulations also suggest that
$\Theta(t)\sim \rho_0(t)^{1-2/\alpha_{\rm max}}$ does not diverge
at $t_{coll}$ (see the insert where a line of slope
$1-2/\alpha_{\rm max}\approx 0.09491...$ has been drawn as a guide
to the eye), so that the asymptotic scaling should correspond to
$\alpha=2$. This apparent ``paradox'' clearly shows that the
convergence to the limit value $\alpha=\alpha_n$ is extremely
slow, suggesting an intermediate pseudo-scaling regime with $\alpha_{n}\le
\alpha\le \alpha_{\rm max}$.} \label{microsca}
\end{figure}

As explained in the previous section, the situation in the
microcanonical ensemble (where $\Theta=\Theta(t)$ evolves with
time in order to conserve energy) is less clear. Like in the case
$n=\infty$ studied in \cite{sc}, the scaling equation admits a
physical solution for any $\alpha_{n}\le \alpha\leq\alpha_{\rm
max}(D,n)$.  In Fig.~\ref{amaxn}, we plot $\alpha_{\rm
max}(D=3,n)$, as well as the corresponding value of $S(0)$, as a
function of $n$. As explained previously, it is doubtful that a
scaling actually develops with $\alpha>\alpha_n$ when the
temperature is uniform. However, a {\it pseudo-scaling} should be
observed with $\alpha\simeq \alpha_{\rm max}$. In Fig.~\ref{microsca},
we present new simulations ($D=3$, $n=\infty$) confirming that the
observed scaling dynamics is better described by
$\alpha=\alpha_{\rm max}$ than by $\alpha=2$, in the time/density
range achieved. Such a value of $\alpha$ implies that the
temperature would diverge with a small exponent
($\Theta(t)\sim\rho_0(t)^{1-2/\alpha_{{\rm max}}}$, with
$1-2/\alpha_{\rm max}=0.09491...$). However, in the range of
accessible densities $\rho_0=2.10^{-1}\sim 10^{6}$, numerical data
tend to suggest that the temperature converges to a finite value
with an infinite derivative $\frac{d\Theta}{dt}(t_{coll})=+\infty$
as $t\to t_{coll}$. This convergence of the temperature has been
observed independently by \cite{guerra}. Thus, we conclude that
the system first develops an apparent scaling with
$\alpha\lesssim\alpha_{\rm max}$, before slowly approaching the
asymptotic scaling regime with $\alpha=2$. In any case, it is
clear that if the $\alpha=2$ scaling is the relevant one, the
scaling regime is approached much more slowly than in the
canonical ensemble (compare Fig.~\ref{microsca} and
Fig.~\ref{cano4}). This is a new aspect of the inequivalence of
statistical ensembles for self-gravitating systems.

\section{Conclusion}

In this paper, we have discussed the structure and the stability of
self-gravitating polytropic spheres by using a formalism of {\it
generalized thermodynamics} \cite{Ctsallis}.  This formalism allows us
to present and organize the results in an original manner. What we
mean by generalized thermodynamics is the extension of the usual
variational principle of ordinary thermodynamics (maximization of the
Boltzmann entropy $S_{B}$ at fixed mass $M$ and energy $E$) to a
larger class of functionals (playing the role of ``generalized
entropies''). This variational problem can arise in various domains of
physics (or biology, economy,...) for different reasons. In any case,
it is relevant to develop a {\it thermodynamical analogy} and use a
vocabulary borrowed from thermodynamics (entropy, temperature,
chemical potential, caloric curve, free energy, microcanonical and
canonical ensembles,...)  even if the initial problem giving rise to
this variational problem is not directly connected to
thermodynamics. Thus, we can directly transpose the methods developed
in the context of ordinary thermodynamics (e.g., Legendre transforms,
turning point arguments, bifurcations,...) to a new context. For
example, in the present study, the maximization of Tsallis entropy at
fixed mass and energy is a condition of nonlinear dynamical stability
for stellar polytropes via the Vlasov equation and for polytropic vortices via
the Euler equation. On the other hand, the minimization of Tsallis
free energy at fixed mass is connected to the nonlinear dynamical
stability of polytropic stars via the Euler-Jeans equations. It is
also a condition of thermodynamical stability (in a generalized sense)
for self-gravitating Langevin particles experiencing anomalous
diffusion and a condition of dynamical stability for bacterial
populations. Although the formalism is the same for all these systems,
the results have a very different physical interpretation. Our results
may also have unexpected applications in other domains of physics that
we are not aware of.

On a technical point of view, we have provided the complete
equilibrium phase diagram of self-gravitating polytropic spheres for
arbitrary value of the polytropic index $n$ and space dimension
$D$. Our study, generalizing the classical study of Emden \cite{emden}
and Chandrasekhar \cite{chandra}, shows how the phase portraits
previously reported in the literature (for particular dimensions and
particular polytropic indices) connect to each other in the full
parameter space. From the geometrical structure of the generalized
caloric curves, we can immediately determine the domains of stability
of the polytropic spheres by using the turning point method
\cite{katz78}. These stability results have been confirmed by
explicitly evaluating the second order variations of entropy and free
energy. This eigenvalue method provides, in addition, the form of the
density profile that triggers the instability at the critical
points. Interestingly, this study can be performed analytically or by
using simple graphical constructions in the Milne plane. We have found
that complete stellar polytropes (with $n<n_5=(D+2)/(D-2)$ if $D>2$)
are stable for $D\le 4$ and unstable for $D>4$. On the other hand,
complete gaseous polytropes are stable for $D\le 2$ and for ($D>2$,
$n\le n_3=D/(D-2)$) and unstable for ($D>2$, $n>n_3$).  Polytropes
with index $n_{3/2}=D/2$ correspond to classical white dwarf stars
(i.e., a self-gravitating Fermi gas at $T=0$). They are self-confined
only for $D<2(1+\sqrt{2})$ and they are stable only for $D<4$. For
$D>4$, quantum mechanics is not able to prevent gravitational
collapse, even in the non-relativistic regime. In this sense, $D=4$ is
a critical dimension. Therefore, our dimension of space $2<D=3<4$ is
bounded by two critical dimensions. It seems that this remark has
never been made before.  The description of phase transitions in the
self-gravitating Fermi gas at non-zero temperature in dimension $D$
will be considered in a future article \cite{FermiD}. Other
possible extensions of our work would be to consider different
equations of state such as the modified isothermal $p=-T\ln
(1-\rho/\rho_{0})$ associated with an ``entropy'' functional
$S[\rho]=-\int \lbrace
\rho\ln\rho+(\rho_{0}-\rho)\ln (\rho_{0}-\rho)\rbrace d^{D}{\bf r}$ or
the logotropic equation of state $p=p_{c}\lbrack 1+A\ln
(\rho/\rho_{c})\rbrack$ \cite{logo} associated with
$S[\rho]=p_{c}A\int \ln\rho d^{D}{\bf r}$.

The concept of generalized thermodynamics is rigorously justified
in the case of stochastic (Langevin) particles experiencing
anomalous diffusion. This happens when the diffusion coefficient
in the Fokker-Planck equation depends on the density of particles
while the friction or drift is constant. In this paper, we have
explicitly studied the nonlinear Smoluchowski-Poisson system (for
self-gravitating Langevin particles) corresponding to a power-law
dependance of the diffusion coefficient. This particular situation
is connected to Tsallis generalized thermodynamics 
but more general Fokker-Planck equations can be constructed and
studied \cite{Ctsallis}. The connection between thermodynamical
and dynamical stability for this type of generalized Fokker-Planck
equations has been established in the general case in
\cite{Ctsallis}. The nonlinear Smoluchowski-Poisson system can
have physical applications for the chemotaxis of bacterial
populations. The collapse and aggregation of bacterial populations is
similar, in some respect, to the phenomenon of {\it core collapse} in
globular clusters (or to the Jeans instability in molecular clouds) and the
neglect of inertia is rigorously justified in biology at variance with
astrophysics. In addition, biological systems are likely to experience
anomalous diffusion so that the NSP system can provide an interesting
and relevant model for the problem of chemotaxis.  We have shown that
the solutions of the NSP system can either converge toward a complete
polytrope or an incomplete polytrope restrained by the box, or lead to
a situation of collapse.  The determination of the scaling exponent
$\alpha$ in the microcanonical ensemble (constant energy) is difficult
due to the extremely slow entry of the system in the scaling
regime. However, it seems to be given asymptotically by
$\alpha_n=2n/(n-1)$ ($\alpha=2$ for isothermal spheres) as in the
canonical ensemble (constant temperature). We expect that an exponent
$\alpha>\alpha_{n}$ will be selected when the temperature is allowed
to vary in space and time. This problem will be considered in a future study
\cite{sc3}.

\newpage

\appendix

\section{Gravitational force  in $D$ dimensions}
\label{sec_gfD}

The gravitational field produced in ${\bf r}$ by a distribution of
$N$ particles with mass $m$ in a space of dimension $D$ is
\begin{equation}
{\bf F}=-\sum_{i=1}^{N} Gm{{\bf r}-{\bf r}_{i}\over |{\bf r}-{\bf
r}_{i}|^{D}}.\label{force1}
\end{equation}
For $D=1$, the gravitational field created by a particle is
independent on the distance. Thus, an object located in $x$
experiences a force (by unit of mass) $F=Gm(N^{+}-N^{-})$, where
$N^{+}$ is the number of particles in its right ($x_{i}>x$) and
$N^{-}$ the number of particles in its left ($x_{i}<x$).

The external gravitational field created by a spherically symmetric
distribution of matter with mass $M$ is
\begin{equation}
{\bf F}=-\nabla\Phi=-{GM\over r^{D-1}}{\bf e}_{r}. \label{force2}
\end{equation}
For $D\neq 2$, the gravitational potential is
\begin{equation}
\Phi=-{GM\over (D-2) r^{D-2}}, \label{force3}
\end{equation}
where the constant of integration has been taken equal to zero
(this implies $\Phi=0$ at infinity for $D>2$). For $D=2$, we have,
\begin{equation}
\Phi=GM\ln (r/R), \label{force4}
\end{equation}
where we have taken $\Phi(R)=0$.

\section{Virial theorem in $D$ dimensions}
\label{sec_vtD}

We define the Virial of the gravitational force in dimension $D$
by
\begin{equation}
{\cal V}_{D}=\int \rho \ {\bf r}\cdot \nabla\Phi \ d^{D}{\bf r}.
\label{vtD1}
\end{equation}
For a spherically symmetric system, the Gauss theorem can be
written
\begin{equation}
{d\Phi\over dr}={GM(r)\over r^{D-1}}, \qquad M(r)=\int_{0}^{r}\rho
S_{D}r^{'D-1}dr'. \label{vtD2}
\end{equation}
Therefore, the Virial is equivalent to
\begin{equation}
{\cal V}_{D}=\int_{0}^{R}{dM\over dr}{GM(r)\over
r^{D-2}}dr={G\over 2}\int_{0}^{R}{dM^{2}\over dr}{1\over
r^{D-2}}dr. \label{vtD3}
\end{equation}
In $D=2$, one has directly
\begin{equation}
{\cal V}_{2}={GM^{2}\over 2}. \label{vtD4}
\end{equation}
If now $D\neq 2$, we obtain after an integration by parts
\begin{equation}
{\cal V}_{D}={GM^{2}\over 2R^{D-2}}+{1\over 2}(D-2)\int_{0}^{R}
{GM(r)^{2}\over r^{D-1}}dr, \label{vtD5}
\end{equation}
or, using Eq. (\ref{vtD2}),
\begin{equation}
{\cal V}_{D}={GM^{2}\over 2R^{D-2}}+{1\over
2G}(D-2)\int_{0}^{R}{\biggl ({d\Phi\over dr}\biggr
)^{2}}r^{D-1}dr. \label{vtD5a}
\end{equation}
Now, using the Poisson equation (\ref{maxent4}), the potential
energy can be written
\begin{equation}
W={1\over 2}\int \rho\Phi d^{D}{\bf r}={1\over 2 S_{D} G}\int
\Phi\Delta\Phi d^{D}{\bf r}. \label{vtD6}
\end{equation}
Integrating by parts, we obtain
\begin{equation}
W={1\over 2 S_{D} G}\biggl\lbrace \Phi(R){d\Phi\over
dr}(R)S_{D}R^{D-1}-\int_{0}^{R} \biggl ({d\Phi\over dr}\biggr
)^{2}r^{D-1}S_{D}dr\biggr\rbrace. \label{vtD7}
\end{equation}
The gravitational force and the gravitational potential at the edge of the box  are given by
Eqs. (\ref{force2}) and (\ref{force3}). Introducing these results in Eq. (\ref{vtD7}) and
comparing with Eq. (\ref{vtD5a}), we obtain
\begin{equation}
{\cal V}_{D}=-(D-2)W, \qquad D\neq 2. \label{vtD9}
\end{equation}
By using the Virial tensor method introduced by Chandrasekhar \cite{bt}, we can show that the foregoing relation remains
valid if the system is not spherically symmetric.

If now the system is in hydrostatic equilibrium, we have
\begin{equation}
\nabla p=-\rho\nabla\Phi. \label{vtD10}
\end{equation}
Inserting this relation in the Virial (\ref{vtD1}) and integrating
by parts, we get
\begin{equation}
{\cal V}_{D}=-\oint p\ {\bf r}\cdot d{\bf S}+2K, \label{vtD11}
\end{equation}
where we have used $K=(D/2)\int p d^{D}{\bf r}$. This is the
expression of the Virial theorem in its general form. Assuming now
that $p_{b}$ is uniform on the domain boundary (which is true at
least for a spherically symmetric system), we have
\begin{equation}
\oint p \ {\bf r}\cdot d{\bf S}=p_{b}\oint  {\bf r}\cdot d{\bf
S}=p_{b}\int \nabla\cdot {\bf r}\  d^{D}{\bf r}=p_{b}DV_{D}R^{D}.
\label{vtD12}
\end{equation}
Therefore, for a spherically symmetric system, the Virial theorem
reads
\begin{equation}
2K-{\cal V}_{D}=p(R)DV_{D}R^{D}, \label{vtD13}
\end{equation}
where ${\cal V}_{D}$ is given by Eqs. (\ref{vtD4}) and
(\ref{vtD9}). 

It is interesting to consider a direct application
of these results. In $D=2$, the Virial theorem reads
\begin{equation}
2K-{GM^{2}\over 2}=2\pi R^{2}p(R). \label{vtD14}
\end{equation}
For an isothermal gas $K=MT$ so that
\begin{equation}
2MT-{GM^{2}\over 2}=2\pi R^{2}p(R). \label{vtD15}
\end{equation}
From this relation, we conclude that there exists an equilibrium
solution with $p(R)=0$ at $T_{c}=GM/4$. We therefore recover the
critical temperature of an isothermal self-gravitating gas in
two-dimensions (see, e.g., \cite{sc}). At $T=T_{c}$, the density
profile is a Dirac peak so that $p(R)=0$. For $T<T_{c}$, there is no
static solution and the system undergoes a gravitational
collapse. This collapse was studied in \cite{sc} with the
Smoluchowski-Poisson system.

\section{Special properties of $n_{3/2}$ polytropes}
\label{sec_wd}

In this Appendix, we consider polytropes with index $n_{3/2}=D/2$ in a
space of dimension $D\ge 4$.  They correspond to $D$-dimensional
``white dwarf'' stars. According to Eq. (\ref{min1}), the curve
$\eta(\alpha)$ is extremum for
\begin{equation}
u_0={D(D-4)\over D-2}.\label{wd1}
\end{equation}
It is easy to check that this particular value is also solution of
Eq. (\ref{min5}) determining maxima of $\Lambda(\alpha)$. We conclude
therefore that the functions $\eta(\alpha)$ and $\Lambda(\alpha)$
achieve extremal values for the same values of $\alpha$ in the series
of equilibria. This implies that the generalized caloric curve
$\eta(\Lambda)$ of $n_{3/2}$ polytropes displays {\it angular points}
(see Figs. \ref{elD4.5part1} and \ref{LHpD5.1}).

\section{Second order variations of generalized entropy and free energy}
\label{sec_so}

According to Eq. (\ref{maxent17}), the variations of entropy up to
second order are
\begin{equation}
\delta S={D-2n\over 2}\biggl (\beta\int \delta p d^{D}{\bf
r}+\delta\beta\int pd^{D}{\bf r}+\int \delta\beta\delta p
d^{D}{\bf r}\biggr ). \label{so1}
\end{equation}
On the other hand, according to the polytropic equation of state
(\ref{maxent12}), we have
\begin{equation}
\delta p=\gamma{\delta\rho\over\rho}p+{\gamma p\over
n}{(\delta\rho)^{2}\over 2\rho^{2}}+{\delta K\over K}p+\gamma
{\delta K\over K}{\delta\rho\over\rho}p. \label{so2}
\end{equation}
From Eqs. (\ref{maxent9}) and (\ref{maxent13}), 
\begin{equation}
K\sim \beta^{D-2n\over 2n}, \label{so3}
\end{equation}
so that to second order
\begin{equation}
{\delta K\over K}={D-2n\over
2n}{\delta\beta\over\beta}+{(D-2n)(D-4n)\over 8n^{2}}\biggl
({\delta\beta\over\beta}\biggr )^{2}. \label{so4}
\end{equation}
Inserting Eqs. (\ref{so4}) and (\ref{so2}) in Eq. (\ref{so1}), we
get
\begin{eqnarray}
\delta S={D-2n\over 2}\biggl\lbrack \beta\gamma
\int{\delta\rho\over\rho}pd^{D}{\bf r}+{\beta\gamma\over 2n} \int
{p\over\rho^{2}}(\delta\rho)^{2}d^{D}{\bf r}+{D\over 2n}
\int p\delta\beta d^{D}{\bf r}\nonumber\\
+{\gamma D\over 2n}\delta\beta\int {\delta\rho\over\rho}pd^{D}{\bf
r}+{D(D-2n)\over 8n^{2}}{(\delta\beta)^{2}\over \beta}\int p
d^{D}{\bf r}\biggr\rbrack. \label{so5}
\end{eqnarray}
Now, the conservation of energy (\ref{maxent16}) implies that
\begin{equation}
0=\delta E={D\over 2}\int \delta p d^{D}{\bf r}+{1\over 2}\int
\delta\rho\delta\Phi d^{D}{\bf r}+\int\Phi \delta\rho d^{D}{\bf
r}. \label{so6}
\end{equation}
Inserting Eqs. (\ref{so2}) and (\ref{so4}) in Eq. (\ref{so6}), we
obtain
\begin{eqnarray}
{D-2n\over 2}\biggl \lbrack{\gamma D\over D-2n}\beta \int
{\delta\rho\over\rho}p d^{D}{\bf r}+{\gamma D\over 2n (D-2n)}
\beta \int p{(\delta\rho)^{2}\over \rho^{2}}d^{D}{\bf r}+
{D\over 2n}\int \delta\beta p d^{D}{\bf r}\nonumber\\
+{D(D-4n)\over 8n^{2}}{(\delta\beta)^{2}\over\beta}\int p
d^{D}{\bf r}+ {\gamma D\over 2n}\delta\beta\int {\delta\rho\over
\rho}p d^{D}
{\bf r}\biggr \rbrack \nonumber\\
+{1\over 2}\beta \int \delta\rho\delta\Phi d^{D}{\bf r}+\beta
\int\Phi \delta\rho d^{D}{\bf r}=0. \label{so7}
\end{eqnarray}
Subtracting this relation from Eq. (\ref{so5}), we get
\begin{eqnarray}
\delta S=-\beta\gamma n\int{\delta\rho\over\rho}pd^{D}{\bf r}-
{1\over 2}\beta\gamma\int
p{(\delta\rho)^{2}\over\rho^{2}}d^{D}{\bf r}+
{D(D-2n)\over 8n}{(\delta\beta)^{2}\over \beta}\int p d^{D}{\bf r}
\nonumber\\
-{1\over 2}\beta \int \delta\rho\delta\Phi d^{D}{\bf r}-\beta
\int\Phi \delta\rho d^{D}{\bf r}. \label{so8}
\end{eqnarray}
Now, to first order, Eq. (\ref{so7}) yields
\begin{equation}
{\delta\beta\over\beta}=-{4n\over D(D-2n)}{\int \Phi \delta\rho
d^{D}{\bf r}+{D\over 2}\gamma \int {\delta\rho\over\rho}p
d^{D}{\bf r}\over \int p d^{D}{\bf r}}. \label{so9}
\end{equation}
Substituting this relation in Eq. (\ref{so8}), we find that the
second order variations of entropy are given by Eq.
(\ref{stab10}). To compute the second order variations of free
energy $F=E-TS$, we can use Eqs. (\ref{so5}) and (\ref{so6})
with $\delta\beta=0$. This  yields Eq. (\ref{stab1}).

\section{Some useful identities}
\label{sec_id}

In this Appendix, we establish the identities
(\ref{stab19})-(\ref{stab20b}) that are needed in the stability
analysis of Sec. \ref{sec_stab}. Using an integration by parts, we
have
\begin{equation}
\int_{0}^{\alpha}\theta'\xi^{D}\theta^{n}d\xi=\int_{0}^{\alpha}\xi^{D}{d\over
d\xi}\biggl ({\theta^{n+1}\over n+1}\biggr )d\xi={1\over
n+1}\alpha^{D}\theta^{n+1}(\alpha)-{D\over
n+1}\int_{0}^{\alpha}\xi^{D-1}\theta^{n+1}d\xi. \label{id1}
\end{equation}
Using the Lane-Emden equation (\ref{emden2}) and integrating by
parts, we obtain
\begin{equation}
\int_{0}^{\alpha}\theta^{n+1}\xi^{D-1}d\xi=-\alpha^{D-1}
\theta(\alpha)\theta'(\alpha)+\int_{0}^{\alpha}(\theta')^{2}\xi^{D-1}d\xi.
\label{id2}
\end{equation}
Using the relation
\begin{equation}
\int_{0}^{\alpha}{\xi^{1+D\over 2}\theta'\over \xi}{d\over
d\xi}\biggl (\xi^{1+D\over 2}\theta'\biggr
)d\xi=\alpha^{D}\theta'(\alpha)^{2}-\int_{0}^{\alpha}{\xi^{1+D\over
2}\theta'\over \xi}{d\over d\xi}\biggl (\xi^{1+D\over
2}\theta'\biggr )d\xi+\int_{0}^{\alpha}\xi^{D-1}(\theta')^{2}d\xi,
\label{id3}
\end{equation}
which results from a simple integration by parts, we get
\begin{equation}
\int_{0}^{\alpha}\xi^{D-1}(\theta')^{2}d\xi=
-\alpha^{D}\theta'(\alpha)^{2}+2\int_{0}^{\alpha}\xi^{D-1\over
2}{d\over d\xi}\biggl (\xi^{1+D\over 2}\theta'\biggr )\theta'd\xi,
\label{id4}
\end{equation}
or, equivalently,
\begin{equation}
D\int_{0}^{\alpha}\xi^{D-1}(\theta')^{2}d\xi=\alpha^{D}
\theta'(\alpha)^{2}-2\int_{0}^{\alpha}\xi^{D}\theta''\theta'
d\xi. \label{id5}
\end{equation}
Using the Lane-Emden equation (\ref{emden2}), we find that
\begin{equation}
(D-2)\int_{0}^{\alpha}\xi^{D-1}(\theta')^{2}d\xi=-\alpha^{D}
\theta'(\alpha)^{2}-2\int_{0}^{\alpha}\xi^{D}\theta'
\theta^{n} d\xi. \label{id6}
\end{equation}
We have three equations (\ref{id1}), (\ref{id2}) and (\ref{id6})
for three unknown integrals. Solving this system of algebraic
equations and introducing the Milne variables (\ref{milne1}), we
obtain the identities (\ref{stab19})-(\ref{stab20b}).

\section{Dynamical stability of gaseous spheres}
\label{sec_stabilo}

In this section, we assume $D>2$. According to Eq. (\ref{newnrj}), the
energy of a polytropic star at equilibrium (${\bf u}={\bf 0}$) can be
written
\begin{equation}
{\cal W}={2n\over D}K+W, \label{add1}
\end{equation}
where $K$ is the kinetic energy and $W$ the potential energy.
Now, for a complete polytrope ($p_{b}=0$), the Virial theorem reads
\begin{equation}
2K+(D-2)W=0. \label{add2}
\end{equation}
Combining the foregoing relations, we get
\begin{equation}
{\cal W}=\biggl (1-{n\over n_3}\biggr )W. \label{add3}
\end{equation}
According to Poincar\'e's theorem, a gaseous star with ${\cal W}>0$ is
unstable \cite{chandra}. For polytropic stars, this condition is equivalent to
$n>n_3$. 

More generally, the internal energy of a mass $dm$ of gas at
temperature $T$ is $dU=C_v dm T$. Its kinetic energy is $dK={D\over
2}dm RT={D\over 2}(C_p-C_v)dm T$ where $R$ is the constant of perfect
gases and $C_{v}$, $C_{p}$ are the specific heats at constant volume
and constant pressure, respectively. Thus, we get
\begin{equation}
U={2\over D(\gamma-1)}K, \label{add4}
\end{equation}
where $\gamma=C_p/C_v$. For a monoatomic gas,
$\gamma=(D+2)/D$ and $U=K$. Using the Virial theorem (\ref{add2}), the total
energy of the star ${\cal W}=U+W$ can be written
\begin{equation}
{\cal W}={D\gamma-2(D-1)\over D(\gamma-1)}W. \label{add5}
\end{equation}
The star is unstable for $\gamma<\gamma_{crit}=2(D-1)/D$. For $D=3$, we
recover the well-known result $\gamma_{crit}=4/3$. For a polytropic gas,
we recover the result  $\gamma_{crit}=1+1/n_{3}$.

\end{document}